%% file: main.tex
\documentclass{lmcs}
\pdfoutput=1
\usepackage[utf8]{inputenc}

\usepackage{lastpage}
\lmcsdoi{20}{2}{4}
\lmcsheading{}{\pageref{LastPage}}{}{}%
{Jan.~15,~2021}{Apr.~18,~2024}{}

\input{macros}

\title{A Strong Bisimulation for a Classical Term Calculus}

\author[E.~Bonelli]{Eduardo Bonelli\lmcsorcid{0000-0003-1856-2856}}[a]
\address{Stevens Institute of Technology, USA}
\email{eabonelli@gmail.com}

\author[D.~Kesner]{Delia Kesner\lmcsorcid{0000-0003-4254-3129}}[b]
\address{Universit\'e Paris Cit\'e, CNRS, IRIF}
\email{kesner@irif.fr}

\author[A.~Viso]{Andr\'es Viso\lmcsorcid{0000-0002-6822-8453}}[c]
\address{Inria, France}
\email{andres-ezequiel.viso@inria.fr}

\begin{document}

\input{abstract}
\maketitle

\input{introduction-v4}
\input{preliminary}
\input{lambda-mu}
\input{lambda-m-v2}
\input{equivalence}

\input{new}
\input{conclusion}

\bibliographystyle{alphaurl}
\bibliography{biblio}

\newpage
\appendix
\input{app-confluencia}

\input{appendix-meaningful-termina}
\input{appendix-equivalencia-v2}
\input{appendix-bisimulation}
\input{appendix-correspondence}

\end{document}


%% file: macros.tex
\usepackage[english]{babel}
\usepackage{diagbox}

\usepackage{amssymb}
\usepackage{appendix}
\usepackage{booktabs}
\usepackage{color}
\usepackage{dashbox}
\usepackage{multirow}
\usepackage{prooftree}
\usepackage{stmaryrd}
\usepackage{tikz}
\usetikzlibrary{arrows}
\usetikzlibrary{automata}
\usetikzlibrary{calc}
\usetikzlibrary{cd}
\usetikzlibrary{decorations.markings}
\usetikzlibrary{decorations.pathmorphing}
\usetikzlibrary{positioning}
\usepackage{tikz-qtree}

\usepackage{xargs}
\usepackage{xifthen}

\tikzset{
  negated/.style={decoration={markings, mark= at position 0.5 with {\node[transform shape] (tempnode) {$/$};}}, postaction={decorate}},
  Rightarrow/.style={double equal sign distance, double distance=4pt, >={Implies}, ->},
  RightarrowDashed/.style={double equal sign distance, double distance=4pt, >={Implies}, ->, dashed},
  Rrightarrow/.style={-, preaction={draw,Rightarrow}},
  RrightarrowDashed/.style={-, dashed, preaction={draw,RightarrowDashed}}
}

\newcommand{\dashnode}[2]{\dashbox[#1][c]{\parbox{#1}{\centering {#2}}}}

\capturecounter{conj}
\capturecounter{cor}
\capturecounter{defi}
\capturecounter{exa}
\capturecounter{lem}
\capturecounter{prop}
\capturecounter{rem}
\capturecounter{thm}

\definecolor{metacolor}{rgb}{1,0.5,0}
\definecolor{explcolor}{rgb}{0,0.5,1}

\newcommand{\cf}{\emph{cf.}~}
\newcommand{\eg}{\emph{e.g.}~}
\newcommand{\ie}{\emph{i.e.}~}
\newcommand{\ih}{\emph{i.h.}~}

\newcommand{\coloneq}{\ensuremath{:=}}
\newcommand{\Coloneq}{\ensuremath{:\coloneq}}

\newcommand{\eqalpha}{\ensuremath{=_{\alpha}}}

\newcommand{\eqdef}{\ensuremath{\triangleq}}

\newcommand{\llbrace}{\{\!\!\{}
\newcommand{\rrbrace}{\}\!\!\}}

\newcommand{\set}[1]{\ensuremath{\{{#1}\}}}

\newcommand{\sequ}[2][]{\ensuremath{{{#1}\vdash{#2}}}}

\newcommand{\calcLambda}{\ensuremath{\lambda}}

\newcommand{\calcLambdaM}{\texorpdfstring{\ensuremath{\Lambda M}}{LM}}
\newcommand{\calcLambdaMu}{\texorpdfstring{\ensuremath{\lambda\mu}}{lm}}
\newcommand{\calcLambdaMuR}{\ensuremath{\lambda\mu\mathtt{s}}}

\newcommand{\calcLambdaMuT}{\ensuremath{\lambda\mu\widetilde{\mu}}}

\newcommand{\formneg}[1]{\ensuremath{{#1}^\iformneg}}

\newcommand{\iformneg}{\ensuremath{\bot}}

\newcommand{\rB}{\ensuremath{\mathtt{dB}}}
\newcommand{\rBeta}{\ensuremath{\mathtt{B}}}
\newcommand{\rbeta}{\ensuremath{\beta}}
\newcommand{\rC}{\ensuremath{\mathtt{C}}}
\newcommand{\rcan}{\ensuremath{\mathcal{P}}}

\newcommand{\rCnl}{\ensuremath{\mathtt{Cnl}}}
\newcommand{\rcomp}{\ensuremath{\mathtt{Comp}}}

\newcommand{\rlm}{\ensuremath{\lambda\mu}}
\newcommand{\rLM}{\ensuremath{\Lambda M}}

\newcommand{\rM}{\ensuremath{\mathtt{dM}}}
\newcommand{\rmu}{\ensuremath{\mu}}

\newcommand{\rN}{\ensuremath{\mathtt{N}}}
\newcommand{\rNnl}{\ensuremath{\mathtt{Nnl}}}
\newcommand{\rname}{\ensuremath{\mathtt{Name}}}

\newcommand{\rR}{\ensuremath{\mathtt{R}}}
\newcommand{\rRm}{\ensuremath{\mathtt{R}^{\bullet}}}
\newcommand{\rRnl}{\ensuremath{\mathtt{Rnl}}}
\newcommand{\rS}{\ensuremath{\mathtt{S}}}

\newcommandx{\reduce}[2][1=,2=]{\ensuremath{\rightarrow_{#1}^{#2}}}

\newcommand{\reducemany}[1][]{\ensuremath{\twoheadrightarrow_{#1}}}
\newcommand{\reducemean}[1][]{\ensuremath{\rightsquigarrow_{#1}}}

\newcommand{\rrule}[1]{\ensuremath{\mathrel{\mapsto_{#1}}}}

\newcommand{\ruleEqlauexAbs}{\eqlauex[1]}
\newcommand{\ruleEqlauexApp}{\eqlauex[2]}
\newcommand{\ruleEqlauexCom}{\eqlauex[9]}
\newcommand{\ruleEqlauexCont}{\eqlauex[3]}
\newcommand{\ruleEqlauexPopPop}{\eqlauex[6]}
\newcommand{\ruleEqlauexPushPop}{\eqlauex[5]}
\newcommand{\ruleEqlauexPushPush}{\eqlauex[4]}
\newcommand{\ruleEqlauexRho}{\eqlauex[8]}
\newcommand{\ruleEqlauexTheta}{\eqlauex[7]}

\newcommand{\ruleEqnewAbs}{\eqpure[1]}
\newcommand{\ruleEqnewApp}{\eqpure[2]}
\newcommand{\ruleEqnewCont}{\eqpure[3]}
\newcommand{\ruleEqnewPopPop}{\eqpure[6]}
\newcommand{\ruleEqnewPushPop}{\eqpure[5]}
\newcommand{\ruleEqnewPushPush}{\eqpure[4]}
\newcommand{\ruleEqnewRenPop}{\eqpure[9]}
\newcommand{\ruleEqnewRenPush}{\eqpure[10]}
\newcommand{\ruleEqnewRenRen}{\eqpure[8]}

\newcommand{\ruleEqnewTheta}{\eqpure[7]}

\newcommand{\ruleEqregAbs}{\eqregnier[1]}
\newcommand{\ruleEqregApp}{\eqregnier[2]}

\newcommand{\ruleEqregexAbs}{\eqregex[1]}
\newcommand{\ruleEqregexApp}{\eqregex[2]}
\newcommand{\ruleEqregexCom}{\eqregex[3]}

\newcommand{\ruleEqsigExSubs}{\eqsigma[\mathtt{exsubs}]}
\newcommand{\ruleEqsigExRepl}{\eqsigma[\mathtt{exrepl}]}
\newcommand{\ruleEqsigExRen}{\eqsigma[\mathtt{exren}]}
\newcommand{\ruleEqsigPopPop}{\eqsigma[\mathtt{ppop}]}

\newcommand{\ruleEqsigTheta}{\eqsigma[\theta]}

\newcommand{\Command}[1]{\ensuremath{{\mathbb{C}_{#1}}}}
\newcommand{\CommandSort}{\ensuremath{\mathsf{c}}}
\newcommand{\genericobjects}{\ensuremath{\mathcal{O}}}

\newcommand{\Object}[1]{\ensuremath{{\mathbb{O}_{#1}}}}

\newcommandx{\PhiEqup}[1][1=\emptyset]{\ensuremath{\varPhi_{\eqtypeup[#1]}}}

\newcommandx{\PhiSubup}[1][1=\emptyset]{\ensuremath{\varPhi_{\subtypeup[#1]}}}

\newcommand{\Term}[1]{\ensuremath{{\mathbb{T}_{#1}}}}

\newcommand{\TermSort}{\ensuremath{\mathsf{t}}}

\newcommand{\TermName}{\ensuremath{{\mathbb{N}}}}
\newcommand{\TermVariable}{\ensuremath{{\mathbb{V}}}}

\newcommand{\R}{\ensuremath{\mathcal{R}}}
\renewcommand{\S}{\ensuremath{\mathcal{S}}}

\newcommand{\setNF}[1][]{\ensuremath{\mathcal{N\!F}_{#1}}}

\newcommand{\exrepl}[3][]{\ensuremath{{\color{explcolor}\llbracket}\change[#1]{#2}{#3}{\color{explcolor}\rrbracket}}}

\newcommand{\exsubs}[2]{\ensuremath{{\color{explcolor}[}\change{#1}{#2}{\color{explcolor}]}}}

\newcommand{\itermabs}{\ensuremath{\lambda}}

\newcommand{\itermapp}{\ensuremath{\,}}

\newcommand{\itermconc}{\ensuremath{::}}
\newcommand{\itermcont}{\ensuremath{\mu}}

\newcommand{\itermpush}{\ensuremath{\cdot}}

\newcommand{\termabs}[3][]{\ensuremath{\itermabs_{#1}{#2}.{#3}}}

\newcommand{\termapp}[2]{\ensuremath{{#1}\itermapp{#2}}}

\newcommand{\termconc}[2]{\ensuremath{{#1}\itermconc{#2}}}

\newcommand{\termcont}[2]{\ensuremath{\itermcont{#1}.{#2}}}

\newcommand{\termname}[2]{\ensuremath{[{#1}]\,{#2}}}

\newcommand{\termpush}[2]{\ensuremath{{#1}\itermpush{#2}}}
\newcommand{\termrepl}[4][]{\ensuremath{{#4}\exrepl[#1]{#2}{#3}}}
\newcommand{\ntermren}[3]{{\ensuremath{[{#2}]\,\itermcont{#1}.}}{#3}}

\newcommand{\termsubs}[3]{\ensuremath{{#3}\exsubs{#1}{#2}}}
\newcommand{\termvar}[1]{\ensuremath{{#1}}}

\newcommand{\ctxt}[1]{\ensuremath{\mathtt{#1}}}
\newcommand{\ctxtapply}[2]{\ensuremath{{#1}\langle{#2}\rangle}}

\newcommand{\itypedata}{\ensuremath{\mathbin@}}

\newcommandx{\deriv}[3][1=,3=]{\ensuremath{{#1}\rhd_{#3}{#2}}}

\newcommand{\envdis}[2]{\ensuremath{{#1};{#2}}}

\newcommandx{\envsum}[3][1=,3=]{\ensuremath{{#1}\mathrel{+_{#3}}{#2}}}

\newcommand{\bn}[1]{\ensuremath{\funcapply{\mathsf{bn}}{#1}}}
\newcommand{\bv}[1]{\ensuremath{\funcapply{\mathsf{bv}}{#1}}}

\newcommand{\fcan}[1]{\ensuremath{\funcapply{\mathcal{P}}{#1}}}

\newcommand{\change}[3][]{\ensuremath{{#2}\setminus^{\!{#1}}{#3}}}

\newcommand{\fexp}[1]{\ensuremath{\funcapply{\mathsf{e}}{#1}}}

\newcommand{\fsize}[1]{\ensuremath{\funcapply{\mathsf{sz}}{#1}}}

\newcommand{\cmeas}[1]{\ensuremath{\llparenthesis{#1}\rrparenthesis}}

\newcommand{\funcapply}[2]{\ensuremath{{#1}({#2})}}

\newcommand{\fc}[2]{\ensuremath{\funcapply{\mathsf{fc}}{{#1},{#2}}}}
\newcommand{\fn}[2][]{\ensuremath{\funcapply{\mathsf{fn}_{#1}}{#2}}}
\newcommand{\fv}[2][]{\ensuremath{\funcapply{\mathsf{fv}_{#1}}{#2}}}

\newcommand{\nf}[2][]{\ensuremath{\funcapply{\mathsf{nf}_{#1}}{#2}}}

\newcommand{\repl}[3][]{\ensuremath{{\color{metacolor}\llbrace}\change[#1]{#2}{#3}{\color{metacolor}\rrbrace}}}
\newcommand{\replapply}[2]{\ensuremath{{#2}}{#1}}

\newcommand{\ren}[2]{\ensuremath{{\color{metacolor}\{}\change{#1}{#2}{\color{metacolor}\}}}}
\newcommand{\renapply}[2]{\ensuremath{{#2}{#1}}}

\newcommand{\subs}[2]{\ensuremath{{\color{metacolor}\{}\change{#1}{#2}{\color{metacolor}\}}}}
\newcommand{\subsapply}[2]{\ensuremath{{#2}{#1}}}

\newcommand{\toLMR}[1]{\ensuremath{{\vphantom{(}{#1}\vphantom{)}}^{\downarrow}}}

\newcommandx{\toPPN}[4][1=,3=,4=]{\ensuremath{{\vphantom{(}{#2}\vphantom{)}}^{\lozenge\ifthenelse{\isempty{#3}}{}{,{#3}}\ifthenelse{\isempty{#4}}{}{,{#4}}}_{#1}}}

\newcommand{\ipncut}{\ensuremath{\mathit{cut}}}

\newlength{\PNdist}
\newlength{\PNmarg}
\newlength{\PNsize}
\tikzstyle{PNcir}=[circle,draw=black,inner sep=0pt,text width=\PNsize]
\tikzstyle{PNsub}=[rectangle,draw=black,inner sep=5pt,minimum height=\PNsize,rounded corners=8pt]

\newcommand{\pncut}[2][]{\pnnode[#1]{#2}{\ipncut}}

\newcommand{\pnnode}[3][]{\node[PNcir] (#2) [#1] {}; \node[#1] {#3}}

\newcommandx{\pnbox}[7][4=0,5=0,6=0,7=0]{\draw[dashed] let \p1=(#1), \p2=(#2), \p3=(#3) in (\x2-\PNmarg+#4,\y1+#5) -- (\x3+\PNmarg+#6,\y1+#5) -- (\x3+\PNmarg+#6,\y3+\PNmarg+#7) -- (\x2-\PNmarg+#4,\y2+\PNmarg+#7) -- cycle}

\tikzstyle{PNardot}=[dashed,rounded corners=8pt]
\tikzstyle{PNarrow}=[rounded corners=8pt]
\tikzstyle{PNarout}=[-{triangle 45},rounded corners=8pt]
\tikzstyle{PNarold}=[-{open triangle 45},rounded corners=8pt]

\newcommand{\sigmalauex}{\tilde{\sigma}}
\newcommand{\sigmalaurent}{\texorpdfstring{\ensuremath{\sigma}}{sigma}}
\newcommand{\sigmalaurentrho}{\sigmalaurent_8}
\newcommand{\sigmalaurenttheta}{\sigmalaurent_7}

\newcommand{\sigmaregnier}{\sigma}
\newcommand{\sigmaregex}{\tilde{\sigma}}

\newcommand{\eqlauex}[1][]{\ensuremath{\simeq_{\sigmalauex_{#1}}}}
\newcommand{\eqlaurent}[1][]{\ensuremath{\simeq_{\sigmalaurent_{#1}}}}
\newcommand{\eqpure}[1][]{\ensuremath{\simeq_{\tau_{#1}}}}
\newcommand{\eqnew}[1][]{\ensuremath{\simeq_{\tau_{#1}}}}
\newcommand{\eqregex}[1][]{\ensuremath{\simeq_{\sigmaregex_{#1}}}}
\newcommand{\eqregnier}[1][]{\ensuremath{\simeq_{\sigmaregnier_{#1}}}}
\newcommand{\eqsigma}[1][]{\ensuremath{\simeq_{#1}}}

\newcommand{\eqname}[1]{{\scriptstyle (\mathsf{#1})}}

\newcommand{\subtype}{\ensuremath{\preceq}}
\newcommand{\eqtype}{\ensuremath{\simeq}}

\newcommandx{\subtypeup}[1][1=\emptyset]{\ensuremath{\subtype^{#1}_{\mathfrak{n}}}}
\newcommandx{\eqtypeup}[1][1=\emptyset]{\ensuremath{\eqtype^{#1}_{\mathfrak{n}}}}

\newcommand{\al}{\alpha}
\newcommand{\Gam}{\Gamma}

\newcommand{\id}{I}
\newcommand{\deft}[1]{{\bf #1}}

\newcommand{\plain}{plain}
\newcommand{\Plain}{Plain}


%% file: abstract.tex
\begin{abstract}
When translating a term calculus into a graphical formalism many inessential
details are abstracted away. In the case of $\calcLambda$-calculus translated
to proof-nets, these inessential details are captured by a notion of
equivalence on $\calcLambda$-terms known as $\eqregnier$-equivalence, in both
the intuitionistic (due to Regnier) and classical (due to Laurent) cases. The
purpose of this paper is to uncover a strong bisimulation behind
$\eqregnier$-equivalence, as formulated by Laurent for Parigot's
$\calcLambdaMu$-calculus. This is achieved by introducing a
relation $\eqsigma$, defined over a revised presentation of
$\calcLambdaMu$-calculus we dub $\calcLambdaM$.

More precisely, we first identify the reasons behind Laurent's
$\eqlaurent$-equivalence on $\calcLambdaMu$-terms failing to be a strong
bisimulation. Inspired by Laurent's \emph{Polarized Proof-Nets}, this leads us
to distinguish multiplicative and exponential reduction steps on terms. Second,
we enrich the syntax of $\calcLambdaMu$ to allow us to track the exponential
operations. These technical ingredients pave the way towards a
strong bisimulation for the classical case. We introduce a calculus
$\calcLambdaM$ and a relation $\eqsigma$ that we show to be a strong
bisimulation with respect to reduction in $\calcLambdaM$, \ie two
$\eqsigma$-equivalent terms have the exact same reduction semantics, a result
which fails for Regnier's $\eqregnier$-equivalence in $\calcLambda$-calculus as
well as for Laurent's $\eqlaurent$-equivalence in $\calcLambdaMu$. Although
$\eqsigma$ is formulated over an enriched syntax and hence is not strictly
included in Laurent's $\eqlaurent$, we show how it can be seen as a restriction
of it.
\end{abstract}


%% file: introduction-v4.tex
\section{Introduction}
\label{s:control:introduction}

An important topic in the study of
programming language theories is unveiling structural similarities
between expressions. They are widely known as
\textit{structural equivalences}; equivalent expressions behaving
exactly in the same way. Process calculi are a rich source of
examples. In CCS, expressions stand for processes in a concurrent
system. For example, $P\parallel Q$ denotes the parallel composition
of processes $P$ and $Q$. Structural equivalence includes equations
such as the one stating that $P\parallel Q$ and $Q\parallel P$ are
equivalent. This minor reshuffling of subexpressions has little
impact on the behavior of the overall expression: structural
equivalence is a \emph{strong bisimulation} for process
reduction.

This paper is concerned with such notions of reshuffling of
expressions in \textit{$\lambda$-calculi with control operators}. The
induced notion of structural equivalence, in the sequel $\simeq$,
should identify terms having exactly the same reduction semantics
too.  Stated equivalently, $\simeq$ 
should be a strong bisimulation with respect to reduction in
these calculi. This means that $\simeq$ should be symmetric  and
moreover  $o \simeq p$ and $o \rightsquigarrow o'$ should imply the
existence of $p'$ such that $p \rightsquigarrow p'$ and $o' \simeq
p'$, where $\rightsquigarrow$ denotes some given notion of reduction
for control operators. Graphically,
\begin{equation}
\begin{tikzcd}[ampersand replacement=\&]
o \arrow[rightsquigarrow]{d}[anchor=north,left]{}
  \&[-25pt] \simeq
  \&[-25pt] p \arrow[densely dashed,rightsquigarrow]{d}[anchor=north,left]{} \\ 
o'  
  \&[-25pt] \simeq
  \&[-25pt] p'
\end{tikzcd}
\label{eq:example:bisimulation}
\end{equation}
It is worth mentioning that we are not after a general theory
  of program equivalence. On the one hand, not all terms having the same reduction semantics are
  identified, only those resulting from reshuffling in the sense made
  precise below. On the other hand, there are terms that do not have the same
  reduction semantics but would still be considered to ``behave in the
  same way'' (\eg~(\ref{eq:example:permutation:lambda}) below).  In particular, our proposed notion of equivalence is not a bisimilarity:  there are terms that have the same reduction behavior
  but are not related by our $\eqsigma$-equivalence.

Before addressing $\lambda$-calculi with control operators, we
  comment on the state of affairs in the $\lambda$-calculus.
Formulating structural equivalences
for  the $\lambda$-calculus is hindered by the
sequential (left-to-right) orientation in which expressions are
written. Consider for example the terms 
$\termapp{(\termabs{x}{\termapp{(\termabs{y}{t})}{u}})}{v}$
and $\termapp{\termapp{(\termabs{x}{\termabs{y}{t}})}{v}}{u}$.
They seem to have the same redexes, only permuted, similar to the situation captured by the
above mentioned CCS equation. A closer look, however, reveals that
this is not entirely correct. The former has two redexes (one
indicated below by underlining and another by overlining) and the latter has only
one (underlined):
\begin{equation}
\underline{\termapp{(\termabs{x}{\overline{\termapp{(\termabs{y}{t})}{u}}})}{v}}
\mbox{ and }
\termapp{\underline{\termapp{(\termabs{x}{(\termabs{y}{t})})}{v}}}{u}
\label{eq:example:permutation:lambda}
\end{equation}
The overlined redex on the left-hand side is not visible on the
right-hand side; it will only reappear, as a newly \emph{created}
redex, once the underlined redex is computed. Despite the fact that
the syntax gets in the way, Regnier~\cite{Regnier94} proved
that these terms behave in \emph{essentially} the same way. More
precisely, he introduced a structural equivalence for $\lambda$-terms,
known as \emph{$\sigma$-equivalence} and proved that
$\sigma$-equivalent terms have head, leftmost, perpetual and, more
generally, maximal reductions of the same length. However, the
mismatch between the terms in (\ref{eq:example:permutation:lambda}) is
unsatisfying since there clearly seems to be an underlying strong
bisimulation, which is not showing itself due to a notational
shortcoming. It turns out that through the graphical intuition
provided by linear logic \emph{proof-nets} (PN), one can
define an enriched $\lambda$-calculus with explicit
  substitutions (ES) that unveils a strong bisimulation for the
  intuitionistic case~\cite{AccattoliBKL14}.  
  In this paper, we
resort to this same intuition to explore whether it is possible to
uncover a strong bisimulation behind the notion of $\sigma$-equivalence formulated by Laurent~\cite{Laurent02,Laurent03} in the setting of classical logic.
Thus, we will not only capture structural equivalence on pure
functions, but also on \emph{programs with control operators}. We next briefly revisit
proof-nets and discuss how they help unveil structural equivalence as
strong bisimulation for $\lambda$-calculi. An explanation of the
challenges that we face in addressing the classical case will follow.

\paragraph{Proof-nets.} A proof-net is a graph-like structure whose
nodes denote logical inferences and whose edges or wires denote the formula
they operate on.
Proof-nets were introduced in the setting of linear logic~\cite{Girard87},
a logic that provides a mechanism to explicitly control the use of resources by
restricting the application of the \emph{structural} rules of weakening and
contraction. Proof-nets are equipped with an operational semantics specified
by graph transformation rules which captures cut elimination in sequent
calculus. The resulting cut elimination rules on proof-nets are split into two
different kinds: \emph{multiplicative}, that essentially reconfigure wires, and
\emph{exponential}, which are the only ones that are able to erase or duplicate
(sub)proof-nets. Most notably, proof-nets abstract away the order
in which certain rules occur in a sequent derivation. As an example, assume
three derivations of the judgements $\vdash \envdis{\Gam}{A}$,
$\vdash \envdis{\Delta}{\envdis{A^\bot}{B}}$ and $\vdash \envdis{\Pi}{B^\bot}$,
resp. The order in which these derivations are composed via cuts into a single
derivation is abstracted away in the resulting proof-net: \[
\begin{array}{c}
\input{proofnets/examples/intro}
\end{array} \]

In other words, \emph{different} terms/derivations are represented by the
\emph{same} proof-net.  Hidden structural similarity between terms can thus be
studied by translating them to proof-nets. Moreover, following the
Curry--Howard isomorphism which relates computation and logic, this
correspondence between a term language and a graphical formalism can also be
extended to their reduction
behavior~\cite{Accattoli18,Kesner22}. In this paper we
focus on defining one such structural equality that is a strong bisimulation
for a classical lambda calculus based on Parigot's
$\calcLambdaMu$-calculus~\cite{Parigot92,Parigot93}. Although we rely on
intuitions provided by Laurent's Polarized Proof
Nets~\cite{Laurent02,Laurent03}, knowledge about Polarized Proof Nets is not
required to read this work and is not further discussed. We begin with an
overview of a similar program carried out in the intuitionistic case.

\paragraph{Intuitionistic $\sigma$-Equivalence.}\label{p:intuitionistic}
Regnier introduced a notion of \emph{$\sigma$-equivalence} on $\lambda$-terms
(written $\eqregnier$ and depicted in Figure~\ref{f:sigma-equivalence-lambda}),
and proved that $\sigma$-equivalent terms behave in essentially identical way.
This equivalence relation involves permuting certain redexes, and was unveiled
through the study of proof-nets. In particular, following Girard's encoding of
intuitionistic logic into linear logic~\cite{Girard87}, $\sigma$-equivalent
terms are mapped to the same proof-net (modulo multiplicative cuts and
structural equivalence of PN).
\begin{figure}[!h]
\centering
\[
\begin{array}{rcll}
\termapp{(\termabs{x}{\termabs{y}{t}})}{u} & \ruleEqregAbs & \termabs{y}{\termapp{(\termabs{x}{t})}{u}} & y \notin u \\
\termapp{(\termabs{x}{\termapp{t}{v}})}{u} & \ruleEqregApp & \termapp{\termapp{(\termabs{x}{t})}{u}}{v} & x \notin v
\end{array}
\]
\caption{$\sigmaregnier$-Equivalence for $\lambda$-terms}
\label{f:sigma-equivalence-lambda}
\end{figure}

The reason why Regnier's result is not immediate is that redexes present on one
side of an equation may disappear on the other side of it, as illustrated with 
the terms in (\ref{eq:example:permutation:lambda}).
One might rephrase this observation by stating that $\eqregnier$ is \emph{not a
strong bisimulation} over the set of $\lambda$-terms. If it were, then
establishing that $\sigma$-equivalent terms behave essentially in the same way
would be trivial.

Adopting a more refined view of $\lambda$-calculus as suggested by linear
logic, which splits cut elimination on logical derivations into multiplicative
and exponential steps, yields a decomposition of $\beta$-reduction on terms
into multiplicative/exponential steps. The theory of \emph{explicit
substitutions} (a survey can be found in~\cite{Kesner09}) provides a convenient
syntax to reflect this splitting at the term level. Indeed, $\beta$-reduction
can be decomposed into two steps, namely $\rBeta$ (for {\tt B}eta), and $\rS$ (for
{\tt S}ubstitution):
\begin{equation}
\begin{array}{rll}
\termapp{(\termabs{x}{t})}{u}  & \rrule{\rBeta} & \termsubs{x}{u}{t} \\
\termsubs{x}{u}{t}             & \rrule{\rS} & \subsapply{\subs{x}{u}}{t}
\end{array}
\end{equation}
or, more generally, the \emph{reduction at a $\mathtt{d}$istance} version for $\rBeta$, introduced in~\cite{AccattoliK10}, and written $\rB$:

\begin{equation}
\begin{array}{rll}
\termapp{\termsubs{x_n}{v_n}{\termsubs{x_1}{v_1}{(\termabs{x}{t})} \ldots}}{u}  & \rrule{\rB} & \termsubs{x_n}{v_n}{\termsubs{x_1}{v_1}{\termsubs{x}{u}{t}} \ldots} \\
\termsubs{x}{u}{t}                                                              & \rrule{\rS} & \subsapply{\subs{x}{u}}{t}
\end{array}
\label{eq:split_of_beta}
\end{equation}
Firing the $\rB$-rule creates a new \emph{explicit substitution}
operator, written $\exsubs{x}{u}$, so that $\rB$ essentially
reconfigures symbols (it is in some sense an innocuous or \emph{\plain} rule), and indeed reads as a multiplicative cut
in proof-nets. The $\rS$-rule executes
the substitution by performing
  a replacement of  all free occurrences of $x$ in $t$ with $u$, written $\subsapply{\subs{x}{u}}{t}$, so that it is $\rS$ that performs
  interesting or \emph{meaningful} computation in the sense that it performs 
    exponential cut steps in proof-nets.  We write $\reduce[\rS]$ for $\rrule{\rS}$ steps inside an arbitrary context and similarly for $\reduce[\rB]$.

Decomposition of $\beta$-reduction by means
of the reduction rules in~(\ref{eq:split_of_beta}) prompts one to
replace Regnier's $\eqregnier$
(Figure~\ref{f:sigma-equivalence-lambda}) with a new relation~\cite{AK12LMCS} that we write here  $\eqregex$
(Figure~\ref{eq:sigma:es}). The latter is formed essentially by taking
the \emph{$\rB$-normal form} of each side of the $\eqregnier$
equations.  Also included in $\eqregex$ is a third equation
$\ruleEqregexCom$ allowing commutation of orthogonal (independent)
substitutions.  Notice however that the $\rB$-expansion of
$\ruleEqregexCom$ results in $\sigmaregnier$-equivalent terms, since
$\termsubs{x}{u}{\termsubs{y}{v}{t}} \eqregex
\termsubs{y}{v}{\termsubs{x}{u}{t}}$, with $x \notin v$ and $y \notin
u$, $\rB$-expands to
$\termapp{(\termabs{y}{\termapp{(\termabs{x}{t})}{u}})}{v} \eqregnier
\termapp{(\termabs{x}{\termapp{(\termabs{y}{t})}{v}})}{u}$, both of
which are $\sigmaregnier$-equivalent by $\ruleEqregAbs$ and
$\ruleEqregApp$.
\begin{figure}[h]
\centering
\[
\begin{array}{rcll}
\termsubs{x}{u}{(\termabs{y}{t})}   & \ruleEqregexAbs & \termabs{y}{\termsubs{x}{u}{t}}     & y \notin u \\
\termsubs{x}{u}{(\termapp{t}{v})}   & \ruleEqregexApp & \termapp{\termsubs{x}{u}{t}}{v}     & x \notin v \\
\termsubs{x}{u}{\termsubs{y}{v}{t}} & \ruleEqregexCom & \termsubs{y}{v}{\termsubs{x}{u}{t}} & x \notin v, y \notin u
\end{array}
\]
\caption{Strong Bisimulation $\eqregex$ for $\lambda$-Terms}
\label{eq:sigma:es}
\end{figure}

Through $\eqregex$ it is possible to unveil a strong
bisimulation for the intuitionistic case by working on $\lambda$-terms with ES and the notion of $\beta$-reduction at a distance.
Indeed, the following holds:
\begin{thm}[Strong Bisimulation for the Intuitionistic Case I]
\label{thm:strong_bisimulation_int_case_i}
Let $t \eqregex t'$. If $t \reduce[\rB,\rS] u$, then there exists $u'$ such that
$t' \reduce[\rB,\rS] u'$ and $t'  \eqregex u'$. Graphically,
\begin{center}
\begin{tikzcd}[->,ampersand replacement=\&]
t \arrow[->]{d}[left]{\rB,\rS}
  \&[-25pt] \eqregex
  \&[-25pt] t' \arrow[->]{d}{\rB,\rS} \arrow[phantom]{d}{} \\
   u
  \&[-25pt] \eqregex
  \&[-25pt] u'
\end{tikzcd} 
\end{center}
\end{thm}

\noindent While any two $\eqregex$-equivalent $\lambda$-terms with ES
translate to the same proof-net,
the converse is not true. For example, the terms
$\termapp{(\termabs{x}{\termapp{t}{v}})}{u}$ and
$\termapp{\termapp{(\termabs{x}{t})}{u}}{v}$, where $x \notin v$,
translate to the same proof-net~\cite{Regnier94} (indeed, they are
$\sigmaregnier$-equivalent), however they are not
$\eqregex$-equivalent. Still, as remarked in~\cite{AK12LMCS}, the
$\rB$-normal forms of those terms, namely
$\termsubs{x}{u}{(\termapp{t}{v})}$ and
$\termapp{\termsubs{x}{u}{t}}{v}$, are $\eqregex$-equivalent, thus
suggesting an alternative equivalence relation that we define only on
$\rB$-normal forms.  Indeed, \emph{\plain\ forms} are $\lambda$-terms with ES that are in $\rB$-normal form (which are
intuitively, those terms that are in multiplicative normal form from
the proof-net point of view).  Similarly, \emph{\plain\
  computation} is given by $\rB$-reduction;
\plain\  computation to normal form produces \plain\  forms.
As mentioned above our notion of \emph{meaningful computation} is taken 
to be the reduction relation $\reduce[\rS]$.
We write
$\eqsigma$ for this refined equivalence notion on \plain\ forms,
which will in fact be included in the strong bisimulation that we
propose in this paper
(\cf Definition~\ref{d:control:structural:equivalence}). The following
variation of the theorem mentioned
above is obtained:
\begin{thm}[Strong Bisimulation for the Intuitionistic Case II]
  \label{t:second-bisimulatio-intuitionistic}
  Let $\reducemean$ be $\reduce[\rS]$ followed by
  $\reduce[\rB]$-reduction to $\rB$-normal form.
  Let $t,t'$ be two terms in $\rB$-normal form such
    that $t \eqsigma t'$. If $t \reducemean u$, 
    then there exists
    $u'$ such that $t' \reducemean u'$ and
    $t' \eqsigma u'$. Graphically, 
\begin{center}
  \[ \begin{array}{l@{\hspace{1cm}\mbox{or equivalently}\hspace{1cm}}l}
    \begin{tikzcd}[->,ampersand replacement=\&]
t \arrow[->]{d}[left]{\rS}
  \&[-25pt] \eqsigma
  \&[-25pt] t' \arrow[->]{d}{\rS} \arrow[phantom]{d}{} \\
  v \arrow[->>]{d}[left]{\rB}  \&[-25pt]   \&[-25pt] v'\arrow[->>]{d}[left]{\rB} \\
  u 
  \&[-25pt] \eqsigma 
  \&[-25pt] u'
    \end{tikzcd} &
    \begin{tikzcd}[->,ampersand replacement=\&]
t \arrow[rightsquigarrow]{d}[left]{}
  \&[-25pt] \eqsigma
  \&[-25pt] t' \arrow[rightsquigarrow]{d}{} \arrow[phantom]{d}{} \\
   u
  \&[-25pt] \eqsigma 
  \&[-25pt] u'
    \end{tikzcd}
    \end{array} \] 
\end{center}
  \end{thm}
Thus, a strong bisimulation can be defined on a
  set of \emph{\plain\  forms} (here $\lambda$-terms with ES 
  which are in $\rB$-normal form), with respect to a
  \emph{meaningful computation} relation (here $\reduce[\rS]$)
  followed by $\rB$-reduction to $\rB$-normal form.

  Summing up, a strong bisimulation was obtained by
  decomposing $\beta$-reduction as follows:
  \[
\input{diagram-beta}
\]
This methodology consisting in identifying an appropriate notion of
\plain\ form and meaningful computation, both over terms, allows for
a strong bisimulation to surface. This requires establishing a
corresponding distinction between multiplicative and exponential
steps in the underlying term semantics. We propose following this
same methodology for the classical case.  However, as we will see,
the notion of \plain\ computation as well as that of meaningful one
are not so easy to construct for Parigot's
$\calcLambdaMu$-calculus. We next briefly introduce
this calculus as well
as the notion of $\sigma$-equivalence as presented by Laurent.

\paragraph{Classical $\sigma$-Equivalence.} \emph{$\lambda$-calculi with
control operators} include operations to manipulate the context in which a
program is executed. We focus here on Parigot's
$\calcLambdaMu$-calculus, which extends the $\lambda$-calculus
with two new operations: $\termname{\alpha}{t}$ (\emph{command}) and
$\termcont{\alpha}{c}$ (\emph{$\mu$-abstraction}). The former may informally be
understood as ``call continuation $\alpha$ with $t$ as argument'' and the
latter as ``record the current continuation as $\alpha$ and continue as $c$''.
Reduction in $\calcLambdaMu$ consists of the $\beta$-rule together with: \[
\begin{array}{rcl}
\termapp{(\termcont{\alpha}{c})}{u} & \rrule{\rmu} & \termcont{\alpha}{\replapply{\repl{\alpha}{u}}{c}}
\end{array}
\] where $\replapply{\repl{\alpha}{u}}{c}$, called here \emph{replacement},
replaces all subexpressions of the form $\termname{\alpha}{t}$ in $c$
with $\termname{\alpha}{(\termapp{t}{u})}$.

Regnier's notion of $\sigma$-equivalence
for $\lambda$-terms was extended to  $\calcLambdaMu$ by
Laurent~\cite{Laurent03}
(\cf Figure~\ref{f:sigma-laurent} in Section~\ref{f:sigma-laurent}). Here is an
example of terms related by this extension, where the redexes are
underlined/overlined and $\eqlaurent$ denotes Laurent's
  aforementioned relation: \[
\termapp{(\underline{\termapp{(\termabs{x}{\termcont{\alpha}{\termname{\gamma}{u}}})}{w}})}{v}
\eqlaurent
\overline{\termapp{(\termcont{\alpha}{\termname{\gamma}{\underline{\termapp{(\termabs{x}{u})}{w}}}})}{v}} 
\] Once again, the fact that a harmless permutation of redexes has taken place
is not obvious. The term on the right has two redexes ($\mu$ and $\beta$) but
the one on the left only has one ($\beta$) redex. Another, more subtle, example
of terms related by Laurent's extension clearly suggests that operational
indistinguishability cannot rely on relating arbitrary $\mu$-redexes; the
underlined $\mu$-redex on the left does not appear at all on the right:
\begin{equation}
\underline{\termapp{(\termcont{\al}{\termname{\al}{x}})}{y}}
\eqlaurent
\termapp{x}{y}
\label{eq:thetaAndSBisim}
\end{equation}
Clearly, Laurent's $\sigma$-equivalence on $\calcLambdaMu$-terms \emph{fails to be a
strong bisimulation}.

\paragraph{Towards a Strong Bisimulation for $\calcLambdaMu$.}  We seek to
formulate a similar notion of equivalence for calculi with control operators in
the sense that it is concerned with harmless permutation of redexes possibly
involving control operators and induces a strong bisimulation. A first step
towards our goal involves decomposing the $\mu$-rule as was done for the
$\beta$-rule in (\ref{eq:split_of_beta}):
\begin{equation}
\begin{array}{rll}
\termapp{(\termcont{\alpha}{c})}{u}  & \rrule{} & \termcont{\alpha'}{\termrepl[\alpha']{\alpha}{u}{c}} \\
\termrepl[\alpha']{\alpha}{u}{c} & \rrule{} & \replapply{\repl[\alpha']{\alpha}{u}}{c}
\end{array}
 \label{eq:split_of_mu_without_distance}
\end{equation}
where $\replapply{\repl[\alpha']{\alpha}{u}}{c}$ denotes the \emph{fresh}
replacement changing all subexpressions of the form
$\termname{\alpha}{t}$ in $c$ to $\termname{\alpha'}{(tu)}$. A brief discussion
on our choice of notation for replacement, may be found at the end of this
section. We still need to add the notion of distance to this operational
semantics, as done for substitution. This produces a rule $\rM$ (for
$\mathtt{M}u$ at a $\mathtt{d}istance$), to introduce an \emph{explicit
replacement}, and another rule $\rR$ (for $\rR$eplacement), that executes
explicit replacements:
\begin{equation}
\begin{array}{rll}
\termapp{\termsubs{x_n}{v_n}{\termsubs{x_1}{v_1}{(\termcont{\alpha}{c})} \ldots}}{u}  & \rrule{\rM} & \termsubs{x_n}{v_n}{\termsubs{x_1}{v_1}{(\termcont{\alpha'}{\termrepl[\alpha']{\alpha}{u}{c}})} \ldots} \\
\termrepl[\alpha']{\alpha}{u}{c}                                                      & \rrule{\rR} & \replapply{\repl[\alpha']{\alpha}{u}}{c}
\end{array}
\label{eq:split_of_mu}
\end{equation}
where $\replapply{\repl[\alpha']{\alpha}{u}}{c}$ replaces each sub-expression
of the form $\termname{\al}{t}$ in $c$ by $\termname{\al'}{(t u)}$.

Following our analogy with the intuitionistic case, our plain rule is $\rM$ and meaningful
computation is performed by $\rR$.

Therefore,  we tentatively fix our notion of
meaningful computation to be $\rS\cup \rR$ over the set of \plain\
forms, the latter now obtained by taking \emph{both} $\rB$ and
$\rM$-normal forms.  
However, in contrast to the intuitionistic case where
 the decomposition of $\beta$  into a multiplicative
  rule $\rB$ and an exponential rule $\rS$ suffices for unveiling
    the strong bisimulation behind Regnier's $\sigma$-equivalence
    in $\lambda$-calculus,
  it turns out that splitting
  the $\mu$-rule into $\rM$ and
  $\rR$ is not enough in the classical case.  We face two obstacles:

    \begin{description}
    \item[Decomposing Meaningful Steps]  Consider~(\ref{eq:thetaAndSBisim}) from above. The methodology that led~\cite{AK12LMCS} to obtain the theory of Figure~\ref{eq:sigma:es} by taking the plain forms of Regnier's $\sigma$-equivalence, would lead us to the equation $\termcont{\al'}{\termrepl[\al']{\alpha}{y}{(\termname{\al}{x})}} \simeq
\termapp{x}{y}$, where \plain\ forms
      are terms in $\rB\cup\rM$-normal form.  However, there is clearly an $\rR$ step on the left term which is not present on the right one:
      \begin{equation}
\begin{tikzcd}[->,ampersand replacement=\&]
\termcont{\al'}{\termrepl[\al']{\alpha}{y}{(\termname{\al}{x})}} \arrow[->]{d}[left]{\rR}
  \&[-25pt] \simeq
  \&[-25pt] \termapp{x}{y} \arrow[->]{d}{\rR} \arrow[phantom,negated]{d}{} \\ 
\termcont{\al'}{\termname{\al'}{\termapp{x}{y}}}
  \&[-25pt] \simeq
  \&[-25pt] \termapp{x}{y}
\end{tikzcd}
\label{eq:theta}
\end{equation}
In its full generality, the $\rR$ rule in (\ref{eq:split_of_mu}) can certainly duplicate or
  erase $u$. However, it may also be the case that  there is a unique
  occurrence of $\alpha$ in $c$. If, furthermore, this occurrence
  cannot be duplicated or erased any time later, then this instance of
  $\rR$ may quite reasonably be catalogued as non-meaningful. Indeed,
  these \emph{linear} replacements will form part of our notion of
  plain computation rather than that of the meaningful one. With this
  revised notion of plain computation, the plain normal form of
  $\termapp{(\termcont{\al}{\termname{\al}{x}})}{y}$ becomes
  $\termcont{\al'}{\termname{\al'}{\termapp{x}{y}}}$ and thus 
  equation~(\ref{eq:thetaAndSBisim}) now reads: 
$\termcont{\al'}{\termname{\al'}{\termapp{x}{y}}}
  \simeq
  \termapp{x}{y}$; moreover both terms are indeed related in
  the strong bisimulation relation that we propose in this paper.

  \item[Name Renaming] In $\calcLambdaMu$, an expression such as
  $\termname{\alpha}{\termcont{\beta}{c}}$ may be simplified to
  another expression $\renapply{\ren{\beta}{\alpha}}{c}$ where all
  occurrences of the name $\beta$ in $c$ are replaced with $\alpha$. The effect is thus to rename $\beta$ with $\alpha$ in $c$.
  Such an equation, dubbed $\eqsigma[\rho]$, is included in Laurent's
  $\eqlaurent$ and also breaks strong bisimulation.
\begin{equation}
\begin{tikzcd}[->,ampersand replacement=\&]
\termapp{(\termcont{\alpha}{\termname{\alpha}{\termcont{\beta}{\termname{\gamma}{x}}}})}{u} \arrow[->]{d}[left]{\rmu}
  \&[-25pt] \eqsigma[\rho]
  \&[-25pt] \termapp{(\termcont{\alpha}{\termname{\gamma}{x}})}{u} \arrow[->]{d}[left]{\rmu} \\
\termcont{\alpha'}{\termname{\alpha'}{\underline{\termapp{(\termcont{\beta}{\termname{\gamma}{x}})}{u}}}}
  \&[-25pt] \not\eqlaurent
  \&[-25pt] \termcont{\alpha'}{\termname{\gamma}{x}}
\end{tikzcd}
\label{e:control:preliminaries:sigma-laurent:rho}
\end{equation}
 An additional $\rmu$-step will be needed on the left (the underlined one), which is not present on the right, to be able to obtain a term equivalent to $\termcont{\alpha'}{\termname{\gamma}{x}}$ on the right. Hence, this does not constitute a strong bisimulation diagram.
 However, completely dropping
  $\eqsigma[\rho]$ is not possible since it is required to be able to swap renamings. For example, the following identity
  \begin{center}
    $ \ntermren{\alpha}{\alpha'}{\ntermren{\beta}{\beta'}{c}}                                                      \simeq     \ntermren{\beta}{\beta'}{\ntermren{\alpha}{\alpha'}{c}}$
  \end{center} 
  where $\beta \neq \alpha', \alpha \neq \beta'$, can be deduced using  $\eqsigma[\rho]$ twice. This swapping identity (and two others, see Section~\ref{s:control:correspondence} for details)
are necessary to be able to close other strong bisimulation diagrams. Example~\ref{e:rho-necessario} in Section~\ref{s:control:equivalence} illustrates this point.
As it turns out, if one drops  $\eqsigma[\rho]$ but retains such swapping equations, just ``enough of  $\eqsigma[\rho]$'' is preserved to obtain our strong bisimulation result.
\end{description}

\paragraph{Contributions.} Our contributions may be summarised as follows:

\begin{enumerate}
\item A refinement of $\calcLambdaMu$, called $\calcLambdaM$-calculus,
including explicit substitutions for variables, and explicit replacement for names. The $\calcLambdaM$-calculus is proved to be confluent
(Theorem~\ref{t:control:lambda-m:semantics:confluence}); 

\item A notion of structural equivalence $\eqsigma$ for $\calcLambdaM$
that is a strong
  bisimulation with respect to meaningful computation on the set of
  plain normal forms.
    (Theorem~\ref{t:control:bisimulation}).

\item A precise correspondence result between our bisimulation $\eqsigma$ on $\calcLambdaM$-objects and
  Laurent's original $\sigma$-equivalence on
  $\calcLambdaMu$-objects.
\end{enumerate}

This paper is an extended and revised version of~\cite{KesnerBV20}.

\paragraph{Structure of the paper.} After some preliminaries on notation
introduced in Section~\ref{s:preliminary}, Section~\ref{s:control:preliminaries} and
Section~\ref{s:control:lambda-m} present $\calcLambdaMu$ and $\calcLambdaM$,
respectively. Section~\ref{s:control:equivalence} discusses the difficulties in
formulating a strong bisimulation and presents the proposed solution.
Section~\ref{s:control:correspondence} addresses the correspondence proof between
our bisimulation $\eqsigma$ on $\calcLambdaM$-objects and Laurent's original
$\sigma$-equivalence on $\calcLambdaMu$-objects. Finally,
Section~\ref{s:conclusion} concludes and describes related work. Most proofs are
relegated to the Appendix.


%% file: proofnets/examples/intro.tex
\begin{tikzpicture}[baseline=(current bounding box.center),node distance=\PNdist]
\node[PNsub] (P1) [] {$\sequ{\envdis{\Gamma}{A}}$};
\node[PNsub] (P2) [right of=P1, minimum width=\PNdist*2, xshift=\PNdist*2] {$\sequ{\envdis{\envdis{\Delta}{\formneg{A}}}{B}}$};
\node[PNsub] (P3) [right of=P2, xshift=\PNdist*2.2] {$\sequ{\envdis{\Pi}{\formneg{B}}}$};
\node (N1) [below of=P1, xshift=-\PNdist*0.5, yshift=-\PNdist*0.25] {$A$};
\node (NG) [below of=P1, xshift= \PNdist*0.5, yshift=-\PNdist*0.25] {$\Gamma$};
\pncut[below of=N1, xshift=\PNdist*1.25]{C1};
\node (N2) [below of=P2, xshift=-\PNdist, yshift=-\PNdist*0.25] {$\formneg{A}$};
\node (N3) [below of=P2, yshift=-\PNdist*0.25] {$B$};
\node (ND) [below of=P2, xshift= \PNdist, yshift=-\PNdist*0.25] {$\Delta$};
\pncut[below of=N3, xshift=\PNdist*1.35]{C2};
\node (N4) [below of=P3, xshift=-\PNdist*0.5, yshift=-\PNdist*0.25] {$\formneg{B}$};
\node (NP) [below of=P3, xshift= \PNdist*0.5, yshift=-\PNdist*0.25] {$\Pi$};

\draw[PNarrow] ([xshift=-\PNdist*0.5]P1.south) -- node{} (N1.north)
               ([xshift= \PNdist*0.5]P1.south) -- node{} (NG.north)
               ([xshift=-\PNdist]P2.south)     -- node{} (N2.north)
               (P2.south)                      -- node{} (N3.north)
               ([xshift= \PNdist]P2.south)     -- node{} (ND.north)
               ([xshift=-\PNdist*0.5]P3.south) -- node{} (N4.north)
               ([xshift= \PNdist*0.5]P3.south) -- node{} (NP.north)
               (N1.south)                      |- node{} (C1.west)
  	  	       (N2.south)                      |- node{} (C1.east)
               (N3.south)                      |- node{} (C2.west)
  	  	       (N4.south)                      |- node{} (C2.east);
\end{tikzpicture}


%% file: diagram-beta.tex
\begin{tikzcd}
& \rB &[-20pt] \dashnode{6pc}{\plain\ computation}   \\[-20pt]
\beta \arrow{ur}{}\arrow{dr}{}                      \\[-20pt]
& \rS &[-20pt] \dashnode{6pc}{meaningful computation}
\end{tikzcd}

%% file: preliminary.tex
\section{Some Basic Preliminary Notions}
\label{s:preliminary}
We start this section by some generic notations. Let $\R$ be any
reduction relation on a set of elements $\genericobjects$. We write
$\reducemany[\R]$ for the reflexive-transitive closure of
$\reduce[\R]$. If $\S$ is another reduction relation on $\genericobjects$, we use
$\reduce[\R,\S]$ to denote $\reduce[\R] \cup \reduce[\S]$.
We say there is an $\R$-reduction sequence starting at
$t_0\in\genericobjects$ if there exists
$t_1\ldots,t_n\in\genericobjects$ with $n\geq 0$, such that
$t_0\reduce[\R]t_1$, $t_1\reduce[\R]t_2$, $\ldots$, $t_{n-1}\reduce[\R]t_n$. 
We occasionally refer to $\R$-reduction as $\R$-computation.
An element
$t \in\genericobjects$ enjoys the \deft{$\R$-diamond property } iff for
every $u \neq v \in\genericobjects$ such that $t\reduce[\R]u$ and
$t\reduce[\R]v$, there exists $t'$ such that $u\reduce[\R]t'$
and $v\reduce[\R]t'$.  An element
$t \in\genericobjects$ is said to be \deft{$\R$-confluent} iff for
every $u,v \in\genericobjects$ such that $t\reducemany[\R]u$ and
$t\reducemany[\R]v$, there exists $t'$ such that $u\reducemany[\R]t'$
and $v\reducemany[\R]t'$.
A reduction relation $\R$
  has the 
\deft{diamond property } iff every $t \in\genericobjects$ has the $\R$-diamond property.  A reduction relation $\R$ is
\deft{confluent} iff every $t \in\genericobjects$ is $\R$-confluent.
If $\R$ has the diamond property, then $\R$ is confluent~\cite{BaaderN98}. 
An element $t \in\genericobjects$ is said to be
\deft{$\R$-terminating} iff there is no infinite $\R$-reduction
sequence starting at $t$.  A reduction relation $\R$ is
\deft{terminating} iff every $t \in\genericobjects$ is
$\R$-terminating.  An element $t\in\genericobjects$ is said to be in
\deft{$\R$-normal form} iff there is no $t'$ such that $t \reduce[\R]
t'$. 
We use $\setNF[\R]$ to denote all the normal forms of $\R$, \ie the set
of all the elements in $\genericobjects$ which
are in $\R$-normal form. 
Given $t\in\genericobjects$, we say that $u$ is \deft{an $\R$-normal form of $t$}
if $t \reducemany[\R] u$ and $u$ is in $\R$-normal form. We
denote by $\nf[\R]{t}$ the set of all $\R$-normal forms of $t$;  this
set  is always  a singleton
when $\R$ is confluent and terminating.  


%% file: lambda-mu.tex
\section{The \calcLambdaMu-Calculus}
\label{s:control:preliminaries}

In this section we introduce the untyped $\calcLambdaMu$-calculus.

\subsection{The Untyped \calcLambdaMu-Calculus}
\label{s:control:preliminaries:lambda-mu}

Given a countably infinite set of variables $\TermVariable$ ($\termvar{x},
\termvar{y}, \ldots$) and names  $\TermName$ ($\alpha, \beta, \ldots$),
the set of  \deft{objects} $\Object{\calcLambdaMu}$, \deft{terms}
$\Term{\calcLambdaMu}$, \deft{commands} $\Command{\calcLambdaMu}$
and \deft{contexts} of the  $\calcLambdaMu$-calculus are
defined by means of the following grammar:  \[
\begin{array}{rrcl}
\textbf{(Objects)}          & o         & \Coloneq  & t \mid c \\
\textbf{(Terms)}            & t         & \Coloneq  & x  \mid \termapp{t}{t} \mid \termabs{x}{t} \mid \termcont{\alpha}{c} \\
\textbf{(Commands)}         & c         & \Coloneq  & \termname{\alpha}{t} \\
\textbf{(Contexts)}         & \ctxt{O}  & \Coloneq  & \ctxt{T} \mid \ctxt{C} \\
\textbf{(Term Contexts)}    & \ctxt{T}  & \Coloneq  & \Box \mid \termapp{\ctxt{T}}{t} \mid \termapp{t}{\ctxt{T}} \mid \termabs{x}{\ctxt{T}} \mid \termcont{\alpha}{\ctxt{C}} \\
\textbf{(Command Contexts)} & \ctxt{C}  & \Coloneq  & \boxdot \mid \termname{\alpha}{\ctxt{T}}
\end{array}
\]
The grammar extends the terms of the $\calcLambda$-calculus with two
new constructors: \deft{commands} $\termname{\alpha}{t}$ and
\deft{$\itermcont$-abstractions} $\termcont{\alpha}{c}$.  The
combination of a command and
a $\mu$-abstraction will be coined \deft{explicit renaming}, as in
$\termname{\alpha}{\termcont{\beta}{c}}$. The term
$\termapp{(\ldots(\termapp{(\termapp{t}{u_1})}{u_2})\ldots)}{u_n}$
abbreviates as $\termapp{\termapp{\termapp{t}{u_1}}{u_2}\ldots}{u_n}$
or $\termapp{t}{\vec{u}}$ when $n$ is clear from the context.
Regarding contexts, there are two holes $\Box$ and $\boxdot$ of sort
\deft{term} ($\TermSort$) and \deft{command}
($\CommandSort$) respectively. We write $\ctxtapply{\ctxt{O}}{o}$ to
denote the replacement of the hole $\Box$ (resp. $\boxdot$) by a term
(resp. by a command).  We often decorate contexts or functions over
expressions with one of the sorts $\TermSort$ and $\CommandSort$ to be
more clear. For example, $\ctxt{O}_\TermSort$ is a context $\ctxt{O}$
with a hole of sort $\deft{term}$. The subscript is omitted if it is
clear from the context.

\deft{Free} and
\deft{bound variables} of objects are defined as expected, in
particular  $\fv{\termcont{\alpha}{c}} \eqdef
\fv{c}$ and  $\fv{\termname{\alpha}{t}} \eqdef \fv{t}$. 
\deft{Free} and \deft{bound names} are defined as follows:  \[
\begin{array}{c@{\qquad}c}
\begin{array}{rcl}
\fn{x}                    & \eqdef  & \emptyset \\
\fn{\termapp{t}{u}}       & \eqdef  & \fn{t} \cup \fn{u} \\
\fn{\termabs{x}{t}}       & \eqdef  & \fn{t} \\
\fn{\termcont{\alpha}{c}} & \eqdef  & \fn{c} \setminus \set{\alpha} \\
\fn{\termname{\alpha}{t}} & \eqdef  & \fn{t} \cup \set{\alpha}
\end{array}
&
\begin{array}{rcl}
\bn{x}                    & \eqdef  & \emptyset \\
\bn{\termapp{t}{u}}       & \eqdef  & \bn{t} \cup \bn{u} \\
\bn{\termabs{x}{t}}       & \eqdef  & \bn{t} \\
\bn{\termcont{\alpha}{c}} & \eqdef  & \bn{c} \cup \set{\alpha} \\
\bn{\termname{\alpha}{t}} & \eqdef  & \bn{t}
\end{array}
\end{array}
\] We use $\fv[x]{o}$ and $\fn[\alpha]{o}$ to denote the number of free
occurrences of the variable $x$ and the name $\alpha$ in the object $o$
respectively. Additionally, we write $x \notin o$ ($\alpha \notin o$) when $x
\notin \fv{o} \cup \bv{o}$ (respectively $\alpha \notin \fn{o} \cup \bn{o}$).
This notion is naturally extended to contexts.

We work with the standard notion of $\alpha$-conversion, \ie renaming of bound
variables and names, thus for example
$\termname{\delta}{\termapp{(\termcont{\alpha}{\termname{\alpha}{\termabs{x}{x}}})}{z}}
\eqalpha
\termname{\delta}{\termapp{(\termcont{\beta}{\termname{\beta}{\termabs{y}{y}}})}{z}}$.
In particular, when using two different symbols to denote bound variables or
names, we assume that they are different without explicitly mentioning it. 

\medskip
\deft{Application} of the \deft{implicit substitution} $\subs{x}{u}$ to the
object $o$, written $\subsapply{\subs{x}{u}}{o}$, may require
$\alpha$-conversion in order to avoid capture  of free variables/names, and it
is defined as expected.

\deft{Application} of the \deft{implicit replacement}
$\repl[\alpha']{\alpha}{u}$ to an object $o$, written
$\replapply{\repl[\alpha']{\alpha}{u}}{o}$, passes the term $u$ as an argument
to any sub-command of $o$ of the form $\termname{\alpha}{t}$ and changes the
name of $\alpha$ to $\alpha'$. This operation is also defined modulo
$\alpha$-conversion in order to avoid the capture of free variables/names.
Formally: \[
\begin{array}{rcll}
\replapply{\repl[\alpha']{\alpha}{u}}{x}                      & \eqdef  & x \\
\replapply{\repl[\alpha']{\alpha}{u}}{(\termapp{t}{v})}       & \eqdef  & \termapp{\replapply{\repl[\alpha']{\alpha}{u}}{t}}{\replapply{\repl[\alpha']{\alpha}{u}}{v}} \\
\replapply{\repl[\alpha']{\alpha}{u}}{(\termabs{x}{t})}       & \eqdef  & \termabs{x}{\replapply{\repl[\alpha']{\alpha}{u}}{t}}       & x \notin u \\
\replapply{\repl[\alpha']{\alpha}{u}}{(\termcont{\beta}{c})}  & \eqdef  & \termcont{\beta}{\replapply{\repl[\alpha']{\alpha}{u}}{c}}  & \beta \notin u, \beta \neq \alpha' \\
\replapply{\repl[\alpha']{\alpha}{u}}{(\termname{\alpha}{c})} & \eqdef  & \termname{\alpha'}{(\termapp{\replapply{\repl[\alpha']{\alpha}{u}}{c}}{u})} \\
\replapply{\repl[\alpha']{\alpha}{u}}{(\termname{\beta}{c})}  & \eqdef  & \termname{\beta}{\replapply{\repl[\alpha']{\alpha}{u}}{c}}  & \beta \neq \alpha
\end{array}
\]
For example, if  $\id = \termabs{w}{w}$, then \[
\begin{array}{lll}
\subsapply{\subs{x}{\id}}{(\termapp{(\termcont{\alpha}{\termname{\alpha}{x}})}{(\termabs{z}{\termapp{z}{x}})})}
& = &
\termapp{(\termcont{\alpha}{\termname{\alpha}{\id}})}{(\termabs{z}{\termapp{z}{\id}})} \\
\replapply{\repl[\alpha']{\alpha}{\id}}{(\termname{\alpha}{\termapp{x}{(\termcont{\beta}{\termname{\alpha}{y}})}})}
& = & 
\termname{\alpha'}{\termapp{\termapp{x}{(\termcont{\beta}{\termname{\alpha'}{\termapp{y}{\id}}})}}{\id}}
\end{array} \] 

\medskip
Parigot's original formulation of
$\calcLambdaMu$-calculus~\cite{Parigot92,Parigot93} uses a binary replacement
operation $\replapply{\repl[]{\alpha}{u}}{c}$ rather than the ternary one we
introduced above. Details on our choice of notation, which are related to
explicit replacements, are developed in Section~\ref{s:on_choice_of_notation}.
  
\medskip
The \deft{one-step reduction relation} $\reduce[\rlm]$ is given by the closure
by \emph{all} contexts $\ctxt{O}_\TermSort$ of the following rewriting rules
$\rbeta$ and $\rmu$, \ie $\mathbin{\reduce[\rlm]} \eqdef
\ctxtapply{\ctxt{O}_\TermSort}{\rrule{\rbeta} \cup \rrule{\rmu}}$: \[
\begin{array}{rcl}
\termapp{(\termabs{x}{t})}{u}       & \rrule{\rbeta} & \subsapply{\subs{x}{u}}{t} \\
\termapp{(\termcont{\alpha}{c})}{u} & \rrule{\rmu}   & \termcont{\alpha'}{\replapply{\repl[\alpha']{\alpha}{u}}{c}}
\end{array}
\]
Given $X \in \set{\rbeta, \rmu}$, we define an \deft{$X$-redex} to be a term
having the form of the left-hand side of the rule $\rrule{X}$. A similar notion
will be used for all the rewriting rules used in this paper. It is worth
noticing that Parigot's~\cite{Parigot92} $\rmu$-rule of the
$\calcLambdaMu$-calculus relies on a binary implicit replacement operation
$\repl{\alpha}{u}$ assigning
$\termname{\alpha}{\termapp{(\replapply{\repl{\alpha}{u}}{t})}{u}}$ to each
sub-expression of the form $\termname{\alpha}{t}$ (thus not changing the name
of the command). We remark that
$\termcont{\alpha}{\replapply{\repl{\alpha}{u}}{c}} \eqalpha
\termcont{\alpha'}{\replapply{\repl[\alpha']{\alpha}{u}}{c}}$; thus \eg
$\termcont{\alpha}{\replapply{\repl{\alpha}{u}}{(\termname{\alpha}{x})}}
= \termcont{\alpha}{\termname{\alpha}{\termapp{x}{u}}} \eqalpha
\termcont{\gamma}{\termname{\gamma}{\termapp{x}{u}}} =
\termcont{\gamma}{\replapply{\repl[\gamma]{\alpha}{u}}{(\termname{\alpha}{x})}}$.
We adopt here the ternary presentation~\cite{KesnerV19} of the implicit
replacement operator, because it naturally extends to that of the
$\calcLambdaM$-calculus in Section~\ref{s:control:lambda-m}.

\medskip
Various control operators can be expressed in the $\calcLambdaMu$-calculus~\cite{Groote94,Laurent03}. A typical example   is the
control operator {\bf call-cc}~\cite{Griffin90}, specified by the term
$\termabs{x}{\termcont{\alpha}{\termname{\alpha}{\termapp{x}{(\termabs{y}{\termcont{\delta}{\termname{\alpha}{y}}})}}}}$.

\subsection{The notion of \sigmalaurent-equivalence for \calcLambdaMu-terms}
\label{s:control:preliminaries:sigma:lambda-mu}

As in $\lambda$-calculus, structural equivalence for the
$\calcLambdaMu$-calculus captures inessential permutation of redexes,
but this time also involving the control constructs.

\begin{defi}
\label{d:sigma-laurent}
Laurent's notion of $\sigma$-equivalence for
$\calcLambdaMu$-objects~\cite{Laurent03} (written here also $\eqlaurent$) is
depicted in Figure~\ref{f:sigma-laurent}, where $\ren{\beta}{\alpha}$ denotes the \deft{implicit renaming} of
  all the free occurrences of the name  $\beta$ by $\alpha$ (a formal
  definition is given in Section~\ref{s:control:lambda-m:semantics}).
\end{defi}

\begin{figure}[h]$$
\begin{array}{rcll}
\termapp{(\termabs{y}{\termabs{x}{t}})}{v}                                                                  & \simeq_{\sigmalaurent_1} & \termabs{x}{\termapp{(\termabs{y}{t})}{v}} & x \notin {v} \\
\termapp{(\termabs{x}{\termapp{t}{v}})}{u}                                                                  & \simeq_{\sigmalaurent_2} & \termapp{\termapp{(\termabs{x}{t})}{u}}{v} & x \notin {v} \\
\termapp{(\termabs{x}{\termcont{\alpha}{\termname{\beta}{u}}})}{w}                                          & \simeq_{\sigmalaurent_3} & \termcont{\alpha}{\termname{\beta}{\termapp{(\termabs{x}{u})}{w}}} & \al \notin {w} \\
\termname{\alpha'}{\termapp{(\termcont{\alpha}{\termname{\beta'}{\termapp{(\termcont{\beta}{c})}{w}}})}{v}} & \simeq_{\sigmalaurent_4} & 
\termname{\beta'}{\termapp{(\termcont{\beta}{\termname{\alpha'}{\termapp{(\termcont{\alpha}{c})}{v}}})}{w}} & \al \notin {w}, \beta \notin {v}, \beta \neq \al', \al \neq \beta' \\
\termname{\alpha'}{\termapp{(\termcont{\alpha}{\termname{\beta'}{\termabs{x}{\termcont{\beta}{c}}}})}{v}}   & \simeq_{\sigmalaurent_5} & 
\termname{\beta'}{\termabs{x}{\termcont{\beta}{\termname{\alpha'}{\termapp{(\termcont{\alpha}{c})}{v}}}}} & x \notin {v}, \beta \notin {v}, \beta \neq \al', \al \neq \beta' \\
\termname{\alpha'}{\termabs{x}{\termcont{\alpha}{\termname{\beta'}{\termabs{y}{\termcont{\beta}{c}}}}}}     & \simeq_{\sigmalaurent_6} & 
\termname{\beta'}{\termabs{y}{\termcont{\beta}{\termname{\alpha'}{\termabs{x}{\termcont{\alpha}{c}}}}}} & \beta \neq \al', \al \neq \beta' \\
\termcont{\alpha}{\termname{\alpha}{v}}                                                                     & \simeq_{\sigmalaurenttheta} & v & \alpha \notin {v} \\
\termname{\alpha}{\termcont{\beta}{c}}                                                                      & \simeq_{\sigmalaurentrho} & c\ren{\beta}{\alpha} \\
\end{array} $$
\caption{$\sigmalaurent$-equivalence for $\calcLambdaMu$-objects}
\label{f:sigma-laurent}
\end{figure}

The first two equations are exactly those of Regnier (hence
$\eqlaurent$ on $\calcLambdaMu$-terms strictly extends $\eqregnier$ on
$\calcLambda$-terms); the remaining ones involve $\mu$-abstractions. It is worth noticing that our equations 
$\simeq_{\sigmalaurent_7}$ and $\simeq_{\sigmalaurent_8}$ are called, respectively,
  $\eqsigma[\theta]$  and $\eqsigma[\rho]$ in~\cite{Laurent03}. 

Laurent proved properties for $\eqlaurent$ on $\calcLambdaMu$-terms
similar to those of Regnier for $\eqregnier$ on $\calcLambda$-terms.
More precisely, $u \eqlaurent v$ implies that $u$ is normalisable
(resp. is head normalisable, strongly normalisable) iff $v$ is
normalisable (resp. is head normalisable, strongly
normalisable)~\cite[Proposition~35]{Laurent03}. Based on Girard's encoding
of classical into intuitionistic logic~\cite{Girard91}, he also proved that
the translation of the left and right-hand sides of the equations of
$\eqlaurent$, in a typed setting, yield structurally equivalent
(polarised) proof-nets~\cite[Theorem~41]{Laurent03}. These results are
non-trivial because the left and right-hand side of the equations in
Figure~\ref{f:sigma-laurent} do not have the same $\beta$ and $\mu$
redexes.  For example,
$\termapp{(\termcont{\al}{\termname{\al}{x}})}{y}$ and
$\termapp{x}{y}$ are related by equation $\sigmalaurent_7$, however
the former has a $\mu$-redex (more precisely it has a \emph{linear}
$\mu$-redex) and the latter has none. Indeed, $\eqlaurent$ is not a
strong bisimulation with respect to $\calcLambdaMu$-reduction, as
mentioned in the introduction (\cf the terms
in~(\ref{eq:thetaAndSBisim})):
\begin{equation}
  \begin{tikzcd}[->,ampersand replacement=\&]
\termapp{(\termcont{\al}{\termname{\al}{x}})}{y} \arrow{d}[left]{\mu}
  \&[-25pt] \simeq_{\sigmalaurent_8}
  \&[-25pt] \termapp{x}{y} \arrow{d}{\mu} \arrow[phantom,negated]{d}{} \\ 
\termcont{\al}{\termname{\al}{\termapp{x}{y}}}  
\&[-25pt] \simeq_{\sigmalaurent_8}
  \&[-25pt] \termapp{x}{y}
\end{tikzcd}
\label{eq:intro:theta}
\end{equation}
The above diagram shows, moreover, that an analogue of Theorem~\ref{thm:strong_bisimulation_int_case_i}  does not hold for $\calcLambdaMu$. There are other examples illustrating that $\eqlaurent$ is not a strong
bisimulation (\cf Section~\ref{s:control:equivalence}). It seems natural to wonder
whether, just like in the intuitionistic case, a more refined notion of
$\calcLambdaMu$-reduction could change this state of affairs; a challenge we
take up in this paper.


%% file: lambda-m-v2.tex
\section{The \calcLambdaM-calculus}
\label{s:control:lambda-m}

As a first step towards the definition of an adequate strongly
bisimilar structural equivalence for the
$\calcLambdaMu$-calculus, we extend its
syntax and operational semantics to a term calculus with explicit operators
for substitution and replacement. 

\subsection{Terms for \calcLambdaM}

We consider again a countably infinite set of
\deft{variables} $\TermVariable$ ($\termvar{x}, \termvar{y}, \ldots$)
and \deft{names} $\TermName$ ($\alpha, \beta,
\ldots$). The set of \deft{objects} $\Object{\calcLambdaM}$,
\deft{terms} $\Term{\calcLambdaM}$, \deft{commands}
$\Command{\calcLambdaM}$, \deft{stacks} and \deft{contexts} of the
$\calcLambdaM$-calculus are given by the following grammar: \[
\begin{array}{rrcl}
\textbf{(Objects)}                & o         & \Coloneq  & t \mid c \mid s \\
\textbf{(Terms)}                  & t         & \Coloneq  & x \mid \termapp{t}{t} \mid \termabs{x}{t} \mid \termcont{\alpha}{c} \mid \termsubs{x}{t}{t} \\
\textbf{(Commands)}               & c         & \Coloneq  & \termname{\alpha}{t} \mid \termrepl[\alpha']{\alpha}{s}{c}  \\
\textbf{(Stacks)}                 & s         & \Coloneq  & t \mid \termpush{t}{s} \\
\textbf{(Contexts)}               & \ctxt{O}  & \Coloneq  & \ctxt{T} \mid \ctxt{C} \mid \ctxt{S}\\
\textbf{(Term Contexts)}          & \ctxt{T}  & \Coloneq  & \Box \mid \termapp{\ctxt{T}}{t} \mid \termapp{t}{\ctxt{T}} \mid \termabs{x}{\ctxt{T}} \mid \termcont{\alpha}{\ctxt{C}} \mid \termsubs{x}{t}{\ctxt{T}} \mid \termsubs{x}{\ctxt{T}}{t} \\
\textbf{(Command Contexts)}       & \ctxt{C}  & \Coloneq  & \boxdot \mid \termname{\alpha}{\ctxt{T}} \mid \termrepl[\alpha']{\alpha}{s}{\ctxt{C}} \mid \termrepl[\alpha']{\alpha}{\ctxt{S}}{c} \\
\textbf{(Stack Contexts)}         & \ctxt{S}  & \Coloneq  & \ctxt{T} \mid \termpush{\ctxt{T}}{s} \mid \termpush{t}{\ctxt{S}} \\
\textbf{(Substitution Contexts)}  & \ctxt{L}  & \Coloneq  & \Box \mid \termsubs{x}{t}{\ctxt{L}} \\
\textbf{(Repl./Ren. Contexts)}    & \ctxt{R}  & \Coloneq  & \boxdot \mid \termrepl[\alpha']{\alpha}{s}{\ctxt{R}} \mid \ntermren{\alpha}{\beta}{\ctxt{R}}
\end{array}
\]

Terms are those of the $\calcLambdaMu$-calculus enriched with
\deft{explicit substitutions (ES)} of the form $\exsubs{x}{u}$.  The
subterm $u$ in a term of the form $\termapp{t}{u}$ (resp.  the ES
$\termsubs{x}{u}{t}$) is called the \deft{argument} of the application
(resp. substitution).  Commands are enriched with \deft{explicit
  replacements} of the
form $\exrepl[\alpha']{\alpha}{s}$ (where the stack $s$ is to
be considered as list of arguments, as \eg in~\cite{Herbelin94}).
Notice that  stacks inside explicit replacements
  are required to be \emph{non-empty}.

Stacks can be concatenated as expected (denoted $\termpush{s}{s'}$ by
abuse of notation): if $s = \termpush{t_0}{\termpush{\ldots}{t_n}}$,
then $\termpush{s}{s'} \eqdef
\termpush{t_0}{\termpush{\ldots}{\termpush{t_n}{s'}}}$; where
$\termpush{\_}{\_}$ is right associative. Given a term $u$, we use the
abbreviation $\termconc{u}{s}$ for the term resulting from the
application of $u$ to all the terms of the stack $s$, \ie if $s =
\termpush{t_0}{\termpush{\ldots}{t_n}}$, then $\termconc{u}{s} \eqdef
\termapp{\termapp{\termapp{u}{t_0}}{\ldots}}{t_n}$. Recall that
application is left associative, so that this operation also is; hence
$\termconc{\termconc{u}{s}}{s'}$ means
$\termconc{(\termconc{u}{s})}{s'}$. The use of stacks in the new calculus is motivated with the forthcoming example just after the definition of the implicit replacement in Section~\ref{s:control:lambda-m:substitution:renaming:and:replacement}.

\deft{Free} and \deft{bound variables} of $\calcLambdaM$-objects
are defined as expected, having the new explicit operators
binding symbols: \ie $\fv{\termsubs{x}{u}{t}} \eqdef (\fv{t} \setminus
\set{x}) \cup \fv{u}$ and  $\bv{t} = \bv{t} \cup \bv{u} \cup \set{x}$.
Concerning \deft{free} and \deft{bound names} of $\calcLambdaM$-objects,
we remark in particular that the occurrences of
$\alpha'$ in the explicit replacements $\termrepl[\alpha']{\alpha}{s}{c}$
are not bound:
\[
\begin{array}{cc}
\begin{array}{r@{\enskip}c@{\enskip}l}
\fn{\termrepl[\alpha']{\alpha}{s}{c}} & \eqdef  & (\fn{c} \setminus \set{\alpha}) \cup \fn{s} \cup \set{\alpha'} \\
\end{array}
&
\begin{array}{r@{\enskip}c@{\enskip}l}
\bn{\termrepl[\alpha']{\alpha}{s}{c}} & \eqdef  & \bn{c} \cup \bn{s} \cup \set{\alpha} \\
\end{array}
\end{array}
\] We work, as usual, modulo $\alpha$-conversion so that bound variables and names
can be renamed. Thus \eg\
$\termsubs{x}{u}{x} \eqalpha \termsubs{y}{u}{y}$, and
$\termrepl[\alpha]{\gamma}{u}{(\termname{\gamma}{x})} \eqalpha
\termrepl[\alpha]{\beta}{u}{(\termname{\beta}{x})}$.  In particular,
we  assume by 
$\alpha$-conversion that $x \notin \fv{u}$ in  $\termsubs{x}{u}{t}$,
and $\alpha
\notin \fn{s}$ in $\termrepl[\alpha']{\alpha}{s}{c}$.

The notions of free and bound variables and names are extended to
contexts by defining $\fv{\Box} = \fv{\boxdot} = \fn{\Box} =
\fn{\boxdot} = \emptyset$.  Then \eg $x$ is bound in
$\termabs{x}{\Box}$, $\termapp{(\termabs{x}{x})}{\Box}$, and $\alpha$
is bound in $\termrepl[\alpha']{\alpha}{s}{\boxdot}$.
Bound names whose scope includes a hole
  $\Box$ or $\boxdot$ cannot be $\alpha$-renamed. An object $o$ is \deft{free for a
  context} $\ctxt{O}$, written $\fc{o}{\ctxt{O}}$,
 if $\fv{o}$ are not captured by binders of $\ctxt{O}$ in $\ctxtapply{\ctxt{O}}{o}$. 
 Thus \eg
 $\fc{\termapp{z}{y}}{\termabs{x}{\termsubs{x'}{w}{\Box}}}$
 and $\fc{x}{\termapp{(\termabs{x}{x})}{\Box}}$ hold but
$\fc{\termapp{x}{y}}{\termabs{x}{\Box}}$ does not hold.
This notion
is naturally extended to sets of objects, \ie $\fc{\S}{\ctxt{O}}$
iff $\fc{o}{\ctxt{O}}$ holds for every $o \in \S$.

\subsection{On Choice of Notation for \calcLambdaM}
\label{s:on_choice_of_notation}
The decomposition (\ref{eq:split_of_mu_without_distance}) of Parigot's
$\calcLambdaMu$-calculus \cite{Parigot92,Parigot93} mentioned in the
introduction, is based on Andou's formalization~\cite{Andou03}. Alternatively,
adopting an explicit formulation of Parigot's original replacement operation,
results in:
\begin{equation}
\begin{array}{rll}
\termapp{(\termcont{\alpha}{c})}{u} & \rrule{} & \termcont{\alpha}{\termrepl[]{\alpha}{u}{c}} \\
\termrepl[]{\alpha}{u}{c}           & \rrule{} & \replapply{\repl[]{\alpha}{u}}{c}
\end{array}
\label{eq:split_of_mu_original}
\end{equation}
\noindent This alternative has the advantage of being relatively simple.
However, it is not without its subtleties. Most notable is determining the
status of names.  Consider an expression such as
$\termcont{\alpha}{\termrepl[]{\alpha}{v}{\termrepl[]{\alpha}{u}{c}}}$,
resulting from reducing $\termapp{\termapp{(\termcont{\alpha}{c})}{u}}{v}$.
One might understand that names are bound by multiple binders. Occurrences of
$\alpha$ in $c$ would thus be bound by three operators: the outermost
$\mu\alpha$ and the two explicit replacements $\termrepl[]{\alpha}{u}{}$ and
$\termrepl[]{\alpha}{v}{}$. Alternatively, the occurrences of $\alpha$ in
$\termrepl[]{\alpha}{u}{}$ and $\termrepl[]{\alpha}{v}{}$ could be understood
as free. In this case, the outermost $\mu\alpha$ binds all free occurrences of
$\alpha$ in $c$ and the two occurrences of $\alpha$ in
$\termrepl[]{\alpha}{u}{}$ and $\termrepl[]{\alpha}{v}{}$. Beyond settling for
one of these two approaches, there is the additional issue that firing
$\termrepl[]{\alpha}{v}{}$ actually doesn't affect $\alpha$ in $c$ at all, for
otherwise the ordering of $u$ and $v$ should be confused. The notion of scope
is lost, and hence the ordering between  $\termrepl[]{\alpha}{u}{}$ and
$\termrepl[]{\alpha}{v}{}$. Presentation
(\ref{eq:split_of_mu_without_distance}) is somewhat heavier but crisper in
terms of meaning.  The previously mentioned term would be recast as
$\termcont{\alpha''}{\termrepl[\alpha'']{\alpha'}{v}{\termrepl[\alpha']{\alpha}{u}{c}}}$.
Here $\alpha$ in $c$ is bound to just one operator, namely
$\termrepl[\alpha']{\alpha}{u}{c}$. Moreover, the dependency between $u$ and
$v$ is now readily apparent: $\alpha'$ is bound by
$\termrepl[\alpha'']{\alpha'}{v}{}$ and $\alpha''$ is bound by $\mu\alpha''$.
In particular, $\mu \alpha''$ does not bind any name in $c$. Presentation
(\ref{eq:split_of_mu}) which we recall below and is the one used in this paper:
\begin{center}
$  \begin{array}{rll}
\termapp{\termsubs{x_n}{v_n}{\termsubs{x_1}{v_1}{(\termcont{\alpha}{c})} \ldots}}{u}  & \rrule{\rM} & \termsubs{x_n}{v_n}{\termsubs{x_1}{v_1}{(\termcont{\alpha'}{\termrepl[\alpha']{\alpha}{u}{c}})} \ldots} \\
\termrepl[\alpha']{\alpha}{u}{c}                                                      & \rrule{\rR} & \replapply{\repl[\alpha']{\alpha}{u}}{c}
   \end{array}$
 \end{center}
has an additional benefit that we shall not get to exploit here but that may be
done so in a continuation of this work (which originally sparked it, in fact).
It has to do with \emph{single replacement}, where one would like to perform
replacement of one occurrence of $\termname{\alpha}{t}$ at a time. Consider a
term such as
$t_0 = \termrepl{\alpha}{u}{(\ldots\termname{\alpha}{x} \ldots
\termname{\alpha}{y} \ldots)}$. We could first  pass the argument $u$ to $x$,
yielding $t_1= (\ldots\termname{\alpha}{xu} \ldots \termname{\alpha}{y} \ldots)
\ldots$, and then to $y$, yielding $t_2 =
(\ldots\termname{\alpha}{xu} \ldots \termname{\alpha}{yu} \ldots) \ldots$. Note
that the partially replaced term $t_1$ must be represented somehow, and in the
simplified syntax we would write
$\termrepl{\alpha}{u}{(\ldots\termname{\alpha}{xu} \ldots \termname{\alpha}{y}
\ldots)}$ which does not make any sense. Indeed, the first argument named
$\alpha$ in $t_1$ has already been replaced while the second one is still
waiting for an argument, a fact not reflected in the syntax. It is exactly in
this framework that  the ternary notion of replacement inherited from
Andou~\cite{Andou03} makes sense.  Our example now reads $t_0 =
\termrepl[\alpha']{\alpha}{u}{(\ldots\termname{\alpha}{x} \ldots
\termname{\alpha}{y} \ldots)}$ which first reduces to
$t_1 = \termrepl[\alpha']{\alpha}{u}{(\ldots\termname{\alpha'}{xu} \ldots
\termname{\alpha}{y} \ldots)}$, then to
$\termrepl[\alpha']{\alpha}{u}{(\ldots\termname{\alpha'}{xu} \ldots
\termname{\alpha'}{yu} \ldots)}$.

\subsection{Substitution, Renaming and Replacement in \calcLambdaM}
\label{s:control:lambda-m:substitution:renaming:and:replacement}

The \deft{application of the implicit substitution} $\subs{x}{u}$ to an
$\calcLambdaM$-object $o$ is defined as the natural extension of that of the
$\calcLambdaMu$-calculus (Section~\ref{s:control:preliminaries:lambda-mu}). We now
detail the applications of implicit replacements and renamings, which are more
subtle. The \deft{application of the implicit replacement}
$\repl[\alpha']{\alpha}{s}$ to an $\calcLambdaM$-object, is defined as the following capture-avoiding operation
(recall that by $\alpha$-conversion $\alpha \notin \fn{s}$ and $\alpha \neq
\alpha'$): \[
\begin{array}{rcll}
\replapply{\repl[\alpha']{\alpha}{s}}{x}                                  & \eqdef  & x \\
\replapply{\repl[\alpha']{\alpha}{s}}{(\termapp{t}{u})}                   & \eqdef  & \termapp{\replapply{\repl[\alpha']{\alpha}{s}}{t}}{\replapply{\repl[\alpha']{\alpha}{s}}{u}} \\
\replapply{\repl[\alpha']{\alpha}{s}}{(\termabs{x}{t})}                   & \eqdef  & \termabs{x}{\replapply{\repl[\alpha']{\alpha}{s}}{t}}                                                                           & x \notin s \\
\replapply{\repl[\alpha']{\alpha}{s}}{(\termcont{\beta}{c})}              & \eqdef  & \termcont{\beta}{\replapply{\repl[\alpha']{\alpha}{s}}{c}}                                                                      & \beta \notin s,  \beta \neq \alpha' \\
\replapply{\repl[\alpha']{\alpha}{s}}{(\termsubs{x}{u}{t})}               & \eqdef  & \termsubs{x}{\replapply{\repl[\alpha']{\alpha}{s}}{u}}{(\replapply{\repl[\alpha']{\alpha}{s}}{t})}                              & x \notin s \\
\replapply{\repl[\alpha']{\alpha}{s}}{(\termname{\alpha}{t})}             & \eqdef  & \termname{\alpha'}{\termconc{(\replapply{\repl[\alpha']{\alpha}{s}}{t})}{s}} \\
\replapply{\repl[\alpha']{\alpha}{s}}{(\termname{\beta}{t})}              & \eqdef  & \termname{\beta}{\replapply{\repl[\alpha']{\alpha}{s}}{t}}                                                                      & \alpha \neq \beta \\
\replapply{\repl[\alpha']{\alpha}{s}}{(\termrepl[\alpha]{\gamma}{s'}{c})} & \eqdef  & \termrepl[\alpha']{\gamma}{\termpush{\replapply{\repl[\alpha']{\alpha}{s}}{s'}}{s}}{(\replapply{\repl[\alpha']{\alpha}{s}}{c})} & \gamma \notin s,  \gamma \neq \alpha' \\
\replapply{\repl[\alpha']{\alpha}{s}}{(\termrepl[\beta]{\gamma}{s'}{c})}  & \eqdef  & \termrepl[\beta]{\gamma}{\replapply{\repl[\alpha']{\alpha}{s}}{s'}}{(\replapply{\repl[\alpha']{\alpha}{s}}{c})}                 & \alpha \neq \beta, \gamma \notin s, \gamma \neq \alpha' \\
\replapply{\repl[\alpha']{\alpha}{s}}{(\termpush{t}{s'})}                 & \eqdef  & \termpush{\replapply{\repl[\alpha']{\alpha}{s}}{t}}{\replapply{\repl[\alpha']{\alpha}{s}}{s'}}
\end{array}
\] Most of the cases in the definition above are straightforward, we only
comment on the interesting ones.  When the implicit replacement
affects an explicit replacement, \ie in the case
$\replapply{\repl[\alpha']{\alpha}{s}}{(\termrepl[\alpha]{\gamma}{s'}{c})}$,
the explicit replacement is \emph{blocking} the implicit replacement
operation over $\gamma$. This means that $\gamma$ and $\alpha$ denote
the same command, but the arguments of $\alpha$ must not be passed to
$\gamma$ yet. This is why the resulting explicit replacement will
accumulate all these arguments in a \emph{stack}, which explains the
need for this data structure inside explicit
replacements.  Examples of these operations are
$\replapply{\repl[\gamma]{\alpha}{\termpush{y_0}{y_1}}}{(\termname{\alpha}{x})}
= \termname{\gamma}{\termapp{\termapp{x}{y_0}}{y_1}}$, and
$\replapply{\repl[\gamma]{\alpha}{y_0}}{(\termrepl[\alpha]{\beta}{z_0}{(\termname{\alpha}{x})})}
=
\termrepl[\gamma]{\beta}{\termpush{z_0}{y_0}}{(\termname{\gamma}{\termapp{x}{y_0}})}$.

\medskip
The \deft{application of the implicit renaming} $\ren{\alpha}{\beta}$ to an
$\calcLambdaM$-object is defined as: \[
\begin{array}{rcll}
\renapply{\ren{\alpha}{\beta}}{x}                                 & \eqdef  & x \\
\renapply{\ren{\alpha}{\beta}}{(\termapp{t}{u})}                  & \eqdef  & \termapp{\renapply{\ren{\alpha}{\beta}}{t}}{\renapply{\ren{\alpha}{\beta}}{u}} \\
\renapply{\ren{\alpha}{\beta}}{(\termabs{x}{t})}                  & \eqdef  & \termabs{x}{\renapply{\ren{\alpha}{\beta}}{t}} \\
\renapply{\ren{\alpha}{\beta}}{(\termcont{\gamma}{c})}            & \eqdef  & \termcont{\gamma}{\renapply{\ren{\alpha}{\beta}}{c}}                                              & \gamma \neq \beta \\
\renapply{\ren{\alpha}{\beta}}{(\termsubs{x}{u}{t})}              & \eqdef  & \termsubs{x}{\renapply{\ren{\alpha}{\beta}}{u}}{(\renapply{\ren{\alpha}{\beta}}{t})} \\
\renapply{\ren{\alpha}{\beta}}{(\termname{\alpha}{t})}            & \eqdef  & \termname{\beta}{\renapply{\ren{\alpha}{\beta}}{t}} \\
\renapply{\ren{\alpha}{\beta}}{(\termname{\delta}{t})}            & \eqdef  & \termname{\delta}{\renapply{\ren{\alpha}{\beta}}{t}}                                              & \alpha \neq \delta \\
\renapply{\ren{\alpha}{\beta}}{(\termrepl[\alpha]{\gamma}{s}{c})} & \eqdef  & \termrepl[\beta]{\gamma}{\renapply{\ren{\alpha}{\beta}}{s}}{(\renapply{\ren{\alpha}{\beta}}{c})}  & \gamma \neq \beta \\
\renapply{\ren{\alpha}{\beta}}{(\termrepl[\delta]{\gamma}{s}{c})} & \eqdef  & \termrepl[\delta]{\gamma}{\renapply{\ren{\alpha}{\beta}}{s}}{(\renapply{\ren{\alpha}{\beta}}{c})} & \alpha \neq \delta, \gamma \neq \beta  \\
\renapply{\ren{\alpha}{\beta}}{(\termpush{t}{s})}                 & \eqdef  & \termpush{\renapply{\ren{\alpha}{\beta}}{t}}{\renapply{\ren{\alpha}{\beta}}{s}}
\end{array}
\]
The three operations  $\subs{x}{u}$, $\repl[\alpha']{\alpha}{s}$ and 
$\ren{\alpha}{\beta}$ are extended to contexts as expected. The table below
summarises the notions of implicit and explicit operations introduced above:
\begin{center}
\begin{tabular}{cll}
\toprule
$\subsapply{\subs{x}{u}}{t}$                & Implicit Substitution & \multirow{2}{*}{Substitute a term for a variable} \\
$\termsubs{x}{u}{t}$                        & Explicit Substitution & \\
\midrule
$\replapply{\repl[\alpha']{\alpha}{s}}{c}$  & Implicit Replacement  & \multirow{2}{*}{Forwards arguments to named terms} \\
$\termrepl[\alpha']{\alpha}{s}{c}$          & Explicit Replacement  & \\
\midrule
$\renapply{\ren{\alpha}{\beta}}{c}$         & Implicit Renaming     & Substitute a name for a name \\
\bottomrule
\end{tabular}
\end{center}

\subsection{Reduction Semantics of \calcLambdaM}
\label{s:control:lambda-m:semantics}

The reduction semantics for $\calcLambdaM$ will be presented in terms of
reduction rules. With an eye placed on the upcoming notion of structural
equivalence $\eqsigma$, we will classify these rules as performing mere
reshuffling of symbols or performing more elaborate work. This classification
is motivated by the multiplicative and exponential nature of the
different redexes of terms of $\calcLambdaM$ as discussed in the introduction.

The first two reduction rules for $\calcLambdaM$ arise from the simple
decomposition of $\beta$-reduction: \[
\begin{array}{rll}
\termapp{\ctxtapply{\ctxt{L}}{\termabs{x}{t}}}{u} & \rrule{\rB} & \ctxtapply{\ctxt{L}}{\termsubs{x}{u}{t}} \\
\termsubs{x}{u}{t}                                & \rrule{\rS} & \subsapply{\subs{x}{u}}{t}
\end{array} \]
where $\rrule{\rB}$ is constrained by the condition $\fc{u}{\ctxt{L}}$. These
have already been studied in the literature, where, as discussed in the
introduction, they suffice to be able to state and prove a strong bisimulation
result for the intuitionistic case
(\cf Theorem~\ref{t:second-bisimulatio-intuitionistic}). As mentioned there, the
first is a simple reshuffling of symbols, which we thus consider to be
\emph{plain} computation, but the second is not. It may substitute $u$ deep
within a term or possibly even erase $u$, involving the use of exponential cuts
in its proof-net semantics and hence considered \emph{meaningful} computation.
Note that $\rB$ operates \emph{at a distance}~\cite{AccattoliK12}, in the sense
that an abstraction and its argument may be separated by an arbitrary number of
explicit substitutions.

The next reduction rules we consider arise from a more subtle decomposition of
$\mu$-reduction. The third reduction rule for $\calcLambdaM$ is:\[
\begin{array}{rll}
\termapp{\ctxtapply{\ctxt{L}}{\termcont{\alpha}{c}}}{u} & \rrule{\rM} & \ctxtapply{\ctxt{L}}{\termcont{\alpha'}{\termrepl[\alpha']{\alpha}{u}{c}}}
\end{array} \]
subject to the constraint $\fc{u}{\ctxt{L}}$, and  $\alpha'$ is a fresh name
(\ie $\alpha' \neq \alpha$, $\alpha' \notin c$ and $\alpha' \notin u$).
This rule is similar in nature to $\rB$ in the sense that it fires a
$\mu$-redex and may be seen to reshuffle symbols. In particular no replacement
actually takes place since it introduces an explicit replacement. Rule $\rM$ is
therefore also considered to perform \emph{plain} computation. Note that $\rM$
operates at a distance too. With the introduction of $\rM$ our notation for
explicit replacement can now be justified. Indeed, following
Parigot~\cite{Parigot92}, one might be tempted to rephrase the reduct of $\rM$
with a binary constructor, writing
$\ctxtapply{\ctxt{L}}{\termcont{\alpha}{\termrepl{\alpha}{u}{c}}}$ on the
right-hand side of the rule $\rM$. This would be incorrect since all free
occurrences of  $\alpha$ in $c$ are bound by the $\alpha$ in
$\termrepl{\alpha}{u}{c}$ which renders the role of ``$\mu\alpha$''
meaningless.

We have not yet finished introducing the reduction rules for $\calcLambdaM$.
All that is missing is a means to execute the explicit replacement introduced
by $\rM$. The natural candidate for executing replacement would be to have just
one reduction rule, namely: $\termrepl[\alpha']{\alpha}{s}{c} \rrule{\rR}
\replapply{\repl[\alpha']{\alpha}{s}}{c}$. However, this is too coarse grained
to be able to obtain our strong bisimulation result (\cf
Section~\ref{s:control:equivalence}) and therefore explicit replacement will be
implemented not by one, but rather by multiple reduction rules. In particular,
these resulting reduction rules can be easily categorised into \emph{plain} and
\emph{meaningful} behavior, according to the following criterion: first, whether  the
bound name $\alpha$ in the command $c$ occurring in the left-hand side
of rule $\rR$ is linear; and, second, if this single occurrence of $\alpha$ in $c$ may be erased or duplicated by reduction or not. We
will next present these rules gradually.

The first of these rules is the
case where $\alpha$ does not occur linearly in $c$ and results in the fourth
reduction rule for $\calcLambdaM$: \[
\begin{array}{rll@{\quad}l}
\termrepl[\alpha']{\alpha}{s}{c}  & \rrule{\rRnl} & \replapply{\repl[\alpha']{\alpha}{s}}{c}  & \fn[\alpha]{c} \neq 1
\end{array}
\]

We still have to address the case where there is a unique occurrence of
$\alpha$ in $c$. Replacing that unique occurrence is not necessarily an act
of mere reshuffling; it depends on where the occurrence of $\alpha$ appears
in $c$. If $\alpha$ appears inside the argument of an application, an explicit
substitution or an explicit replacement, then this single occurrence of
$\alpha$ could be further erased or duplicated. One might say replacing $s$
performs \emph{hereditarily} meaningful work. Let us make this more precise. If
$\alpha$ occurs exactly once in $c$, then the left-hand side
$\termrepl[\alpha']{\alpha}{s}{c}$ of the reduction rule $\rR$
must have one of the following forms: \[
\begin{array}{c}
\termrepl[\alpha']{\alpha}{s}{\ctxtapply{\ctxt{C}}{\termname{\pmb{\alpha}}{t}}} \\
\termrepl[\alpha']{\alpha}{s}{\ctxtapply{\ctxt{C}}{\termrepl[\pmb{\alpha}]{\beta}{s'}{c'}}} \\
\end{array}
\] where the unique free occurrence of $\alpha$ in $c$ has been highlighted in
bold and $\alpha$ does not occur in any of $\ctxt{C}$, $t$, $c'$, $s'$. These
determine the following two instances of the reduction rule
$\termrepl[\alpha']{\alpha}{s}{c} \rrule{\rR}
\replapply{\repl[\alpha']{\alpha}{s}}{c}$ with $\fn[\alpha]{c} = 1$: \[
\begin{array}{rll@{\quad}l}
\termrepl[\alpha']{\alpha}{s}{\ctxtapply{\ctxt{C}}{\termname{\alpha}{t}}}             & \rrule{\rname} & \ctxtapply{\ctxt{C}}{\termname{\alpha'}{\termconc{t}{s}}}              & \alpha \notin \ctxt{C}, \alpha \notin t \\
\termrepl[\alpha']{\alpha}{s}{\ctxtapply{\ctxt{C}}{\termrepl[\alpha]{\beta}{s'}{c'}}} & \rrule{\rcomp} & \ctxtapply{\ctxt{C}}{\termrepl[\alpha']{\beta}{\termpush{s'}{s}}{c'}}  & \alpha \notin \ctxt{C}, \alpha \notin c', \alpha \notin s'
\end{array}
\] The first rule applies the explicit replacement when finding the (only)
occurrence of the name $\alpha$; while the second rule composes the explicit
replacements by concatenating their respective stacks. As mentioned above, we
would like to further identify the case when $\alpha$ appears inside the
argument of an application, an explicit substitution or an explicit replacement
in each of these reduction rules. The ones where it does are called \emph{non-linear} and the ones where it does
not are called \emph{linear}. This results in four (disjoint)
rules. From now on, we call $\rN$ and $\rNnl$ the linear and non-linear
instances of rule $\rN$ame respectively; similarly we have $\rC$ and $\rCnl$
for $\rC$omp:

\begin{center}
\begin{tabular}{l@{\qquad}c@{\qquad}c}
\toprule
\textbf{Rule} & \textbf{Linear Instance} & \textbf{Non-Linear Instance} \\
\midrule
$\rname$ & $\rN$ & $\rNnl$ \\
$\rcomp$ & $\rC$ & $\rCnl$ \\
\bottomrule
\end{tabular}
\end{center}

\medskip
These rules can be formulated with the notion of \emph{linear contexts},
generated by the following grammars: \[
\begin{array}{rrcl}
\textbf{(TT Linear Contexts)}  & \ctxt{LTT}  & \Coloneq  & \Box \mid \termapp{\ctxt{LTT}}{t} \mid \termabs{x}{\ctxt{LTT}} \mid \termcont{\alpha}{\ctxt{LCT}} \mid \termsubs{x}{t}{\ctxt{LTT}} \\
\textbf{(TC Linear Contexts)}  & \ctxt{LTC}  & \Coloneq  & \termapp{\ctxt{LTC}}{t} \mid \termabs{x}{\ctxt{LTC}} \mid \termcont{\alpha}{\ctxt{LCC}} \mid \termsubs{x}{t}{\ctxt{LTC}} \\
\textbf{(CC Linear Contexts)}  & \ctxt{LCC}  & \Coloneq  & \boxdot \mid \termname{\alpha}{\ctxt{LTC}} \mid \termrepl[\alpha']{\alpha}{s}{\ctxt{LCC}} \\
\textbf{(CT Linear Contexts)}  & \ctxt{LCT}  & \Coloneq  & \termname{\alpha}{\ctxt{LTT}} \mid \termrepl[\alpha']{\alpha}{s}{\ctxt{LCT}}
\end{array}
\] where each category $\ctxt{LXY}$ represents the linear context which takes
an object of \emph{sort} $\ctxt{Y}$ and returns another of \emph{sort}
$\ctxt{X}$. For example, $\ctxt{LTC}$ is a linear context taking a command and
generating a term. Indeed, notice that the grammar does not allow the hole
$\boxdot$ to occur inside a parameter (of an application or an ES). With this
definition in place we can, for example, formulate the decomposition of the
reduction rule $\rname$ as follows: \[
\begin{array}{rll@{\quad}l}
\termrepl[\alpha']{\alpha}{s}{\ctxtapply{\ctxt{LCC}}{\termname{\alpha}{t}}} & \rrule{\rN} & \ctxtapply{\ctxt{LCC}}{\termname{\alpha'}{\termconc{t}{s}}} & \alpha \notin \ctxt{LCC}, \alpha \notin t \\
\termrepl[\alpha']{\alpha}{s}{\ctxtapply{\ctxt{C}}{\termname{\alpha}{t}}}   & \rrule{\rNnl} & \ctxtapply{\ctxt{C}}{\termname{\alpha'}{\termconc{t}{s}}} & \ctxt{C}\mbox{ non linear}, \alpha \notin \ctxt{C}, \alpha \notin t
\end{array} \]

The following diagram summarises which instances of $\rR$ are considered plain
and which are considered meaningful:
\[
\input{diagram-r}
\]
Summarizing our analysis, the full set the reduction rules for $\calcLambdaM$ is presented next.

\begin{defi}
\label{def:m:reduction}
Reduction in the \deft{$\calcLambdaM$-calculus} is given by the following
reduction rules closed under arbitrary contexts: \[
\begin{array}{rll@{\quad}l}
\termapp{\ctxtapply{\ctxt{L}}{\termabs{x}{t}}}{u}                                       & \rrule{\rB}   & \ctxtapply{\ctxt{L}}{\termsubs{x}{u}{t}} & \fc{u}{\ctxt{L}} \\
\termsubs{x}{u}{t}                                                                      & \rrule{\rS}   & \subsapply{\subs{x}{u}}{t} \\
\termapp{\ctxtapply{\ctxt{L}}{\termcont{\alpha}{c}}}{u}                                 & \rrule{\rM}   & \ctxtapply{\ctxt{L}}{\termcont{\alpha'}{\termrepl[\alpha']{\alpha}{u}{c}}}  & \fc{u}{\ctxt{L}}, \alpha' \mbox{ fresh } \\
\termrepl[\alpha']{\alpha}{s}{\ctxtapply{\ctxt{LCC}}{\termname{\alpha}{t}}}             & \rrule{\rN}   & \ctxtapply{\ctxt{LCC}}{\termname{\alpha'}{\termconc{t}{s}}}                 & \alpha \notin \ctxt{LCC}, \alpha \notin t \\
\termrepl[\alpha']{\alpha}{s}{\ctxtapply{\ctxt{LCC}}{\termrepl[\alpha]{\beta}{s'}{c'}}} & \rrule{\rC}   & \ctxtapply{\ctxt{LCC}}{\termrepl[\alpha']{\beta}{\termpush{s'}{s}}{c'}}     & \alpha \notin \ctxt{LCC}, \alpha \notin c', \alpha \notin s' \\
\termrepl[\alpha']{\alpha}{s}{\ctxtapply{\ctxt{C}}{\termname{\alpha}{t}}}               & \rrule{\rNnl} & \ctxtapply{\ctxt{C}}{\termname{\alpha'}{\termconc{t}{s}}}                   & \ctxt{C}\mbox{ non linear}, \alpha \notin \ctxt{C}, \alpha \notin t \\
\termrepl[\alpha']{\alpha}{s}{\ctxtapply{\ctxt{C}}{\termrepl[\alpha]{\beta}{s'}{c'}}}   & \rrule{\rCnl} & \ctxtapply{\ctxt{C}}{\termrepl[\alpha']{\beta}{\termpush{s'}{s}}{c'}}       & \ctxt{C}\mbox{ non linear}, \alpha \notin \ctxt{C}, \alpha \notin c', \alpha \notin s' \\
\termrepl[\alpha']{\alpha}{s}{c}                                                        & \rrule{\rRnl} & \replapply{\repl[\alpha']{\alpha}{s}}{c}                                    & \fn[\alpha]{c} \neq 1
\end{array} \]
\end{defi}

Note that $o \reduce[\rR] o'$ iff $o$ reduces to $o'$ using the reduction
obtained from the union of the rules $\rrule{\rN} \cup \rrule{\rC} \cup
\rrule{\rNnl} \cup \rrule{\rCnl} \cup \rrule{\rRnl}$.

\medskip
We can then state that the $\calcLambdaM$-calculus refines the
$\calcLambdaMu$-calculus by decomposing $\beta$ and $\mu$ in more atomic steps.

\begin{lem}
Let $o \in \Object{\calcLambdaMu}$. If $o \reduce[\rlm] o'$, then $o
\reducemany[\rLM] o'$.
\label{l:control:lambda-m:semantics:simulation}
\end{lem}

\begin{proof}
By induction on $o \reduce[\rlm] o'$.
\end{proof}

\medskip
As in the case of the $\calcLambdaMu$-calculus, the $\calcLambdaM$-calculus is
confluent too.

\begin{toappendix}
\begin{thm}
The reduction relation  $\reduce[\rLM]$ is confluent (CR).
\label{t:control:lambda-m:semantics:confluence}
\end{thm}
\end{toappendix}

\begin{proof}
The proof uses the
interpretation method~\cite{CurienHL96}, by projecting the
$\calcLambdaM$-calculus into  the $\calcLambdaMu$-calculus. Details in the
Appendix~\ref{app:confluence}.
\end{proof}


%% file: diagram-r.tex
\begin{tikzpicture}
\node (root)    [] {$\termrepl[\alpha']{\alpha}{s}{c}$};

\node (anq)     [below of=root, xshift=-8.2em, yshift=0.5em]  {$\fn[\alpha]{c} \neq 1$};
\node (aeq)     [below of=root, xshift= 8.2em, yshift=0.5em]  {$\fn[\alpha]{c} = 1$};

\node (name)    [below of=aeq, xshift=-6.5em, yshift=0.5em]     {$c = \ctxtapply{\ctxt{C}}{\termname{\alpha}{t}}$};
\node (repl)    [below of=aeq, xshift= 6.5em, yshift=0.5em]     {$c = \ctxtapply{\ctxt{C}}{\termrepl[\alpha]{\beta}{s'}{c'}}$};

\node (namel)   [below of=name, xshift= 3.25em, yshift=1em]     {\parbox{5pc}{\centering $\ctxt{C}$ linear}};
\node (namenl)  [below of=name, xshift=-3.25em, yshift=1em]     {\parbox{5pc}{\centering $\ctxt{C}$ non linear}};
\node (repll)   [below of=repl, xshift= 3.25em, yshift=1em]     {\parbox{5pc}{\centering $\ctxt{C}$ linear}};
\node (replnl)  [below of=repl, xshift=-3.25em, yshift=1em]     {\parbox{5pc}{\centering $\ctxt{C}$ non linear}};

\node (rb1)     [below of=anq,    yshift=-5em]      {\dashnode{5.5pc}{meaningful computation ($\reduce[\rRnl]$)}};
\node (eqlin)   [below of=namel,  yshift=-1.25em]   {\dashnode{5.5pc}{\plain\ computation ($\reduce[\rN]$)}};
\node (rb2)     [below of=namenl, yshift=-1.25em]   {\dashnode{5.5pc}{meaningful computation ($\reduce[\rNnl]$)}};
\node (comp)    [below of=repll,  yshift=-1.25em]   {\dashnode{5.5pc}{\plain\ computation ($\reduce[\rC]$)}};
\node (rb3)     [below of=replnl, yshift=-1.25em]   {\dashnode{5.5pc}{meaningful computation ($\reduce[\rCnl]$)}};

\draw (root.south)  -- node{} (anq.north)
      (root.south)  -- node{} (aeq.north)
      (aeq.south)   -- node{} (name.north)
      (aeq.south)   -- node{} (repl.north)
      (name.south)  -- node{} (namel.north)
      (name.south)  -- node{} (namenl.north)
      (repl.south)  -- node{} (repll.north)
      (repl.south)  -- node{} (replnl.north);
\draw[->] (anq.south)     -- node{} (rb1.north);
\draw[->] (namel.south)   -- node{} (eqlin.north);
\draw[->] (namenl.south)  -- node{} (rb2.north);
\draw[->] (repll.south)   -- node{} (comp.north);
\draw[->] (replnl.south)  -- node{} (rb3.north);
\end{tikzpicture}

%% file: equivalence.tex
\section{A Strong Bisimulation for \calcLambdaM}
\label{s:control:equivalence}

We now introduce our notion of structural equivalence for $\calcLambdaM$, written
$\eqsigma$, breaking down the presentation into two key tools
on which we have based our development: \plain\ forms and linear
contexts.

\paragraph{\Plain\ Forms}
\label{s:control:equivalence:canonical}

As discussed in Section~\ref{s:control:introduction}, the initial intuition in
defining a strong bisimulation for $\calcLambdaM$ arises from the
intuitionistic case: Regnier's equivalence $\sigmaregnier$ is not a strong
bisimulation, but decomposing the $\rbeta$-rule and taking the $\rB$-normal
form of the left and right hand sides of the equations in
Figure~\ref{f:sigma-equivalence-lambda}, results in the $\sigmaregex$-equivalence
relation terms on $\calcLambda$-terms with explicit substitutions. This
relation turns out to be a strong bisimulation with respect to the notion of
meaningful computation (the relation $\reduce[\rS]$ in the case of
$\calcLambda$-calculus).

One can identify $\rB$ as performing innocuous or plain computation, a fact
that can also be supported by how this step translates as a multiplicative cut
in polarized proof-nets~\cite{Laurent02,Laurent03}. Similarly, one can identify $\rS$ as
performing non-trivial or meaningful work. In the classical case, this leads us
to introduce two restrictions of reduction in $\calcLambdaM$ (\cf
Definition~\ref{def:m:reduction}), one called plain and one called meaningful.
\begin{defi}
\label{def:plain}
The \deft{\plain\ reduction relation} $\reduce[\rcan]$ is defined as the
closure by contexts of the following \plain\ rules: \[
\begin{array}{rll@{\quad}l}
\termapp{\ctxtapply{\ctxt{L}}{\termabs{x}{t}}}{u}                                       & \rrule{\rB} & \ctxtapply{\ctxt{L}}{\termsubs{x}{u}{t}}                                    & \fc{u}{\ctxt{L}} \\
\termapp{\ctxtapply{\ctxt{L}}{\termcont{\alpha}{c}}}{u}                                 & \rrule{\rM} & \ctxtapply{\ctxt{L}}{\termcont{\alpha'}{\termrepl[\alpha']{\alpha}{u}{c}}}  &  \fc{u}{\ctxt{L}}, \alpha' \mbox{ fresh} \\
\termrepl[\alpha']{\alpha}{s}{\ctxtapply{\ctxt{LCC}}{\termname{\alpha}{t}}}             & \rrule{\rN} & \ctxtapply{\ctxt{LCC}}{\termname{\alpha'}{\termconc{t}{s}}}                 & \alpha \notin \ctxt{LCC}, \alpha \notin t \\
\termrepl[\alpha']{\alpha}{s}{\ctxtapply{\ctxt{LCC}}{\termrepl[\alpha]{\beta}{s'}{c'}}} & \rrule{\rC} & \ctxtapply{\ctxt{LCC}}{\termrepl[\alpha']{\beta}{\termpush{s'}{s}}{c'}}     & \alpha \notin \ctxt{LCC}, \alpha \notin c', \alpha \notin s'
\end{array} \]
\ie \[
\mathbin{\reduce[\rcan]} \eqdef
\ctxtapply{\ctxt{O}_\TermSort}{\rrule{\rB} \cup \rrule{\rM}} \cup
\ctxtapply{\ctxt{O}_\CommandSort}{\rrule{\rN} \cup \rrule{\rC}} \]
\end{defi}
The set of \deft{\plain\ forms} of the $\calcLambdaM$-calculus is given by
$\setNF[\rcan] \eqdef \setNF[\rB,\rM,\rN,\rC]$. Moreover, the relation 
$\reduce[\rcan]$ is terminating and confluent.

\begin{toappendix}
\begin{thm}
The relation $\reduce[\rcan]$ is terminating.
\label{t:control:meaningful:can-sn}
\end{thm}
\end{toappendix}

\begin{proof}
We prove this result by resorting to a polynomial interpretation. Details in
Appendix~\ref{app:meaningful}.
\end{proof}

\begin{toappendix}
\begin{thm}
The relation $\reduce[\rcan]$ has the diamond property and hence it is
confluent.
\label{t:control:meaningful:can-confluence}
\end{thm}
\end{toappendix}

\begin{proof}
We prove that the relation $\reduce[\rcan]$ has the diamond property by
inspecting all possible cases. Details can be found in Appendix~\ref{app:meaningful}.
\end{proof}

From now on, we will refer to the (unique) \deft{\plain\  form} of an object
$o$ as $\fcan{o} \eqdef \nf[\rcan]{o}$.

\begin{defi}
\label{d:canonical:reduction}
The \deft{meaningful replacement reduction relation} $\reduce[\rRm]$ is
defined as the closure by contexts of the non-linear rules $\rRnl$, $\rNnl$,
and $\rCnl$, \ie \[
\mathbin{\reduce[\rRm]} \eqdef
\ctxtapply{\ctxt{O}_\CommandSort}{\rrule{\rRnl} \cup \rrule{\rNnl}
\cup \rrule{\rCnl}}
\] The \deft{meaningful reduction relation} $\reducemean$ for the
$\calcLambdaM$-calculus on \plain\ forms is given by: \[
o \reducemean o'
\quad\text{iff}\quad
o \reduce[\rS,\rRm] p \mbox{ and } o' = \fcan{p}
\] We occasionally use $\reducemean[\rS]$ and $\reducemean[\rRm]$ to make
explicit which rule is used in a $\reducemean$-step. 
\end{defi}

For example,
$
\termrepl[\alpha']{\alpha}{u}{(\termname{\alpha}{\termabs{x}{\termcont{\gamma}{\termname{\alpha}{\termabs{y}{\termcont{\delta}{c}}}}}})}
\reduce[\rRm]
\termname{\alpha'}{\termapp{(\termabs{x}{\termcont{\gamma}{\termname{\alpha'}{\termapp{(\termabs{y}{\termcont{\delta}{c'}})}{u}}}})}{u}}
\reducemany[\rcan]
\termname{\alpha'}{\termsubs{x}{u}{(\termcont{\gamma}{\termname{\alpha'}{\termsubs{y}{u}{(\termcont{\delta}{c''})}}})}}
$.

In the classical case, a first attempt to obtaining a strong bisimulation is to
consider the rules that result from taking the $\rcan$-normal form at each side
of those from Laurent's $\sigmalaurent$-equivalence relation (\cf
Figure~\ref{f:sigma-laurent}). The resulting relation $\eqlauex$ on
$\calcLambdaM$-objects is depicted in
Figure~\ref{f:control:equivalence:canonical:eqlaurent-ex}. This equational theory
would be a natural candidate for our strong bisimulation, but unfortunately it is
not the case. As discussed in the introduction, the rule $\ruleEqlauexRho$ breaks
strong bisimulation, so we are thus forced to remove it. But we cannot remove it
completely, as it is required for firing linear
redexes~\cite[Proposition~40]{Laurent03}. It is also required for swapping
explicit renamings in order to close strong bisimulation diagrams
(\cf Theorem~\ref{t:control:bisimulation}), this aspect of $\rho$ is incorporated
as the rule $\ruleEqsigExRen$ in our upcoming $\sigma$ equivalence.

\begin{figure}
\[
\begin{array}{rcl}
\termsubs{x}{u}{(\termabs{y}{t})}                                                                                                                 & \ruleEqlauexAbs       & \termabs{y}{\termsubs{x}{u}{t}}   \\                                   \multicolumn{3}{c}{y \notin u} \\ \\
\termsubs{x}{u}{(\termapp{t}{v})}                                                                                                                 & \ruleEqlauexApp       & \termapp{\termsubs{x}{u}{t}}{v}  \\
           \multicolumn{3}{c}{x \notin v} \\ \\
           \termsubs{x}{u}{(\termcont{\beta}{\termname{\alpha}{t}})}                                                                                         & \ruleEqlauexCont      & \termcont{\beta}{\termname{\alpha}{\termsubs{x}{u}{t}}}\\
           \multicolumn{3}{c}{\beta \notin u} \\ \\
           \termname{\alpha'}{\termcont{\alpha''}{\termrepl[\alpha'']{\alpha}{u}{(\termname{\beta'}{\termcont{\beta''}{\termrepl[\beta'']{\beta}{v}{c}}})}}} & \ruleEqlauexPushPush  & \termname{\beta'}{\termcont{\beta''}{\termrepl[\beta'']{\beta}{v}{(\termname{\alpha'}{\termcont{\alpha''}{\termrepl[\alpha'']{\alpha}{u}{c}}})}}} \\
           \multicolumn{3}{c}{\alpha \notin v, \alpha'' \notin v, \beta \notin u, \beta'' \notin u} \\ \\
           \termname{\alpha'}{\termcont{\alpha''}{\termrepl[\alpha'']{\alpha}{u}{(\termname{\beta'}{\termabs{y}{\termcont{\beta}{c}}})}}}                    & \ruleEqlauexPushPop   & \termname{\beta'}{\termabs{y}{\termcont{\beta}{\termname{\alpha'}{\termcont{\alpha''}{\termrepl[\alpha'']{\alpha}{u}{c}}}}}}                     \\
           \multicolumn{3}{c}{y \notin u, \beta \notin u, \alpha'' \notin u}\\ \\
\termname{\alpha'}{\termabs{x}{\termcont{\alpha}{\termname{\beta'}{\termabs{y}{\termcont{\beta}{c}}}}}}                                           & \ruleEqlauexPopPop    & \termname{\beta'}{\termabs{y}{\termcont{\beta}{\termname{\alpha'}{\termabs{x}{\termcont{\alpha}{c}}}}}}          \\ \\
\termcont{\alpha}{\termname{\alpha}{t}}                                                                                                           & \ruleEqlauexTheta     & t                            \\
                                    \multicolumn{3}{c}{\alpha \notin t} \\ \\
            \termname{\beta}{\termcont{\alpha}{c}}                                                          & \ruleEqlauexRho       & \renapply{\ren{\alpha}{\beta}}{c} \\ \\
\termsubs{x}{u}{\termsubs{y}{v}{t}}                                                                                                               & \ruleEqlauexCom       & \termsubs{y}{v}{\termsubs{x}{u}{t}}  \\
\multicolumn{3}{c}{y \notin u, x \notin v}
\end{array}
\]
\caption{$\sigmalauex$-equivalence on $\calcLambdaM$-objects.}
\label{f:control:equivalence:canonical:eqlaurent-ex}
\end{figure}

\paragraph{Linear Contexts}
\label{s:control:equivalence:linear-contexts}
Linear contexts turn out to be very useful when decomposing the rewriting
rule $\rR$ (\cf Section~\ref{s:control:lambda-m:semantics}). Here they are used
once again to reduce the amount of necessary rules for the equivalence
relation. Note that linear contexts are not only used to support commutation of
explicit substitution but also for explicit replacement.

Indeed, the equations $\ruleEqlauexAbs$, $\ruleEqlauexApp$, $\ruleEqlauexCont$
and $\ruleEqlauexCom$ in
Figure~\ref{f:control:equivalence:canonical:eqlaurent-ex} are generalised into a
single equation reflecting the fact that an explicit substitution commutes with
linear contexts. Something similar can be stated for rules
$\ruleEqlauexPushPush$ and $\ruleEqlauexPushPop$, between linear contexts and
explicit replacement. Moreover, linear contexts can be skipped by any explicit
operator (substitution or replacement) as long as they are independent, \ie no
undesired capture of free variables/names takes place.

Therefore, we introduce into our equivalence relation $\eqsigma$ three equations
that in turn replace rules $\ruleEqlauexAbs$, $\ruleEqlauexApp$,
$\ruleEqlauexCont$, $\ruleEqlauexPushPush$, $\ruleEqlauexPushPop$ and
$\ruleEqlauexCom$ from Figure~\ref{f:control:equivalence:canonical:eqlaurent-ex},
while extending its behavior to explicit renaming as well: \[
\begin{array}{rcl@{\quad}l}
\termsubs{x}{u}{\ctxtapply{\ctxt{LTT}}{t}}                      & \ruleEqsigExSubs  & \ctxtapply{\ctxt{LTT}}{\termsubs{x}{u}{t}}                      & x \notin \ctxt{LTT}, \fc{u}{\ctxt{LTT}} \\
\termrepl[\alpha']{\alpha}{s}{\ctxtapply{\ctxt{LCC}}{c}}        & \ruleEqsigExRepl  & \ctxtapply{\ctxt{LCC}}{\termrepl[\alpha']{\alpha}{s}{c}}        & \alpha \notin \ctxt{LCC}, \fc{\alpha'}{\ctxt{LCC}}, \fc{s}{\ctxt{LCC}} \\
\termname{\beta}{ \termcont{\alpha}{\ctxtapply{\ctxt{LCC}}{c}}} & \ruleEqsigExRen   & \ctxtapply{\ctxt{LCC}}{\termname{\beta}{\termcont{\alpha}{c}}}  & \alpha \notin \ctxt{LCC}, \fc{\beta}{\ctxt{LCC}}
\end{array}
\]

\paragraph{Structural Equivalence}
\label{s:control:equivalence:eqsigma}
We are now ready to present $\eqsigma$.

\begin{defi}[Structural Equivalence over \Plain\ Forms]
\label{d:control:structural:equivalence}
The \emph{structural equivalence relation} $\eqsigma$ is the least congruence relation
over \plain\ forms of the $\calcLambdaM$-calculus generated by the rules in
Figure~\ref{f:control:equivalence:eqsigma}.
\label{d:control:equivalence:eqsigma}
\end{defi}

\begin{figure}[!h]
\[
\begin{array}{rcl@{\quad}l}
\termsubs{x}{u}{\ctxtapply{\ctxt{LTT}}{t}}                                                              & \ruleEqsigExSubs  & \ctxtapply{\ctxt{LTT}}{\termsubs{x}{u}{t}}                                                              & x \notin \ctxt{LTT}, \fc{u}{\ctxt{LTT}} \\
\termrepl[\alpha']{\alpha}{s}{\ctxtapply{\ctxt{LCC}}{c}}                                                & \ruleEqsigExRepl  & \ctxtapply{\ctxt{LCC}}{\termrepl[\alpha']{\alpha}{s}{c}}                                                & \alpha \notin \ctxt{LCC}, \fc{\alpha'}{\ctxt{LCC}}, \fc{s}{\ctxt{LCC}} \\
\ntermren{\alpha}{\beta}{\ctxtapply{\ctxt{LCC}}{c}}                                                     & \ruleEqsigExRen   & \ctxtapply{\ctxt{LCC}}{\ntermren{\alpha}{\beta}{c}}                                                     & \alpha \notin \ctxt{LCC}, \fc{\beta}{\ctxt{LCC}} \\
\termname{\alpha'}{\termabs{x}{\termcont{\alpha}{\termname{\beta'}{\termabs{y}{\termcont{\beta}{c}}}}}} & \ruleEqsigPopPop  & \termname{\beta'}{\termabs{y}{\termcont{\beta}{\termname{\alpha'}{\termabs{x}{\termcont{\alpha}{c}}}}}} & \beta \neq \alpha', \alpha \neq \beta' \\
\termcont{\alpha}{\termname{\alpha}{t}}                                                                 & \ruleEqsigTheta   & t                                                                                                       & \alpha \notin t
\end{array}
\]
\caption{Structural equivalence $\eqsigma$ on $\calcLambdaM$-objects.}
\label{f:control:equivalence:eqsigma}
\end{figure}

First a result on commutation between linear contexts and explicit
operators:

\begin{toappendix}
\begin{lem}\mbox{}
\begin{enumerate}
  \item\label{l:control:equivalence:permute:l} Let $t \in \Term{\calcLambdaM}$,
  $\ctxt{L}$ be a substitution context and $\ctxt{LTT}$ a linear context. \\
  Then, $\fcan{\ctxtapply{\ctxt{L}}{\ctxtapply{\ctxt{LTT}}{t}}} \eqsigma
  \fcan{\ctxtapply{\ctxt{LTT}}{\ctxtapply{\ctxt{L}}{t}}}$ if $\bv{\ctxt{L}}
  \notin \ctxt{LTT}$ and $\fc{\ctxt{L}}{\ctxt{LTT}}$.
  
  \item\label{l:control:equivalence:permute:r} Let $c \in \Command{\calcLambdaM}$,
  $\ctxt{R}$ be a repl./ren. context and $\ctxt{LCC}$ a linear context. \\
  Then, $\fcan{\ctxtapply{\ctxt{R}}{\ctxtapply{\ctxt{LCC}}{c}}} \eqsigma
  \fcan{\ctxtapply{\ctxt{LCC}}{\ctxtapply{\ctxt{R}}{c}}}$ if $\bn{\ctxt{R}}
  \notin \ctxt{LCC}$ and $\fc{\ctxt{R}}{\ctxt{LCC}}$.
\end{enumerate}
\label{l:control:equivalence:permute}
\end{lem}
\end{toappendix}

\begin{proof}
Each case is proved by induction on $\ctxt{L}$ or $\ctxt{R}$ respectively, using
some auxiliary results. Details in the Appendix~\ref{a:equivalence}.  
\end{proof}

Note that $\eqsigma$ is not a congruence on arbitrary terms but on plain forms.
For example, $\termcont{\alpha}{\termname{\alpha}{x}}$ and $x$ are both in
\plain\ form and moreover $\termcont{\alpha}{\termname{\alpha}{x}} \eqsigma x$.
However, $\termapp{(\termcont{\alpha}{\termname{\alpha}{x}})}{y} \not \eqsigma
\termapp{x}{y}$ since $\termapp{(\termcont{\alpha}{\termname{\alpha}{x}})}{y}$
is not a \plain\ form. Nevertheless,
$\fcan{\termapp{(\termcont{\alpha}{\termname{\alpha}{x}})}{y}} =
\termcont{\alpha}{\termname{\alpha}{\termapp{x}{y}}} \eqsigma \termapp{x}{y} =
\fcan{\termapp{x}{y}}$. More generally:
\begin{toappendix}
\begin{lem}
Let $o, o' \in \Object{\calcLambdaM}$. If $o \eqsigma o'$, then for all context
$\ctxt{O}$ of appropriate sort, $\fcan{\ctxtapply{\ctxt{O}}{o}} \eqsigma
\fcan{\ctxtapply{\ctxt{O}}{o'}}$.
\label{l:control:bisimulation:ctxt-closure}
\end{lem}
\end{toappendix}

\begin{proof}
By induction on the size of $\ctxt{O}$. Details in the Appendix~\ref{a:bisimulation}.  
\end{proof}   

Notice that $o \eqsigma o'$ implies $\fv{o} = \fv{o'}$ and $\fn{o} = \fn{o'}$.
Moreover, all the rules from
Figure~\ref{f:control:equivalence:canonical:eqlaurent-ex} except
$\ruleEqlauexRho$ are indeed admissible:
\begin{itemize}
  \item $\ruleEqlauexAbs$: $\termsubs{x}{u}{(\termabs{y}{t})} \ruleEqsigExSubs
  \termabs{y}{\termsubs{x}{u}{t}}$ with $y \notin u$ and $\ctxt{LTT} =
  \termabs{y}{\Box}$.
  
  \item $\ruleEqlauexApp$: $\termsubs{x}{u}{(\termapp{t}{v})} \ruleEqsigExSubs
  \termapp{\termsubs{x}{u}{t}}{v}$ with $x \notin v$ and $\ctxt{LTT} =
  \termapp{\Box}{v}$.
  
  \item $\ruleEqlauexCont$: $\termsubs{x}{u}{(\termcont{\beta}{\termname{\alpha}{t}})} \ruleEqsigExSubs
  \termcont{\beta}{\termname{\alpha}{\termsubs{x}{u}{t}}}$ with $\alpha \notin
  u$ and $\ctxt{LTT} = \termcont{\beta}{\termname{\alpha}{\Box}}$.
  
  \item $\ruleEqlauexPushPush$:
  $\ntermren{\alpha''}{\alpha'}{\termrepl[\alpha'']{\alpha}{u}{(\ntermren{\beta''}{\beta'}{\termrepl[\beta'']{\beta}{v}{c}})}}
  \eqsigma
  \ntermren{\beta''}{\beta'}{\termrepl[\beta'']{\beta}{v}{(\ntermren{\alpha''}{\alpha'}{\termrepl[\alpha'']{\alpha}{u}{c}})}}$
  with $\alpha \notin v, \alpha'' \notin v, \beta \notin u, \beta'' \notin u,
  \beta'' \neq \alpha', \alpha'' \neq \beta'$. This case is particularly
  interesting since rules $\ruleEqsigExRepl$ and $\ruleEqsigExRen$ play a key
  role in it: \[
\begin{array}{lll}
\ntermren{\alpha''}{\alpha'}{\termrepl[\alpha'']{\alpha}{u}{(\ntermren{\beta''}{\beta'}{\termrepl[\beta'']{\beta}{v}{c}})}} &  \ruleEqsigExRepl \\
\ntermren{\alpha''}{\alpha'}{\ntermren{\beta''}{\beta'}{\termrepl[\beta'']{\beta}{v}{\termrepl[\alpha'']{\alpha}{u}{c}}}}   &  \ruleEqsigExRen \\
\ntermren{\beta''}{\beta'}{\ntermren{\alpha''}{\alpha'}{\termrepl[\beta'']{\beta}{v}{\termrepl[\alpha'']{\alpha}{u}{c}}}}   &  \ruleEqsigExRepl \\
\ntermren{\beta''}{\beta'}{\termrepl[\beta'']{\beta}{v}{(\ntermren{\alpha''}{\alpha'}{\termrepl[\alpha'']{\alpha}{u}{c}})}}
\end{array} \]

  \item $\ruleEqlauexPushPop$:
  $\ntermren{\alpha''}{\alpha'}{\termrepl[\alpha'']{\alpha}{u}{(\termname{\beta'}{\termabs{y}{\termcont{\beta}{c}}})}}
  \eqsigma
  \termname{\beta'}{\termabs{y}{\termcont{\beta}{\ntermren{\alpha''}{\alpha'}{\termrepl[\alpha'']{\alpha}{u}{c}}}}}$
  with $y \notin u, \beta \notin u, \beta'' \notin u, \beta'' \neq \alpha',
  \alpha'' \neq \beta'$. As in the previous case, we can conclude thanks to
  rules $\ruleEqsigExRepl$ and $\ruleEqsigExRen$: \[
\begin{array}{lll}
\ntermren{\alpha''}{\alpha'}{\termrepl[\alpha'']{\alpha}{u}{(\termname{\beta'}{\termabs{y}{\termcont{\beta}{c}}})}} & \ruleEqsigExRepl \\
\ntermren{\alpha''}{\alpha'}{\termname{\beta'}{\termabs{y}{\termcont{\beta}{\termrepl[\alpha'']{\alpha}{u}{c}}}}} & \ruleEqsigExRen \\
\termname{\beta'}{\termabs{y}{\termcont{\beta}{\ntermren{\alpha''}{\alpha'}{\termrepl[\alpha'']{\alpha}{u}{c}}}}}
\end{array} \]
  
  \item $\ruleEqlauexPopPop$:
  $\termname{\alpha'}{\termabs{x}{\termcont{\alpha}{\termname{\beta'}{\termabs{y}{\termcont{\beta}{c}}}}}}
  \ruleEqsigPopPop
  \termname{\beta'}{\termabs{y}{\termcont{\beta}{\termname{\alpha'}{\termabs{x}{\termcont{\alpha}{c}}}}}}$.

  \item $\ruleEqlauexTheta$: $\termcont{\alpha}{\termname{\alpha}{t}} \ruleEqsigTheta t$.

  \item $\ruleEqlauexCom$: $\termsubs{x}{u}{\termsubs{y}{v}{t}} \ruleEqsigExSubs
  \termsubs{y}{v}{\termsubs{x}{u}{t}}$ with $y \notin u$, $x \notin v$ and
  $\ctxt{LTT} = \termsubs{y}{v}{\Box}$.
\end{itemize}

\medskip
Hence, commutation rules $\ruleEqlauexAbs$, $\ruleEqlauexApp$,
$\ruleEqlauexCont$ and $\ruleEqlauexCom$ from
Figure~\ref{f:control:equivalence:canonical:eqlaurent-ex} are replaced by
$\ruleEqsigExSubs$, while $\ruleEqsigExRepl$ and $\ruleEqsigExRen$
replace both $\ruleEqlauexPushPush$ and $\ruleEqlauexPushPop$, and
$\ruleEqlauexRho$ is discarded. Only rules $\ruleEqlauexPopPop$ and
$\ruleEqlauexTheta$ remain unaltered, here called $\ruleEqsigPopPop$ (for
$\mathtt{p}$op/$\mathtt{p}$op) and $\ruleEqsigTheta$ respectively
(see~\cite{Laurent03} for the origin of these names). The following table
summarises the correspondence between the different rules:
\medskip
\begin{center}
\begin{tabular}{cc}
\toprule
\textbf{Captured rule from $\sigmalauex$-equivalence}                           & \textbf{New rule} \\
\midrule
$\ruleEqlauexAbs$, $\ruleEqlauexApp$, $\ruleEqlauexCont$ and $\ruleEqlauexCom$  & $\ruleEqsigExSubs$ \\
\midrule
$\ruleEqlauexPushPush$ and $\ruleEqlauexPushPop$                                & $\ruleEqsigExRepl$ and $\ruleEqsigExRen$\\
\midrule
$\ruleEqlauexPopPop$                                                            & $\ruleEqsigPopPop$ \\
\midrule
$\ruleEqlauexRho$                                                               & \\
\midrule
$\ruleEqlauexTheta$                                                             & $\ruleEqsigTheta$ \\
\bottomrule
\end{tabular}
\end{center}

\paragraph{Strong Bisimulation Result}
\label{s:control:bisimulation}

The resulting equivalence relation $\eqsigma$ is in fact a strong bisimulation
with respect to the notion of meaningful computation, as we will show. First a
simple result relating plain reduction and structural equivalence.

\begin{toappendix}
\begin{lem}
Let $u,s,o \in \Object{\calcLambdaM}$ in $\rcan$-normal form such that $o
\eqsigma o'$, $u \eqsigma u'$ and  $s \eqsigma s'$, where $u,u'$ are terms and
$s,s'$ are stacks. Then, 
\begin{enumerate}
  \item\label{l:control:bisimulation:meta-eqsigma:subs}
  $\fcan{\subsapply{\subs{x}{u}}{o}} \eqsigma
  \fcan{\subsapply{\subs{x}{u}}{o'}}$ and  $\fcan{\subsapply{\subs{x}{u}}{o}}
  \eqsigma \fcan{\subsapply{\subs{x}{u'}}{o}}$.
  
  \item\label{l:control:bisimulation:meta-eqsigma:repl}
  $\fcan{\replapply{\repl[\alpha']{\alpha}{s}}{o}} \eqsigma
  \fcan{\replapply{\repl[\alpha']{\alpha}{s}}{o'}}$ and 
  $\fcan{\replapply{\repl[\alpha']{\alpha}{s}}{o}} \eqsigma
  \fcan{\replapply{\repl[\alpha']{\alpha}{s'}}{o}}$.
\end{enumerate}
\label{l:control:bisimulation:meta-eqsigma}
\end{lem}
\end{toappendix}

\begin{proof}
Each case is proved by induction on $o \eqsigma o'$. Details in the
Appendix~\ref{a:bisimulation}.  
\end{proof}   

We are now able to state the promised result, namely, the fact that $\eqsigma$
is a strong bisimulation with respect to the meaningful computation relation.

\begin{toappendix}
\begin{thm}
Let $o, p \in \Object{\calcLambdaM}$. If $o \eqsigma p$ and $o \reducemean o'$,
then there exists $p'$ such that  $p \reducemean p'$  and  $o' \eqsigma p'$.
\label{t:control:bisimulation}
\end{thm}
\end{toappendix}

\begin{proof}
The proof is by induction on  $o \eqsigma p$ and uses
Lemma~\ref{l:control:equivalence:permute},
Lemma~\ref{l:control:bisimulation:ctxt-closure} and 
Lemma~\ref{l:control:bisimulation:meta-eqsigma}. All the details are in the
Appendix~\ref{a:bisimulation}.
\end{proof}

\begin{exa}
\label{e:rho-necessario}
We illustrate the previous theorem with the following example. Let $o \!=\!
\termrepl[\alpha']{\alpha}{u}{o_0} \!\ruleEqsigPopPop\!
\termrepl[\alpha']{\alpha}{u}{p_0} \!=\! p$, where $o_0 \!=\!
\termname{\alpha}{\termabs{x}{\termcont{\gamma}{\termname{\alpha}{\termabs{y}{\termcont{\delta}{c}}}}}}
\!\ruleEqsigPopPop\!
\termname{\alpha}{\termabs{y}{\termcont{\delta}{\termname{\alpha}{\termabs{x}{\termcont{\gamma}{c}}}}}}
\!=\! p_0$. Notice in particular that $o_0$, $o$, $p_0$ and $p$ are all \plain\ 
forms. We have \[
\begin{tikzcd}[->,ampersand replacement=\&]
o = \termrepl[\alpha']{\alpha}{u}{(\termname{\alpha}{\termabs{x}{\termcont{\gamma}{\termname{\alpha}{\termabs{y}{\termcont{\delta}{c}}}}}})} \arrow[->]{d}[left]{\rRm}
  \&[-25pt] \ruleEqsigPopPop
  \&[-25pt] \termrepl[\alpha']{\alpha}{u}{(\termname{\alpha}{\termabs{y}{\termcont{\delta}{\termname{\alpha}{\termabs{x}{\termcont{\gamma}{c}}}}}})} = p \arrow[->]{d}[left]{\rRm} \\
\termname{\alpha'}{\termapp{(\termabs{x}{\termcont{\gamma}{\termname{\alpha'}{\termapp{(\termabs{y}{\termcont{\delta}{c'}})}{u}}}})}{u}} \arrow[->>]{d}[left]{\rcan}
  \&[-25pt]
  \&[-25pt] \termname{\alpha'}{\termapp{(\termabs{y}{\termcont{\delta}{\termname{\alpha'}{\termapp{(\termabs{x}{\termcont{\gamma}{c'}})}{u}}}})}{u}} \arrow[->>]{d}[left]{\rcan} \\
o' = \termname{\alpha'}{\termsubs{x}{u}{(\termcont{\gamma}{\termname{\alpha'}{\termsubs{y}{u}{(\termcont{\delta}{c''})}}})}}
  \&[-25pt] \eqsigma[\mathtt{exsubs},\mathtt{exren}]
  \&[-25pt] \termname{\alpha'}{\termsubs{y}{u}{(\termcont{\delta}{\termname{\alpha'}{\termsubs{x}{u}{(\termcont{\gamma}{c''})}}})}} = p'
\end{tikzcd}
\]
so that $o \reducemean o'$  and $p \reducemean p'$.
We conclude $o' \eqsigma p'$ as follows: 
\[ \begin{array}{ll}
  \termname{\alpha'}{\termsubs{x}{u}{(\termcont{\gamma}{\termname{\alpha'}{\termsubs{y}{u}{(\termcont{\delta}{c''})}}})}} & \ruleEqsigExSubs \\
  \ntermren{\gamma}{\alpha'}{\termname{\alpha'}{\termsubs{x}{u}{\termsubs{y}{u}{(\termcont{\delta}{c''})}}}}              & \ruleEqsigExRen \\
  \termname{\alpha'}{\termsubs{x}{u}{\termsubs{y}{u}{(\termcont{\delta}{\ntermren{\gamma}{\alpha'}{c''}})}}}              & \ruleEqsigExSubs \\
  \termname{\alpha'}{\termsubs{y}{u}{(\termcont{\delta}{\termname{\alpha'}{\termsubs{x}{u}{(\termcont{\gamma}{c''})}}})}}
\end{array} \]
Note the use of $\ruleEqsigExRen$ to swap the occurrences of
$\ntermren{\gamma}{\alpha'}{\ldots}$ and $\ntermren{\delta}{\alpha'}{\ldots}$.
\end{exa}


%% file: new.tex
\section{Correspondence result}
\label{s:control:correspondence}

In this section we relate our bisimulation $\eqsigma$ on $\calcLambdaM$-objects to
  Laurent's original $\sigma$-equivalence on
  $\calcLambdaMu$-objects.  More precisely, we show that
 $\eqsigma$ can be seen as Laurent's $\sigma$-equivalence devoid of axiom
  $\simeq_{\sigmalaurentrho}$ (usually called $\rho$-axiom) but
  enriched with three additional axioms $ \ruleEqnewRenRen$,
  $\ruleEqnewRenPop$, and $\ruleEqnewRenPush $ (also called
  $\eqname{ren/ren} $, $\eqname{ren/pop} $ and $\eqname{ren/push}$ resp.), which are admissible in Laurent's $\sigma$-equivalence. Indeed,  all of $ \ruleEqnewRenRen$,
  $\ruleEqnewRenPop$, and $\ruleEqnewRenPush $ are derivable from $\simeq_{\sigmalaurentrho}$, but the converse does not hold. In this sense, $\eqnew$ can be seen as a restriction of Laurent's $\sigma$-equivalence.

\begin{defi}
\label{d:sigma-new}
The new $\eqnew$-equivalence for
$\calcLambdaMu$-objects  is
depicted in Figure~\ref{f:sigmanew}.
\end{defi}
  
\begin{figure}[h]
\[
\begin{array}{rcll}
\termapp{(\termabs{y}{\termabs{x}{t}})}{v}                                                                  & \ruleEqnewAbs       & \termabs{x}{\termapp{(\termabs{y}{t})}{v}}                                                                  & x \notin v \\
\termapp{(\termabs{x}{\termapp{t}{v}})}{u}                                                                  & \ruleEqnewApp       & \termapp{\termapp{(\termabs{x}{t})}{u}}{v}                                                                  & x \notin v \\
\termapp{(\termabs{x}{\termcont{\alpha}{\termname{\beta}{t}}})}{u}                                          & \ruleEqnewCont      & \termcont{\alpha}{\termname{\beta}{\termapp{(\termabs{x}{t})}{u}}}                                          & \alpha \notin u \\
\termname{\alpha'}{\termapp{(\termcont{\alpha}{\termname{\beta'}{\termapp{(\termcont{\beta}{c})}{v}}})}{u}} & \ruleEqnewPushPush  & \termname{\beta'}{\termapp{(\termcont{\beta}{\termname{\alpha'}{\termapp{(\termcont{\alpha}{c})}{u}}})}{v}} & \alpha \notin v, \beta \notin u, \beta \neq \alpha', \alpha \neq \beta' \\
\termname{\alpha'}{\termapp{(\termcont{\alpha}{\termname{\beta'}{\termabs{y}{\termcont{\beta}{c}}}})}{u}}   & \ruleEqnewPushPop   & \termname{\beta'}{\termabs{y}{\termcont{\beta}{\termname{\alpha'}{\termapp{(\termcont{\alpha}{c})}{u}}}}}   & y \notin u, \beta \notin u, \beta \neq \alpha', \alpha \neq \beta' \\
\termname{\alpha'}{\termabs{x}{\termcont{\alpha}{\termname{\beta'}{\termabs{y}{\termcont{\beta}{c}}}}}}     & \ruleEqnewPopPop    & \termname{\beta'}{\termabs{y}{\termcont{\beta}{\termname{\alpha'}{\termabs{x}{\termcont{\alpha}{c}}}}}}     & \beta \neq \alpha', \alpha \neq \beta' \\
\termcont{\alpha}{\termname{\alpha}{t}}                                                                     & \ruleEqnewTheta     & t                                                                                                           & \alpha \notin t \\
\ntermren{\alpha}{\alpha'}{\ntermren{\beta}{\beta'}{c}}                                                     & \ruleEqnewRenRen    & \ntermren{\beta}{\beta'}{\ntermren{\alpha}{\alpha'}{c}}                                                     & \beta \neq \alpha', \alpha \neq \beta' \\
\ntermren{\alpha}{\alpha'}{\termname{\beta'}{\termabs{y}{\termcont{\beta}{c}}}}                             & \ruleEqnewRenPop    & \termname{\beta'}{\termabs{y}{\termcont{\beta}{\ntermren{\alpha}{\alpha'}{c}}}}                             & \beta \neq \alpha', \alpha \neq \beta' \\
\ntermren{\alpha}{\alpha'}{\termname{\beta'}{\termapp{(\termcont{\beta}{c})}{v}}}                           & \ruleEqnewRenPush   & \termname{\beta'}{\termapp{(\termcont{\beta}{\ntermren{\alpha}{\alpha'}{c}})}{v}}                           & \alpha \notin v, \beta \neq \alpha', \alpha \neq \beta'
\end{array} \]
\caption{New $\eqnew$-equivalence for $\calcLambdaMu$-objects}
\label{f:sigmanew}
\end{figure}

The new $\eqnew$-equivalence is built by
removing axiom $\simeq_{\sigmalaurentrho}$ from Laurent's $\sigma$-equivalence, and by adding $ \ruleEqnewRenRen$,
$\ruleEqnewRenPop$, and $\ruleEqnewRenPush $.  Notice that axiom 
$\simeq_{\tau_i}$ in Figure~\ref{f:sigmanew} is exactly the same as $\simeq_{\sigma_i} $ in Figure~\ref{f:sigma-laurent} for $i=1 \ldots 7$.  Moreover, notice that none of $ \ruleEqnewRenRen$,
  $\ruleEqnewRenPop$, and $\ruleEqnewRenPush$ erase or introduce explicit renamings.

Some axioms of the new relation $\eqnew$ can be  generalized to several arguments. For that, we use the meta-notation $\termconc{u}{s}$
  introduced in Section~\ref{s:control:lambda-m},
in this case denoting a term in $\Term{\calcLambdaMu}$, resulting from the
application of $u$ to a stack $s$ of terms in $\Term{\calcLambdaMu}$, \ie if $s =
\termpush{t_0}{\termpush{\ldots}{t_n}}$ and $t_0 \ldots t_n \in \Term{\calcLambdaMu}$, then $\termconc{u}{s} \eqdef
\termapp{\termapp{\termapp{u}{t_0}}{\ldots}}{t_n}$ denotes a term in $\Term{\calcLambdaMu}$.

\begin{toappendix}
\begin{lem}
  Let $t, u \in \Term{\calcLambdaMu}$ and $c \in \Command{\calcLambdaMu}$. Let $s$ be a stack of terms in $\Term{\calcLambdaMu}$.
Let $x \notin {s}, \alpha \notin {s'}, \beta \notin {s}, \beta \neq \alpha',
\alpha \neq \beta'$. Then, 
\begin{enumerate}
  \item\label{l:eqnew-stack:app}
  $\termconc{(\termapp{(\termabs{x}{t})}{u})}{s} \eqnew \termapp{(\termabs{x}{\termconc{t}{s}})}{u}$

  \item\label{l:eqnew-stack:push}
  $\termname{\alpha'}{\termconc{(\termcont{\alpha}{\termname{\beta'}{\termconc{(\termcont{\beta}{c})}{s'}}})}{s}}
  \eqnew
  \termname{\beta'}{\termconc{(\termcont{\beta}{\termname{\alpha'}{\termconc{(\termcont{\alpha}{c})}{s}}})}{s'}}$

  \item\label{l:eqnew-stack:pop}
  $\termname{\alpha'}{\termconc{(\termcont{\alpha}{\termname{\beta'}{\termabs{x}{\termcont{\beta}{c}}}})}{s}}
  \eqnew
  \termname{\beta'}{\termabs{x}{\termcont{\beta}{\termname{\alpha'}{\termconc{(\termcont{\alpha}{c})}{s}}}}}$

  \item\label{l:eqnew-stack:ren}
  $\termname{\alpha'}{\termconc{(\termcont{\alpha}{\ntermren{\beta}{\beta'}{c}})}{s}}
  \eqnew
  \ntermren{\beta}{\beta'}{\termname{\alpha'}{\termconc{(\termcont{\alpha}{c})}{s}}}$
\end{enumerate}
\label{l:eqnew-stack}
\end{lem}
\end{toappendix}

\begin{proof}
The proof is in Appendix~\ref{app:correspondence}.
\end{proof}

To obtain the desired correspondence between $\eqsigma$ on $\calcLambdaM$-objects and $\eqnew$ on $\calcLambdaMu$-objects, it is necessary to relate the sets $\Object{\calcLambdaM}$ and
$\Object{\calcLambdaMu}$. We do so by means of an
\deft{expansion function}
$\fexp{\_}$, that eliminates the explicit operators of an object by
$\rB\rM$-expansions: \[
\begin{array}{c@{\qquad}c}
\begin{array}{rcll}
\fexp{x}                                & \eqdef  & x \\
\fexp{\termapp{t}{u}}                   & \eqdef  & \termapp{\fexp{t}}{\fexp{u}} \\
\fexp{\termabs{x}{t}}                   & \eqdef  & \termabs{x}{\fexp{t}} \\
\fexp{\termcont{\alpha}{c}}             & \eqdef  & \termcont{\alpha}{\fexp{c}}
\end{array}
&
\begin{array}{rcll}
\fexp{\termsubs{x}{u}{t}}               & \eqdef  & \termapp{(\termabs{x}{\fexp{t}})}{\fexp{u}} \\
\fexp{\termname{\alpha}{t}}             & \eqdef  & \termname{\alpha}{\fexp{t}} \\
\fexp{\termrepl[\alpha']{\alpha}{s}{c}} & \eqdef  & \termname{\alpha'}{\termconc{(\termcont{\alpha}{\fexp{c}})}{\fexp{s}}} \\
\fexp{\termpush{t}{s}}                  & \eqdef  & \termpush{\fexp{t}}{\fexp{s}}
\end{array}
\end{array}
\] Note that $\fexp{\_}$ is not the left-inverse of the plain form $\fcan{\_}$, \ie given
$o \in \Object{\calcLambdaMu}$, it is not necessarily the case $\fexp{\fcan{o}}
= o$. For example, take $o =
\termapp{\termapp{(\termcont{\alpha}{\termname{\alpha}{\termapp{x}{(\termcont{\beta}{\termname{\alpha}{w}})}}})}{y}}{z}$.
Then, $\fcan{o} =
\termcont{\alpha'}{\termrepl[\alpha']{\alpha}{\termpush{y}{z}}{(\termname{\alpha}{\termapp{x}{(\termcont{\beta}{\termname{\alpha}{w}})}})}}$
while $\fexp{\fcan{o}} =
\termcont{\alpha'}{\termname{\alpha'}{\termapp{\termapp{(\termcont{\alpha}{\termname{\alpha}{\termapp{x}{(\termcont{\beta}{\termname{\alpha}{w}})}}})}{y}}{z}}}$.
However, it yields an equivalent object thanks to rule
$\ruleEqnewTheta$.

Some basic properties of the expansion function are stated below.

\begin{lem}
 Let $t, u \in \Term{\calcLambdaMu}$ and $o \in
   \Object{\calcLambdaM}$. Let $\ctxt{L}$ be a substitution context. Then,  
\begin{enumerate}
  \item $\fexp{\ctxtapply{\ctxt{L}}{\termsubs{x}{u}{t}}} \eqnew
  \fexp{\termsubs{x}{u}{\ctxtapply{\ctxt{L}}{t}}}$
  \item $\fexp{\termapp{\ctxtapply{\ctxt{L}}{t}}{u}} \eqnew
  \fexp{\ctxtapply{\ctxt{L}}{\termapp{t}{u}}}$
  \item $\fexp{\fexp{o}} = \fexp{o}$
  \item $\fexp{\ctxtapply{\ctxt{L}}{t}} = \fexp{\ctxtapply{\ctxt{L}}{\fexp{t}}}$
  \item If $t,t' \in \Term{\calcLambdaMu}$, then $t \eqnew t'$ implies
  $\fexp{\ctxtapply{\ctxt{L}}{t}} \eqnew \fexp{\ctxtapply{\ctxt{L}}{t'}}$
\end{enumerate}
\label{l:expan-aux}
\end{lem}

\begin{proof}
The first and second point are by induction on $\ctxt{L}$. The third point is
by straightforward induction on $o$. The fourth point is by induction on
$\ctxt{L}$ using the third point. The last one is by induction on $\ctxt{L}$
using the fact that $\eqnew$ is a congruence.
\end{proof}

The expansion function allows to $\eqnew$-equate $\calcLambdaM$-objects that are related by the reduction $\reduce[\rcan]$ (\cf Definition~\ref{def:plain}). We start with the more subtle  cases $\rrule{\rN}$ and $\rrule{\rC}$.  

\begin{toappendix}
\begin{lem}
Let $t \in \Term{\calcLambdaMu}$ and   $c \in
  \Command{\calcLambdaMu}$. Let $s,s'$ be stacks and
  $\ctxt{LCC}$ be a \textbf{CC} linear context. Then,  
\begin{enumerate}
  \item\label{l:expan-aux-replacement:app}
  $\fexp{\termrepl[\alpha']{\alpha}{s}{\ctxtapply{\ctxt{LCC}}{\termname{\alpha}{t}}}}
  \eqnew \fexp{\ctxtapply{\ctxt{LCC}}{\termname{\alpha'}{\termconc{t}{s}}}}$,
  where $\alpha \notin \ctxt{LCC}$ and $ \alpha \notin t$.

  \item\label{l:expan-aux-replacement:conc}
  $\fexp{\termrepl[\alpha']{\alpha}{s}{\ctxtapply{\ctxt{LCC}}{\termrepl[\alpha]{\beta}{s'}{c}}}}
  \eqnew
  \fexp{\ctxtapply{\ctxt{LCC}}{\termrepl[\alpha']{\beta}{\termpush{s'}{s}}{c}}}
  = o'$, where $ \alpha \notin \ctxt{LCC}$, $\alpha \notin c$ and $\alpha
  \notin s'$.
\end{enumerate}
\label{l:expan-aux-replacement}
\end{lem}
\end{toappendix}

\begin{proof}
The proof is in Appendix~\ref{app:correspondence}.
\end{proof}

\begin{lem}
Let $o \in \Object{\calcLambdaM}$.  Then $o \reduce[\rcan] o'$ implies $\fexp{o} \eqnew \fexp{o'}$
\label{l:rcan-newrel}
\end{lem}

\input{proofs/new/rcan-newrel.tex}

We can then conclude that plain forms do not change
  by expansion.

\begin{cor}
Let $o \in \Object{\calcLambdaM}$. Then $\fexp{\fcan{o}} \eqnew \fexp{o}$.
\label{c:newrel-expan}
\end{cor}

We now show that $\eqnew$-equivalent $\calcLambdaMu$-objects project into
$\eqsigma$ by means of the plain form.

\begin{toappendix}
\begin{lem}
\label{l:newrel-eqsigma}
Let $o, p \in \Object{\calcLambdaMu}$. Then, $o \eqnew p$ implies $\fcan{o} \eqsigma \fcan{p}$.
\end{lem}
\end{toappendix}

\begin{proof}
The proof is in Appendix~\ref{app:correspondence}.
\end{proof}

For the converse we use the expansion function, \ie $\eqsigma$-equivalent
$\calcLambdaM$-objects project into $\eqnew$ by means of the expansion
function.

\begin{toappendix}
\begin{lem}
\label{l:eqsigma-newrel}
Let $o, p \in \Object{\calcLambdaM}$.
Then, $o \eqsigma p$ implies $\fexp{o} \eqnew \fexp{p}$.
\end{lem}
\end{toappendix}

\begin{proof}
The proof is in Appendix~\ref{app:correspondence}.
\end{proof}

The properties above allow to us conclude with the following result.

\begin{thm}
Let $o, p \in \Object{\calcLambdaMu}$.
Then $o \eqnew p$ iff $\fcan{o}\eqsigma\fcan{p}$.
\end{thm}

\begin{proof} \mbox{}
\begin{itemize}
  \item $\Rightarrow$) By Lemma~\ref{l:newrel-eqsigma}.

  \item $\Leftarrow$) $\fcan{o} \eqsigma \fcan{p}$ implies
  $\fexp{\fcan{o}} \eqnew \fexp{\fcan{p}}$ by Lemma~\ref{l:eqsigma-newrel}.
  From $\fexp{\fcan{o}} \eqnew \fexp{\fcan{p}}$ we obtain $\fexp{o} \eqnew
  \fexp{p}$ by Corollary~\ref{c:newrel-expan}. Since $o, p$ are pure
  $\calcLambdaMu$-terms, then $\fexp{o} = o$ and $\fexp{p} = p$. Thus,
  $o \eqnew p$.
  \qedhere
\end{itemize}
\end{proof}

Even if this last theorem relates the new $\eqnew$-equivalence to the
strong bisimulation $\simeq$ presented in
Section~\ref{s:control:equivalence}, the resulting property also
explains the relationship between Laurent's
$\sigma$-equivalence and $\simeq$. 
Indeed, starting from the fact that
$\rho$-equivalence breaks strong bisimulation (\cf example in the
introduction), $\rho$-equivalence is restricted (through our adoption of  $ \ruleEqnewRenRen$,
  $\ruleEqnewRenPop$, and $\ruleEqnewRenPush$ in lieu of $\rho$) to its non-erasing and non-duplicating role in swapping names and $\mu$-binders in the new relation $\eqnew$. In this way, we keep the
strictly necessary renaming operation of Laurent's original $\sigma$-equivalence which is able to materialize a
correspondence with our strong bisimulation.


%% file: proofs/new/rcan-newrel.tex
\begin{proof}
We only show the base cases:
\begin{itemize}
  \item $o = \termapp{\ctxtapply{\ctxt{L}}{\termabs{x}{t}}}{u} \rrule{\rB}
  \ctxtapply{\ctxt{L}}{\termsubs{x}{u}{t}} = o'$, where $\fc{u}{\ctxt{L}}$.
  Then by Lemma~\ref{l:expan-aux} we have \[
\begin{array}{rclcl}
\fexp{\termapp{\ctxtapply{\ctxt{L}}{\termabs{x}{t}}}{u}}
  & \eqnew  & \fexp{\ctxtapply{\ctxt{L}}{\termapp{(\termabs{x}{t})}{u}}}
  & \eqnew  & \fexp{\ctxtapply{\ctxt{L}}{\fexp{\termapp{(\termabs{x}{t})}{u}}}} \\
  & =       & \fexp{\ctxtapply{\ctxt{L}}{\termapp{(\termabs{x}{\fexp{t}})}{\fexp{u}}}}
  & =       & \fexp{\ctxtapply{\ctxt{L}}{\fexp{\termsubs{x}{u}{t}}}} \\
  & \eqnew  & \fexp{\ctxtapply{\ctxt{L}}{\termsubs{x}{u}{t}}}
\end{array} \]

  \item $o = \termapp{\ctxtapply{\ctxt{L}}{\termcont{\alpha}{c}}}{u}
  \rrule{\rM}
  \ctxtapply{\ctxt{L}}{\termcont{\alpha'}{\termrepl[\alpha']{\alpha}{u}{c}}} =
  o'$, where $\fc{u}{\ctxt{L}}$ and $\alpha'$ is fresh. Then by
  Lemma~\ref{l:expan-aux} we have \[
\begin{array}{rclcl}
\fexp{o}
  & =       & \fexp{\termapp{\ctxtapply{\ctxt{L}}{\termcont{\alpha}{c}}}{u}}
  & \eqnew  & \fexp{\ctxtapply{\ctxt{L}}{\termapp{(\termcont{\alpha}{c})}{u}}} \\
  & \eqnew  & \fexp{\ctxtapply{\ctxt{L}}{\fexp{\termapp{(\termcont{\alpha}{c})}{u}}}}
  & =       & \fexp{\ctxtapply{\ctxt{L}}{\termapp{(\termcont{\alpha}{\fexp{c}})}{\fexp{u}}}} \\
  & \eqnew  & \fexp{\ctxtapply{\ctxt{L}}{\termcont{\alpha'}{\termname{\alpha'}{\termapp{(\termcont{\alpha}{\fexp{c}})}{\fexp{u}}}}}}
  & =       & \fexp{\ctxtapply{\ctxt{L}}{\fexp{\termcont{\alpha'}{\termrepl[\alpha']{\alpha}{u}{c}}}}} \\
  & =       & \fexp{o'}
\end{array} \]

  \item $o =
  \termrepl[\alpha']{\alpha}{s}{\ctxtapply{\ctxt{LCC}}{\termname{\alpha}{t}}}
  \rrule{\rN} \ctxtapply{\ctxt{LCC}}{\termname{\alpha'}{\termconc{t}{s}}} =
  o'$, where $\alpha \notin \ctxt{LCC}$ and $ \alpha \notin t$. The result follows from Lemma~\ref{l:expan-aux-replacement}(1).

  \item $o =
  \termrepl[\alpha']{\alpha}{s}{\ctxtapply{\ctxt{LCC}}{\termrepl[\alpha]{\beta}{s'}{c'}}}
  \rrule{\rC}
  \ctxtapply{\ctxt{LCC}}{\termrepl[\alpha']{\beta}{\termpush{s'}{s}}{c'}} =
  o'$, where $ \alpha \notin \ctxt{LCC}$, $\alpha \notin c'$ and $\alpha \notin
  s'$. The result follows from Lemma~\ref{l:expan-aux-replacement}(2).
  \qedhere
\end{itemize}
\end{proof}


%% file: conclusion.tex
\section{Conclusion}
\label{s:conclusion}

\newcommand{\lmex}{\lambda\mu\mathtt{r}}

This paper is about $\sigma$-equivalence in classical logic and the negligible
effect it has on the operational behavior of the terms it relates. It refines
the $\calcLambdaMu$-calculus with explicit operators for substitution and
replacement, by splitting in particular each of the rules $\beta$ and $\mu$ of
$\calcLambdaMu$ into multiplicative and exponential fragments, thus resulting
in the introduction of a new calculus called $\calcLambdaM$. This new
presentation of $\calcLambdaMu$ allows to reformulate
$\sigmalaurent$-equivalence on $\calcLambdaMu$-terms as a strong bisimulation
relation $\eqsigma$ on $\calcLambdaM$-terms. The main obstacle to extract a
bisimulation on $\calcLambdaM$ from the original Laurent's $\sigma$-equivalence
on $\calcLambdaMu$-terms is axiom $\eqsigma[\rho]$, which leads to
$\sigma$-equivalence failing to be a strong bisimulation. We learn that we
cannot remove $\eqsigma[\rho]$ entirely, since it is needed to close several
commutation diagrams in the proof of strong bisimulation. However, a
restriction of $\eqsigma[\rho]$ turns out to suffice.

In~\cite{KesnerV19}, the $\calcLambdaMu$-calculus is refined to a
calculus $\calcLambdaMuR$ with explicit operators, together with a
\emph{linear} substitution/replacement operational semantics \emph{at
  a distance}. In contrast to $\calcLambdaM$, 
$\calcLambdaMuR$ does not support composition of explicit replacements.
In particular,
explicit replacements in $\calcLambdaMuR$ are defined on terms, and
not on stacks, thus the calculus is not able to capture an appropriate notion
of bisimulation such as the one presented in this paper.

Other classical term calculi exist,
\eg~\cite{CurienH00,Audebaud94,Polonovski04,BakelV14}, but none of these
formalisms decomposes term reduction  by means of a fine distinction between
multiplicative and exponential rules. Thus, the main ingredients needed to
build a strong bisimulation are simply not available. Of particular interest
would be obtain a strong bisimulation in the setting of
$\calcLambdaMuT$~\cite{CurienH00}, a calculus inspired from sequent calculus
which is constructed as a perfectly symmetric formalism to deal uniformly with
CBN and CBV.

Explicit Substitutions are a means of modeling sharing in lambda calculi and
hence well suited for capturing call-by-need~\cite{AccattoliBM14}. We believe
$\calcLambdaM$ may prove useful in devising a notion of call-by-need for
classical computation. Our notation for explicit replacement and the notion
of single replacement it supports (\cf Section~\ref{s:on_choice_of_notation}),
would play a crucial role in formulating such a calculus.

A further related reference is~\cite{HondaL10}, where PPNs are used to
interpret processes from the $\pi$-calculus. A precise correspondence is
established between PPN and a typed version of the asynchronous $\pi$-calculus.
Moreover, they show that Laurent's $\eqlaurent$ corresponds exactly to
structural equivalence of $\pi$-calculus processes (Proposition~1 in op.cit).
In~\cite{LaurentR03} Laurent and Regnier show that there is a precise
correspondence between CPS translations from classical calculi (such as
$\calcLambdaMu$) into intuitionistic ones on the one hand, and translations
between LLP and LL on the other.

It would be interesting to analyse other rewriting properties of our term
language such as preservation of $\calcLambdaMu$-strong normalisation of the
reduction relations $\reduce[\rLM]$ and $\reducemean$ or confluence of
$\reducemean$.

Moreover, following the computational
interpretation of \emph{deep inference} provided by the intuitionistic atomic
lambda-calculus~\cite{GundersenHP13}, it would be interesting to investigate a
classical extension and its corresponding notion of strong bisimulation. It is
also natural to wonder what would be an ideal syntax for classical logic, that
is able to capture strong bisimulation by reducing the syntactical axioms to a
small and simple set of equations.

Finally, our notion of $\eqsigma$-equivalence could facilitate proofs of
correctness between abstract machines and $\calcLambdaMu$
(like~\cite{AccattoliBM14} for lambda-calculus) and help establish whether
abstract machines for $\calcLambdaMu$ are ``reasonable''~\cite{AccattoliBM14}.

\section*{Acknowledgment}
To the reviewers for detailed feedback. In particular, for suggesting an
alternative proof of Theorem~\ref{t:control:meaningful:can-sn} and a simplified
syntax for $\calcLambdaM$ regarding explicit renamings.


%% file: app-confluencia.tex
\section{Confluence of the \calcLambdaM-calculus}
\label{app:confluence}

To prove confluence of the $\calcLambdaM$-calculus we use the interpretation
method~\cite{CurienHL96}, where $\calcLambdaM$ is projected into the
$\calcLambdaMu$-calculus.

\begin{defi}
The \deft{projection} $\toLMR{\_}$ from $\calcLambdaM$-objects to
$\calcLambdaMu$-objects is defined as \[
\begin{array}{c@{\qquad}c}
\begin{array}{rcl}
\toLMR{x}                                   & \eqdef  & x \\
\toLMR{(\termapp{t}{u})}                    & \eqdef  & \termapp{\toLMR{t}}{\toLMR{u}} \\
\toLMR{(\termabs{x}{t})}                    & \eqdef  & \termabs{x}{\toLMR{t}} \\
\toLMR{(\termcont{\alpha}{c})}              & \eqdef  & \termcont{\alpha}{\toLMR{c}} \\
\toLMR{(\termsubs{x}{u}{t})}                & \eqdef  & \subsapply{\subs{x}{\toLMR{u}}}{\toLMR{t}}
\end{array}
&
\begin{array}{rcl}
\toLMR{(\termname{\alpha}{t})}              & \eqdef  & \termname{\alpha}{\toLMR{t}} \\
\toLMR{(\termrepl[\alpha']{\alpha}{s}{c})}  & \eqdef  & \replapply{\repl[\alpha']{\alpha}{\toLMR{s}}}{\toLMR{c}} \\
\toLMR{(\termpush{s}{t})}                   & \eqdef  & \termpush{\toLMR{s}}{\toLMR{t}}
\end{array}
\end{array} \]
\label{d:control:lambda-m:semantics:projection}
\end{defi}

\begin{lem}
  Let  $o \in \Object{\calcLambdaM}$, $u \in \Term{\calcLambdaM}$ and $s$
  be a stack. Then,
$\toLMR{(\subsapply{\subs{x}{u}}{o})} =
  \subsapply{\subs{x}{\toLMR{u}}}{\toLMR{o}}$
  and $\toLMR{(\replapply{\repl[\alpha']{\alpha}{s}}{o})} =
\replapply{\repl[\alpha']{\alpha}{\toLMR{s}}}{\toLMR{o}}$.
\label{l:control:lambda-m:semantics:projection-subs-repl}
\end{lem}

\input{proofs/lambda-m/projection-subs-repl}

\begin{lem}
Let  $o,o' \in \Object{\calcLambdaMu}$ and $u,u' \in \Term{\calcLambdaMu}$ and
$s,s'$ be stacks such that $o \reduce[\rlm] o'$, $u \reduce[\rlm] u'$ and
$s \reduce[\rlm] s'$. Then,
\begin{enumerate}
  \item\label{l:control:lambda-m:semantics:reduce-subs:o}
  $\subsapply{\subs{x}{u}}{o} \reduce[\rlm] \subsapply{\subs{x}{u}}{o'}$.
  
  \item\label{l:control:lambda-m:semantics:reduce-subs:u}
  $\subsapply{\subs{x}{u}}{o} \reducemany[\rlm] \subsapply{\subs{x}{u'}}{o}$.
 \item\label{l:control:lambda-m:semantics:reduce-repl:o}
  $\replapply{\repl[\alpha']{\alpha}{s}}{o} \reduce[\rlm]
  \replapply{\repl[\alpha']{\alpha}{s}}{o'}$.
  
  \item\label{l:control:lambda-m:semantics:reduce-repl:s}
  $\replapply{\repl[\alpha']{\alpha}{s}}{o} \reducemany[\rlm]
  \replapply{\repl[\alpha']{\alpha}{s'}}{o}$.
\end{enumerate}
\label{l:control:lambda-m:semantics:reduce-subs-repl}
\end{lem}

\input{proofs/lambda-m/reduce-subs-repl}

Last, to apply the interpretation method we need to relate the relations
$\reduce[\rLM]$ and $\reduce[\rlm]$ by means the projection function
$\toLMR{\_}$.

\begin{lem}\mbox{}
\begin{enumerate}
  \item\label{l:control:lambda-m:semantics:projection-reduce:rLM} Let  $o \in
  \Object{\calcLambdaM}$. Then, $o \reducemany[\rLM] \toLMR{o}$.
  
  \item\label{l:control:lambda-m:semantics:projection-reduce:rlmr-rLM} Let 
  $o,o' \in \Object{\calcLambdaMu}$. If $o \reduce[\rlm] o'$, then 
  $o \reducemany[\rLM] o'$.
  
  \item\label{l:control:lambda-m:semantics:projection-reduce:rLM-rlmr} Let 
  $o,o' \in \Object{\calcLambdaM}$. If   $o \reduce[\rLM] o'$, then   
  $\toLMR{o} \reducemany[\rlm] \toLMR{o'}$.
\end{enumerate}
\label{l:control:lambda-m:semantics:projection-reduce}
\end{lem}

\input{proofs/lambda-m/projection-reduce}

Confluence of $\reduce[\rLM]$ is a consequence of
Lemma~\ref{l:control:lambda-m:semantics:projection-reduce} and confluence of 
$\reduce[\rlm]$~\cite{Parigot92}.

\gettoappendix{t:control:lambda-m:semantics:confluence}
\input{proofs/lambda-m/confluence}

%% file: proofs/lambda-m/projection-subs-repl.tex
\begin{proof}
Both statements are by induction on $o$.
\end{proof}


%% file: proofs/lambda-m/reduce-subs-repl.tex
\begin{proof} \mbox{}
  Items (1) and (3) are by induction on $o \reduce[\rlm] o'$, while
  items (2) and (4) are by induction on $o$.  
\end{proof}


%% file: proofs/lambda-m/projection-reduce.tex
\begin{proof} \mbox{}
  \begin{enumerate}
  \item By induction on $o$.
  \item By induction on $o \reduce[\rlm] o'$.
  \item By induction on $o \reduce[\rLM] o'$ using
    Lemma~\ref{l:control:lambda-m:semantics:projection-subs-repl}, and
    Lemma~\ref{l:control:lambda-m:semantics:reduce-subs-repl}.
    \qedhere
\end{enumerate}
\end{proof}


%% file: proofs/lambda-m/confluence.tex
\begin{proof}
  By the interpretation method,
  using confluence of $\reduce[\rlm]$ and 
Lemma~\ref{l:control:lambda-m:semantics:projection-reduce}:
\begin{center}
\begin{tikzcd}[ampersand replacement=\&]
  \&[-10pt] o \arrow[twoheadrightarrow]{dl}[pos=1.05,inner sep=5pt,right]{\rLM}
              \arrow[twoheadrightarrow]{dr}[pos=1.05,inner sep=5pt,left]{\rLM}
              \arrow[twoheadrightarrow]{d}[pos=.85,inner sep=5pt,right]{\rLM} \\[10pt]
o_0 \arrow[dashed,twoheadrightarrow]{d}[pos=.85,inner sep=5pt,left]{\rLM}
  \&[-10pt] \toLMR{o} \arrow[twoheadrightarrow]{dl}[pos=.85,inner sep=5pt,below]{\rlm}
                      \arrow[twoheadrightarrow]{dr}[pos=.85,inner sep=5pt,below]{\rlm}
  \&[-10pt] o_1 \arrow[dashed,twoheadrightarrow]{d}[pos=.85,inner sep=5pt,right]{\rLM} \\[10pt]
\toLMR{o_0} \arrow[dashed,twoheadrightarrow]{dr}[pos=.65,inner sep=5pt,below]{\rlm}
            \arrow[dashed,twoheadrightarrow,bend right=45]{dr}[pos=.85,inner sep=5pt,below]{\rLM}
  \&[-10pt] \text{\cite{Parigot92}}
  \&[-10pt] \toLMR{o_1} \arrow[dashed,twoheadrightarrow]{dl}[pos=.65,inner sep=5pt,below]{\rlm}
                        \arrow[dashed,twoheadrightarrow,bend left=45]{dl}[pos=.85,inner sep=5pt,below]{\rLM} \\[10pt]
  \&[-10pt] o'
\end{tikzcd}
\end{center}
where $o \reducemany[\rLM] o_0$ and  $o \reducemany[\rLM] o_1$
are the hypothesis of the theorem.
The three vertical reductions are justified by
Lemma~\ref{l:control:lambda-m:semantics:projection-reduce}
(\ref{l:control:lambda-m:semantics:projection-reduce:rLM}),
since  $p
\reducemany[\rLM] \toLMR{p}$ for all  $o \in \Object{\calcLambdaM}$.
The reductions  $\toLMR{o} \reducemany[\rlm] \toLMR{o_0}$ and  $\toLMR{o}
\reducemany[\rlm] \toLMR{o_1}$ come from 
Lemma~\ref{l:control:lambda-m:semantics:projection-reduce}
(\ref{l:control:lambda-m:semantics:projection-reduce:rLM-rlmr}).
The diagram is closed
by~\cite{Parigot92}
thus obtaining $\toLMR{o_0} \reducemany[\rLM] o'$ and  $\toLMR{o_1}
\reducemany[\rLM] o'$ by 
Lemma~\ref{l:control:lambda-m:semantics:projection-reduce}
(\ref{l:control:lambda-m:semantics:projection-reduce:rlmr-rLM}).
\end{proof}


%% file: appendix-meaningful-termina.tex
\section{Strong Normalisation of \Plain\ Computation}
\label{app:meaningful}

To show that \plain\ computation is strongly normalising we define a measure
over objects of the $\calcLambdaM$-calculus. It is worth noticing that using
the standard size of an object (\ie counting all its constructors) is not
enough since it does not strictly decrease under computation due to the
following remarks:
\begin{enumerate}
  \item Rule $\rM$ discards an application while introducing a new explicit
  replacement, thus preserving the number of constructors in the object.
  
  \item Rule $\rN$ discards a linear explicit replacement with a stack of $n$
  elements, replacing it with $n$ applications. The number of stack
  constructors in a stack of $n$ elements turns out to be $n-1$ which, together
  with the discarded explicit replacement, compensates the $n$ introduced
  applications.
  
  \item Rule $\rC$ discards a linear explicit replacement by combining it with
  another one, appending their respective stacks. This introduces a new stack
  constructor, preserving the total number of constructors in the object.
\end{enumerate}

The first remark suggests that the application constructor should have more
weight than the replacement constructor to guarantee normalisation by means of
a polynomial interpretation. However, the second remark suggests exactly the
opposite. On another note, the third remark requires explicit replacements to
have more weight than stacks.

Under these considerations we define the following measure over objects of the
$\calcLambdaM$-calculus which turns out to be decreasing w.r.t. reduction
$\reduce[\rcan]$. \[
\begin{array}{c@{\qquad\qquad}c}
\begin{array}{lll}
\cmeas{x}                                 & \eqdef  & 3 \\
\cmeas{\termapp{t}{u}}                    & \eqdef  & \cmeas{t} * \cmeas{u} \\
\cmeas{\termabs{x}{t}}                    & \eqdef  & \cmeas{t} \\
\cmeas{\termcont{\alpha}{c}}              & \eqdef  & \cmeas{c} + 1 \\
\cmeas{\termsubs{x}{u}{t}}                & \eqdef  & \cmeas{t} + \cmeas{u}
\end{array}
&
\begin{array}{lll}
\cmeas{\termname{\alpha}{t}}              & \eqdef  & \cmeas{t}  \\
\cmeas{\termrepl[\alpha']{\alpha}{s}{c}}  & \eqdef  & \cmeas{c} * \cmeas{s} + 1 \\
\\
\cmeas{\termpush{t}{s}}                   & \eqdef  & \cmeas{t} * \cmeas{s}
\end{array}
\end{array} \]

Notice that $\cmeas{o} \geq 3$ for every $o \in \Object{\calcLambdaM}$. We also
have $\cmeas{o} + \cmeas{o'} < \cmeas{o} * \cmeas{o'}$.

\begin{lem}
\label{l:control:meaningful:can-sn-measure}
For every $\ctxt{O}$, there is $f_{\ctxt{O}} : x \mapsto a * x + b$ with $a > 0$
and $b \geq 0$, such that for every $o$, one has 
$\cmeas{\ctxtapply{\ctxt{O}}{o}} = \funcapply{f_{\ctxt{O}}}{\cmeas{o}}$.
\end{lem}

\input{proofs/meaningful/can-sn-measure.tex}

\gettoappendix{t:control:meaningful:can-sn}
\input{proofs/meaningful/can-sn}

\gettoappendix{t:control:meaningful:can-confluence}
\input{proofs/meaningful/can-confluence}


%% file: proofs/meaningful/can-sn-measure.tex
\begin{proof} By induction on $\ctxt{O}$. \mbox{}
\begin{itemize}
  \item For $\ctxt{O} = \Box$ and $\ctxt{O} = \boxdot$ we take
  $\funcapply{f_{\ctxt{O}}}{x} = 1 * x + 0$.

  \item For $\ctxt{O} = \termsubs{x}{t}{\ctxt{T}}$ and $\ctxt{O} =
  \termsubs{x}{\ctxt{T}}{t}$, we take $\funcapply{f_{\ctxt{O}}}{x} = a' * x +
  (b' + \cmeas{t})$, where $\funcapply{f_{\ctxt{T}}}{x} = a' * x + b'$.

  \item For $\ctxt{O} = \termapp{\ctxt{T}}{t}$ and $\ctxt{O} =
  \termapp{t}{\ctxt{T}}$ we take $\funcapply{f_{\ctxt{O}}}{x} = (a' *
  \cmeas{t}) * x + (b' * \cmeas{t})$, where $\funcapply{f_{\ctxt{T}}}{x} = a' *
  x + b'$.

  \item For $\ctxt{O} = \termabs{x}{\ctxt{T}}$ and $\ctxt{O} =
  \termname{\alpha}{\ctxt{T}}$ we take $\funcapply{f_{\ctxt{O}}}{x} = a' * x +
  b'$, where $\funcapply{f_{\ctxt{T}}}{x} = a' * x + b'$.

  \item For $\ctxt{O} = \termcont{\alpha}{\ctxt{C}}$ we take
  $\funcapply{f_{\ctxt{O}}}{x} = a' * x + (b' + 1)$, where
  $\funcapply{f_{\ctxt{C}}}{x} = a' * x + b'$.
  
  \item For $\ctxt{O} = \termrepl[\alpha']{\alpha}{s}{\ctxt{C}}$ we take
  $\funcapply{f_{\ctxt{O}}}{x} = (a' * \cmeas{s}) * x + (b' * \cmeas{s} + 1)$,
  where $\funcapply{f_{\ctxt{C}}}{x} = a' * x + b'$.

  \item For $\ctxt{O} = \termrepl[\alpha']{\alpha}{\ctxt{S}}{c}$ we take
  $\funcapply{f_{\ctxt{O}}}{x} = (a' * \cmeas{c}) * x + (b' * \cmeas{c} + 1)$,
  where $\funcapply{f_{\ctxt{S}}}{x} = a' * x + b'$.

  \item For $\ctxt{O} = \termpush{\ctxt{T}}{s}$ we take
  $\funcapply{f_{\ctxt{O}}}{x} = (a' * \cmeas{s}) * x + (b' * \cmeas{s} + 1)$,
  where $\funcapply{f_{\ctxt{T}}}{x} = a' * x + b'$.

  \item For $\ctxt{O} = \termpush{t}{\ctxt{S}}$ we take
  $\funcapply{f_{\ctxt{O}}}{x} = (a' * \cmeas{t}) * x + (b' * \cmeas{t} + 1)$,
  where $\funcapply{f_{\ctxt{S}}}{x} = a' * x + b'$.
  \qedhere
\end{itemize}
\end{proof}


%% file: proofs/meaningful/can-sn.tex
\begin{proof}
We prove $o \reduce[\rcan] p$ implies $\cmeas{o} > \cmeas{p}$ by induction on
the relation $\reduce[\rcan]$. We first analyse all the base cases:
\begin{itemize}
  \item $o = \termapp{\ctxtapply{\ctxt{L}}{\termabs{x}{t}}}{u} \rrule{\rB}
  \ctxtapply{\ctxt{L}}{\termsubs{x}{u}{t}} = p$, where $\fc{u}{\ctxt{L}}$. Then
  Lemma~\ref{l:control:meaningful:can-sn-measure} gives
  $\funcapply{f_{\ctxt{L}}}{x} =  a * x + b$, with $a > 0$ and $b \geq 0$. We
  conclude by \[
\begin{array}{lll}
\cmeas{o} & =     & (a * \cmeas{t} + b) * \cmeas{u} \\
          & =     & a * \cmeas{t} * \cmeas{u} + b * \cmeas{u} \\
          & >     & a * (\cmeas{t} + \cmeas{u}) + b * \cmeas{u} \\ 
          & \geq  & a * (\cmeas{t} + \cmeas{u}) + b \\
          & =     & \cmeas{p}
\end{array} \]

  \item $o = \termapp{\ctxtapply{\ctxt{L}}{\termcont{\alpha}{c}}}{u}
  \rrule{\rM}
  \ctxtapply{\ctxt{L}}{\termcont{\alpha'}{\termrepl[\alpha']{\alpha}{u}{c}}} =
  p$, where $\alpha'$ is fresh. Then
  Lemma~\ref{l:control:meaningful:can-sn-measure} gives
  $\funcapply{f_{\ctxt{L}}}{x} = a * x + b$, with $a > 0$ and $b \geq 0$. We
  conclude by \[
\begin{array}{lll}
\cmeas{o} & = & (a * (\cmeas{c} + 1) + b) * \cmeas{u} \\
          & = & a * \cmeas{c} * \cmeas{u} + \cmeas{u} * a + \cmeas{u} * b \\
          & > & a * \cmeas{c} * \cmeas{u} + 2 * a + b \\
          & = & a * ((\cmeas{c} * \cmeas{u} + 1) + 1) + b \\
          & = & \cmeas{p}
\end{array} \]

  \item $o =
  \termrepl[\alpha']{\alpha}{s}{\ctxtapply{\ctxt{LCC}}{\termname{\alpha}{t}}}
  \rrule{\rN} \ctxtapply{\ctxt{LCC}}{\termname{\alpha'}{\termconc{t}{s}}} =
  p$, where $\alpha \notin \ctxt{LCC}$ and $\alpha \notin t$. Assume $s =
  \termpush{u_0}{\termpush{\ldots}{u_n}}$.
  Lemma~\ref{l:control:meaningful:can-sn-measure} gives
  $\funcapply{f_{\ctxt{LCC}}}{x} = a * x + b$, with $a > 0$ and $b \geq 0$. We
  conclude by \[
\begin{array}{lll}
\cmeas{o} & = & (a * \cmeas{t} + b) * \cmeas{u_0} * \ldots * \cmeas{u_n} + 1 \\
          & = & a * \cmeas{t} * \cmeas{u_0} * \ldots * \cmeas{u_n} + b * \cmeas{u_0} * \ldots * \cmeas{u_n} + 1 \\
          & > & a * \cmeas{t} * \cmeas{u_0} * \ldots * \cmeas{u_n} + b \\
          & = & \cmeas{p}
\end{array} \]

  \item $o =
  \termrepl[\alpha']{\alpha}{s}{\ctxtapply{\ctxt{LCC}}{\termrepl[\alpha]{\beta}{s'}{c'}}}
  \rrule{\rC}
  \ctxtapply{\ctxt{LCC}}{\termrepl[\alpha']{\beta}{\termpush{s'}{s}}{c'}} = p$,
  where $\alpha \notin \ctxt{LCC}$, $\alpha \notin c'$ and $\alpha \notin s'$.
  Lemma~\ref{l:control:meaningful:can-sn-measure} gives
  $\funcapply{f_{\ctxt{LCC}}}{x} = a * x + b$, with $a > 0$ and $b \geq 0$. We
  conclude by \[
\begin{array}{lll}
\cmeas{o} & = & (a * (\cmeas{c'} * \cmeas{s'} + 1) + b) * \cmeas{s} + 1 \\
          & = & a * \cmeas{c'} * \cmeas{s'} * \cmeas{s} + (a + b) * \cmeas{s} + 1 \\
          & > & a * \cmeas{c'} * \cmeas{s'} * \cmeas{s} + a + b \\
          & = & a * (\cmeas{c'} * \cmeas{s'} * \cmeas{s} + 1) + b \\
          & = & \cmeas{p}
\end{array} \]
\end{itemize}

For every inductive case of the form $o = \ctxtapply{\ctxt{O}}{o'}
\reduce[\rcan] \ctxtapply{\ctxt{O}}{p'} = p$ where $o' \reduce[\rcan] p'$ is a
base reduction step, we get $\cmeas{o'} > \cmeas{p'}$ by the \ih We then use
Lemma~\ref{l:control:meaningful:can-sn-measure} to get  $\cmeas{o} =
\funcapply{f_{\ctxt{O}}}{\cmeas{o'}}$ and $\cmeas{p} =
\funcapply{f_{\ctxt{O}}}{\cmeas{p'}}$. We conclude $\cmeas{o} > \cmeas{p}$
since $f_{\ctxt{O}}$ is clearly strictly monotone by construction.
\end{proof}


%% file: proofs/meaningful/can-confluence.tex
\begin{proof}
We first remark that the rules $\rB$, $\rM$, $\rN$ and $\rC$ do not
duplicate any subterm. Thus, any trivial one-step divergence between these
rules can be easily closed in one step as well. Then, the only cases left to
be considered are the critical pairs between them. There are only two such
cases:
\begin{enumerate}
  \item $\rN$--$\rC$. We have $o =
  \termrepl[\alpha']{\alpha}{s}{\ctxtapply{\ctxt{LCC}_2}{\termrepl[\alpha]{\delta}{s'}{\ctxtapply{\ctxt{LCC}_1}{\termname{\delta}{t}}}}}$
  with the conditions $\alpha \notin \ctxt{LCC}_2$, $\alpha \notin
  \ctxtapply{\ctxt{LCC}_1}{\termname{\delta}{t}}$, $\alpha \notin s$,
  $\fc{\alpha'}{\ctxt{LCC}_2}$, $\fc{s}{\ctxt{LCC}_2}$ given by rule $\rC$, and
  the conditions $\delta \notin t$, $\delta \notin \ctxt{LCC}_1$,
  $\fc{\alpha}{\ctxt{LCC}_1}$ and $\fc{s'}{\ctxt{LCC}_1}$ given by rule $\rN$.
  We conclude since $\termconc{(\termconc{t}{s'})}{s} =
  \termconc{t}{(\termpush{s'}{s})}$, thus obtaining the diagram: \[
\begin{tikzcd}
\termrepl[\alpha']{\alpha}{s}{\ctxtapply{\ctxt{LCC}_2}{\termrepl[\alpha]{\delta}{s'}{\ctxtapply{\ctxt{LCC}_1}{\termname{\delta}{t}}}}} \arrow[rightarrow]{d}[anchor=north,left]{\rC}
  &[-30pt] \reduce[\rN] &[-30pt] \termrepl[\alpha']{\alpha}{s}{\ctxtapply{\ctxt{LCC}_2}{\ctxtapply{\ctxt{LCC}_1}{\termname{\alpha}{\termconc{t}{s'}}}}} \arrow[rightarrow]{d}[anchor=north,left]{\rN} \\[-5pt]
\ctxtapply{\ctxt{LCC}_2}{\termrepl[\alpha']{\delta}{\termpush{s'}{s}}{\ctxtapply{\ctxt{LCC}_1}{\termname{\delta}{t}}}}
  &[-30pt] \reduce[\rN] &[-30pt] \ctxtapply{\ctxt{LCC}_2}{\ctxtapply{\ctxt{LCC}_1}{\termname{\alpha'}{\termconc{(\termconc{t}{s'})}{s}}}}
\end{tikzcd} \]
  
  \item $\rC$--$\rC$. Then we have $o =
  \termrepl[\alpha']{\alpha}{s}{\ctxtapply{\ctxt{LCC}_2}{\termrepl[\alpha]{\delta}{s'}{\ctxtapply{\ctxt{LCC}_1}{\termrepl[\delta]{\gamma}{s''}{c}}}}}$
  with the conditions $\alpha \notin \ctxt{LCC}_2$, $\alpha \notin
  \ctxtapply{\ctxt{LCC}_1}{\termrepl[\delta]{\gamma}{s''}{c}}$, $\alpha \notin
  s$, $\fc{\alpha'}{\ctxt{LCC}_2}$, $\fc{s}{\ctxt{LCC}_2}$ due to the outermost
  application of the rule, and the conditions $\delta \notin c$, $\delta \notin
  \ctxt{LCC}_1$, $\delta \notin s'$, $\fc{\alpha}{\ctxt{LCC}_1}$ and
  $\fc{s'}{\ctxt{LCC}_1}$ due to the innermost one. We conclude since 
  $\termpush{s''}{(\termpush{s'}{s})} = \termpush{(\termpush{s''}{s'})}{s}$,
  thus obtaining the following diagram: \[
\begin{tikzcd}
\termrepl[\alpha']{\alpha}{s}{\ctxtapply{\ctxt{LCC}_2}{\termrepl[\alpha]{\delta}{s'}{\ctxtapply{\ctxt{LCC}_1}{\termrepl[\delta]{\gamma}{s''}{c}}}}} \arrow[rightarrow]{d}[anchor=north,left]{\rC}
  &[-30pt] \reduce[\rC] &[-30pt] \termrepl[\alpha']{\alpha}{s}{\ctxtapply{\ctxt{LCC}_2}{\ctxtapply{\ctxt{LCC}_1}{\termrepl[\alpha]{\gamma}{\termpush{s''}{s'}}{c}}}} \arrow[rightarrow]{d}[anchor=north,left]{\rC} \\[-5pt]
\ctxtapply{\ctxt{LCC}_2}{\termrepl[\alpha']{\delta}{\termpush{s'}{s}}{\ctxtapply{\ctxt{LCC}_1}{\termrepl[\delta]{\gamma}{s''}{c}}}}
  &[-30pt] \reduce[\rC] &[-30pt] \ctxtapply{\ctxt{LCC}_2}{\ctxtapply{\ctxt{LCC}_1}{\termrepl[\alpha']{\gamma}{\termpush{\termpush{s''}{s'}}{s}}{c}}}
\end{tikzcd} \]\vspace*{(-\baselineskip*2)-3pt}
\end{enumerate}
\end{proof}


%% file: appendix-equivalencia-v2.tex
\section{Structural Equivalence for the \calcLambdaM-Calculus}
\label{a:equivalence}

To prove Lemma~\ref{l:control:equivalence:permute} we introduce two auxiliary
results about contexts $\ctxt{LTT}$ and $\ctxt{LCC}$.

\begin{lem}\hfill
  \begin{enumerate}
    \item Suppose $u = \ctxtapply{\ctxt{L}}{\ctxtapply{\ctxt{LTT}}{t}}$ with
    $\bv{\ctxt{L}} \notin \ctxt{LTT}$ and $\fc{\ctxt{L}}{\ctxt{LTT}}$. If
    $u \reduce[\rcan] u'$, then there exists $\ctxt{LTT}'$, $\ctxt{L}'$ and
    $t'$ such that $u' = \ctxtapply{\ctxt{L}'}{\ctxtapply{\ctxt{LTT}'}{t'}}$.
    Moreover $v = \ctxtapply{\ctxt{LTT}}{\ctxtapply{\ctxt{L}}{t}}
    \reduce[\rcan] \ctxtapply{\ctxt{LTT}'}{\ctxtapply{\ctxt{L}'}{t'}} = v'$.
    In a diagram:
\begin{center}
\begin{tabular}{ccc}
\begin{tikzcd}[ampersand replacement=\&]
  u = \ctxtapply{\ctxt{L}}{\ctxtapply{\ctxt{LTT}}{t}} \arrow{d}[anchor=north,left]{\rcan} \\ 
  u' = \ctxtapply{\ctxt{L}'}{\ctxtapply{\ctxt{LTT}'}{t'}}
\end{tikzcd}
    & implies &
\begin{tikzcd}[ampersand replacement=\&]
  \&[-25pt] \ctxtapply{\ctxt{LTT}}{\ctxtapply{\ctxt{L}}{t}} = v \arrow[densely dashed]{d}[anchor=north,left]{\rcan} \\ 
  \&[-25pt] \ctxtapply{\ctxt{LTT}'}{\ctxtapply{\ctxt{L}'}{t'}} = v'
\end{tikzcd}
\end{tabular}
\end{center}

    \item Suppose $v = \ctxtapply{\ctxt{LTT}}{\ctxtapply{\ctxt{L}}{t}}$ with
    $\bv{\ctxt{L}} \notin \ctxt{LTT}$ and $\fc{\ctxt{L}}{\ctxt{LTT}}$. If
    $v \reduce[\rcan] v'$, then there exists $\ctxt{LTT}'$, $\ctxt{L}'$ and
    $t'$ such that $v' = \ctxtapply{\ctxt{LTT}'}{\ctxtapply{\ctxt{L}'}{t'}}$.
    Moreover, $u = \ctxtapply{\ctxt{L}}{\ctxtapply{\ctxt{LTT}}{t}}
    \reduce[\rcan] \ctxtapply{\ctxt{L}'}{\ctxtapply{\ctxt{LTT}'}{t'}} = u'$.
    On a diagram,
\begin{center}
\begin{tabular}{ccc}
\begin{tikzcd}[ampersand replacement=\&]
  \&[-25pt] v = \ctxtapply{\ctxt{LTT}}{\ctxtapply{\ctxt{L}}{t}} \arrow{d}[anchor=north,left]{\rcan} \\ 
  \&[-25pt] v' = \ctxtapply{\ctxt{LTT}'}{\ctxtapply{\ctxt{L}'}{t'}} 
\end{tikzcd}
    & implies &
\begin{tikzcd}[ampersand replacement=\&]
  \ctxtapply{\ctxt{L}}{\ctxtapply{\ctxt{LTT}}{t}} = u \arrow[densely dashed]{d}[anchor=north,left]{\rcan} \\ 
  \ctxtapply{\ctxt{L}'}{\ctxtapply{\ctxt{LTT}'}{t'}} = u'
\end{tikzcd}
\end{tabular}
\end{center}
\end{enumerate}
\label{l:control:equivalence:ltt-bisimulation}
\end{lem}

\input{proofs/equivalence/ltt-bisimulation-v2}

We recall the definition of Replacement/Renaming Contexts:
\[
\begin{array}{rrcl}
\textbf{(Repl./Ren. Contexts)}  & \ctxt{R}  & \Coloneq  & \boxdot \mid \termrepl[\alpha']{\alpha}{s}{\ctxt{R}} \mid \ntermren{\alpha}{\beta}{\ctxt{R}} 
\end{array}
\]
\pagebreak
\begin{lem}\hfill
  \begin{enumerate}
    \item Suppose $d = \ctxtapply{\ctxt{R}}{\ctxtapply{\ctxt{LCC}}{c}}$ with
    $\bn{\ctxt{R}} \notin \ctxt{LCC}$ and $\fc{\ctxt{R}}{\ctxt{LCC}}$. If
    $d \reduce[\rcan] d''$, then there exists $\ctxt{R}'$, $\ctxt{LCC}'$ and
    $c'$ such that $d'' \reducemany[\rM]
    \ctxtapply{\ctxt{R}'}{\ctxtapply{\ctxt{LCC}'}{c'}}$. Moreover, $e =
    \ctxtapply{\ctxt{LCC}}{\ctxtapply{\ctxt{R}}{c}} \reduce[\rcan] e''
    \reducemany[\rM] \ctxtapply{\ctxt{LCC}'}{\ctxtapply{\ctxt{R}'}{c'}} = e'$.
    In a diagram:
\begin{center}
\begin{tabular}{ccc}
\begin{tikzcd}[ampersand replacement=\&]
  d = \ctxtapply{\ctxt{R}}{\ctxtapply{\ctxt{LCC}}{c}} \arrow{d}[anchor=north,left]{\rcan} \\
  d'' \arrow[densely dashed,two heads]{d}[anchor=north,left]{\rM} \\
  d' = \ctxtapply{\ctxt{R}'}{\ctxtapply{\ctxt{LCC}'}{c'}}
\end{tikzcd}
& implies &
\begin{tikzcd}[ampersand replacement=\&]
  \&[-25pt] \ctxtapply{\ctxt{LCC}}{\ctxtapply{\ctxt{R}}{c}} = e \arrow[densely dashed]{d}[anchor=north,left]{\rcan} \\ 
  \&[-25pt] e'' \arrow[densely dashed,two heads]{d}[anchor=north,left]{\rM} \\
  \&[-25pt] \ctxtapply{\ctxt{LCC}'}{\ctxtapply{\ctxt{R}'}{c'}} = e'
\end{tikzcd}
\end{tabular}
\end{center}

    \item Suppose $e = \ctxtapply{\ctxt{LCC}}{\ctxtapply{\ctxt{R}}{c}}$ with
    $\bn{\ctxt{R}} \notin \ctxt{LCC}$ and $\fc{\ctxt{R}}{\ctxt{LCC}}$. If $e
    \reduce[\rcan] e''$, then there exists $\ctxt{R}'$, $\ctxt{LCC}'$ and $c'$
    such that $e'' \reducemany[\rM]
    \ctxtapply{\ctxt{LCC}'}{\ctxtapply{\ctxt{R}'}{c'}}$, Moreover, $d =
    \ctxtapply{\ctxt{R}}{\ctxtapply{\ctxt{LCC}}{c}} \reduce[\rcan] d''
    \reducemany[\rM] \ctxtapply{\ctxt{R}'}{\ctxtapply{\ctxt{LCC}'}{c'}} = d'$.
    In a diagram,
\begin{center}
  \begin{tabular}{ccc}
\begin{tikzcd}[ampersand replacement=\&]
  e = \ctxtapply{\ctxt{LCC}}{\ctxtapply{\ctxt{R}}{c}} \arrow{d}[anchor=north,left]{\rcan} \\ 
  e'' \arrow[densely dashed,two heads]{d}[anchor=north,left]{\rM} \\
  e' = \ctxtapply{\ctxt{LCC}'}{\ctxtapply{\ctxt{R}'}{c'}}
\end{tikzcd}
& implies &
\begin{tikzcd}[ampersand replacement=\&]
  \&[-25pt] \ctxtapply{\ctxt{R}}{\ctxtapply{\ctxt{LCC}}{c}} = d \arrow[densely dashed]{d}[anchor=north,left]{\rcan} \\
  \&[-25pt] d'' \arrow[densely dashed,two heads]{d}[anchor=north,left]{\rM} \\
  \&[-25pt] \ctxtapply{\ctxt{R}'}{\ctxtapply{\ctxt{LCC}'}{c'}} = d'
\end{tikzcd}
\end{tabular}
\end{center}
\end{enumerate}
\label{l:control:equivalence:lcc-bisimulation}
\end{lem}

\input{proofs/equivalence/lcc-bisimulation-v2}

\gettoappendix{l:control:equivalence:permute}
\input{proofs/equivalence/permute-v2}


%% file: proofs/equivalence/ltt-bisimulation-v2.tex
\begin{proof}
We address the first item, the second one is similar. 
The possible overlap between the $\rcan$-step and 
$\ctxtapply{\ctxt{L}}{\ctxtapply{\ctxt{LTT}}{t}}$ can be broken down as follows:
\begin{itemize}
  \item The step is entirely within $t$, \ie $t \reduce[\rcan]
  t'$. Then it suffices to take $\ctxt{L}' = \ctxt{L}$ and $\ctxt{LTT}' = \ctxt{LTT}$.
  
\item The step is entirely within $\ctxt{L}$, \ie $\ctxt{L}
  \reduce[\rcan] \ctxt{L}'$. Then we take $t' = t$ and $\ctxt{LTT}' =
  \ctxt{LTT}$.
  
  \item The step is entirely within $\ctxt{LTT}$, \ie
  $\ctxt{LTT} \reduce[\rcan] \ctxt{LTT}'$. Then we take $t' = t$ and
  $\ctxt{L}' = \ctxt{L}$.
  
  \item The step overlaps with $\ctxt{LTT}$. There are four cases according to the reduction rule applied:
  \begin{enumerate}
    \item $\rB$. There are two further cases.
    \begin{enumerate}
      \item It overlaps with $t$. Then, $t =
      \ctxtapply{\ctxt{L}_1}{\termabs{x}{v}}$, $\ctxt{LTT} =
      \ctxtapply{\ctxt{LTT}_2}{\termapp{\ctxt{L}_2}{u}}$ and the LHS of the rule $\rB$ is
      $\termapp{\ctxtapply{\ctxt{L}_2}{\ctxtapply{\ctxt{L}_1}{\termabs{x}{v}}}}{u}$.
      We conclude by setting $t' = \ctxtapply{\ctxt{L}_1}{\termsubs{x}{u}{v}}$,
      $\ctxt{L}' = \ctxt{L}$ and $\ctxt{LTT}' = \ctxtapply{\ctxt{LTT}_2}{\ctxt{L}_2}$: \[
\begin{tikzcd}[ampersand replacement=\&]
\ctxtapply{\ctxt{L}}{\ctxtapply{\ctxt{LTT}}{t}} \arrow{d}[anchor=north,left]{\rB}
  \&[-25pt] 
  \&[-25pt] \ctxtapply{\ctxt{LTT}_2}{\termapp{\ctxtapply{\ctxt{L}_2}{\ctxtapply{\ctxt{L}}{\ctxtapply{\ctxt{L}_1}{\termabs{x}{v}}}}
}{u}} \arrow[densely dashed]{d}[anchor=north,left]{\rB} \\ 
\ctxtapply{\ctxt{L}}{\ctxtapply{\ctxt{LTT}_2}{\ctxtapply{\ctxt{L}_2}{t'}}}
  \&[-25pt] 
  \&[-25pt] \ctxtapply{\ctxt{LTT}_2}{\ctxtapply{\ctxt{L}_2}{\ctxtapply{\ctxt{L}}{\ctxtapply{\ctxt{L}_1}{\termsubs{x}{u}{v}}}}}
\end{tikzcd} \]

      \item It does not overlap with $t$. Then, $\ctxt{LTT} =
      \ctxtapply{\ctxt{LTT}_2}{\termapp{\ctxtapply{\ctxt{L}_2}{\termabs{x}{\ctxt{LTT}_1}}}{u}}$.
      We conclude with $t' = t$, $\ctxt{L}' = \ctxt{L}$ and $\ctxt{LTT}' =
      \ctxtapply{\ctxt{LTT}_2}{\ctxtapply{\ctxt{L}_2}{\termsubs{x}{u}{\ctxt{LTT}_1}}}$: \[
\begin{tikzcd}[ampersand replacement=\&]
\ctxtapply{\ctxt{L}}{\ctxtapply{\ctxt{LTT}}{t}} \arrow{d}[anchor=north,left]{\rB}
  \&[-25pt] 
  \&[-25pt] \ctxtapply{\ctxt{LTT}_2}{\termapp{\ctxtapply{\ctxt{L}_2}{\termabs{x}{\ctxtapply{\ctxt{LTT}_1}{\ctxtapply{\ctxt{L}}{t}}}}}{u}} \arrow[densely dashed]{d}[anchor=north,left]{\rB} \\ 
\ctxtapply{\ctxt{L}}{\ctxtapply{\ctxt{LTT}'}{t}}
  \&[-25pt] 
  \&[-25pt] \ctxtapply{\ctxt{LTT}_2}{\ctxtapply{\ctxt{L}_2}{\termsubs{x}{u}{\ctxtapply{\ctxt{LTT}_1}{\ctxtapply{\ctxt{L}}{t}}}}}
\end{tikzcd} \]
    \end{enumerate}
    
    \item $\rM$. There are two further cases.
    \begin{enumerate}
       \item It overlaps with $t$. Then, $t =
      \ctxtapply{\ctxt{L}_1}{\termcont{\alpha}{c}}$, $\ctxt{LTT} =
      \ctxtapply{\ctxt{LTT}_2}{\termapp{\ctxt{L}_2}{u}}$ and the LHS of the step $\rM$ is
      $\termapp{\ctxtapply{\ctxt{L}_2}{\ctxtapply{\ctxt{L}_1}{\termcont{\alpha}{c}}}}{u}$.
      We conclude with $t' = \ctxtapply{\ctxt{L}_1}{\termcont{\alpha'}{\termrepl[\alpha']{\alpha}{u}{c}}}$,
      $\ctxt{L}' = \ctxt{L}$ and $\ctxt{LTT}' = \ctxtapply{\ctxt{LTT}_2}{\ctxt{L}_2}$: \[
\begin{tikzcd}[ampersand replacement=\&]
\ctxtapply{\ctxt{L}}{\ctxtapply{\ctxt{LTT}}{t}} \arrow{d}[anchor=north,left]{\rM}
  \&[-25pt] 
  \&[-25pt] \ctxtapply{\ctxt{LTT}_2}{\termapp{\ctxtapply{\ctxt{L}_2}{\ctxtapply{\ctxt{L}}{\ctxtapply{\ctxt{L}_1}{\termcont{\alpha}{c}}}}
}{u}} \arrow[densely dashed]{d}[anchor=north,left]{\rM} \\ 
\ctxtapply{\ctxt{L}}{\ctxtapply{\ctxt{LTT}'}{t'}}
  \&[-25pt] 
  \&[-25pt] \ctxtapply{\ctxt{LTT}_2}{\ctxtapply{\ctxt{L}_2}{\ctxtapply{\ctxt{L}}{\ctxtapply{\ctxt{L}_1}{\termcont{\alpha'}{\termrepl[\alpha']{\alpha}{u}{c}}}}}}
\end{tikzcd} \]
      
      \item It does not overlap with $t$. Then, $\ctxt{LTT} =
      \ctxtapply{\ctxt{LTT}_2}{\termapp{\ctxtapply{\ctxt{L}_2}{\termcont{\alpha}{\ctxt{LCT}}}}{u}}$.
      We conclude with $t' = t$, $\ctxt{L}' = \ctxt{L}$ and $\ctxt{LTT}' =
      \ctxtapply{\ctxt{LTT}_2}{\ctxtapply{\ctxt{L}_2}{\termcont{\alpha'}{\termrepl[\alpha']{\alpha}{u}{\ctxt{LCT}}}}}$: \[
\begin{tikzcd}[ampersand replacement=\&]
\ctxtapply{\ctxt{L}}{\ctxtapply{\ctxt{LTT}}{t}} \arrow{d}[anchor=north,left]{\rM}
  \&[-25pt] 
  \&[-25pt] \ctxtapply{\ctxt{LTT}_2}{\termapp{\ctxtapply{\ctxt{L}_2}{\termcont{\alpha}{\ctxtapply{\ctxt{LCT}}{\ctxtapply{\ctxt{L}}{t}}}}}{u}} \arrow[densely dashed]{d}[anchor=north,left]{\rM} \\ 
\ctxtapply{\ctxt{L}}{\ctxtapply{\ctxt{LTT}'}{t}}
  \&[-25pt] 
  \&[-25pt] \ctxtapply{\ctxt{LTT}_2}{\ctxtapply{\ctxt{L}_2}{\termcont{\alpha'}{\termrepl[\alpha']{\alpha}{u}{\ctxtapply{\ctxt{LCT}}{\ctxtapply{\ctxt{L}}{t}}}}}}
\end{tikzcd} \]
    \end{enumerate}
    
    \item $\rN$. There are two further cases.
    \begin{enumerate}
     \item It overlaps with $t$. Then, $t =
      \ctxtapply{\ctxt{LTC}_1}{\termname{\alpha}{u}}$,
      $\ctxt{LTT} =
      \ctxtapply{\ctxt{LTC}_2}{\termrepl[\alpha']{\alpha}{s}{\ctxt{LCT}}}$
      and the LHS of the step $\rN$ is
      $\termrepl[\alpha']{\alpha}{s}{\ctxtapply{\ctxt{LCT}}{\ctxtapply{\ctxt{LTC}_1}{\termname{\alpha}{u}}}}$.
      We conclude by setting $t' =
      \ctxtapply{\ctxt{LTC}_1}{\termname{\alpha'}{\termconc{u}{s}}}$,
      $\ctxt{L}' = \ctxt{L}$ and $\ctxt{LTT}' =
      \ctxtapply{\ctxt{LTT}_2}{\termcont{\delta}{\ctxtapply{\ctxt{LCC}_2}{\ctxt{LCT}}}}$: \[
\begin{tikzcd}[ampersand replacement=\&]
\ctxtapply{\ctxt{L}}{\ctxtapply{\ctxt{LTT}}{t}} \arrow{d}[anchor=north,left]{\rN}
  \&[-25pt] 
  \&[-25pt] \ctxtapply{\ctxt{LTC}_2}{\termrepl[\alpha']{\alpha}{s}{\ctxtapply{\ctxt{LCT}}{\ctxtapply{\ctxt{L}}{\ctxtapply{\ctxt{LTC}_1}{\termname{\alpha}{u}}}}}} \arrow[densely dashed]{d}[anchor=north,left]{\rN} \\ 
\ctxtapply{\ctxt{L}}{\ctxtapply{\ctxt{LTT}'}{t'}}
  \&[-25pt] 
  \&[-25pt] \ctxtapply{\ctxt{LTC}_2}{\ctxtapply{\ctxt{LCT}}{\ctxtapply{\ctxt{L}}{\ctxtapply{\ctxt{LTC}_1}{\termname{\alpha'}{\termconc{u}{s}}}}}}
\end{tikzcd} \]
      
      \item It does not overlap with $t$. Then, $\ctxt{LTT} =
      \ctxtapply{\ctxt{LTT}_2}{\termcont{\gamma}{\ctxtapply{\ctxt{LCC}_2}{\termrepl[\alpha']{\alpha}{s}{\ctxtapply{\ctxt{LCC}}{\termname{\alpha}{\ctxt{LCT}}}}}}}$.
      We conclude with $t' = t$, $\ctxt{L}' = \ctxt{L}$ and $\ctxt{LTT}' =
      \ctxtapply{\ctxt{LTT}_2}{\termcont{\gamma}{\ctxtapply{\ctxt{LCC}_2}{\ctxtapply{\ctxt{LCC}}{\termname{\alpha'}{\termconc{\ctxt{LCT}}{s}}}}}}$: \[
\begin{tikzcd}[ampersand replacement=\&]
\ctxtapply{\ctxt{L}}{\ctxtapply{\ctxt{LTT}}{t}} \arrow{d}[anchor=north,left]{\rN}
  \&[-25pt] 
  \&[-25pt] \ctxtapply{\ctxt{LTT}_2}{\termcont{\gamma}{\ctxtapply{\ctxt{LCC}_2}{\termrepl[\alpha']{\alpha}{s}{\ctxtapply{\ctxt{LCC}}{\termname{\alpha}{\ctxtapply{\ctxt{LCT}}{\ctxtapply{\ctxt{L}}{t}}}}}}}} \arrow[densely dashed]{d}[anchor=north,left]{\rN} \\ 
\ctxtapply{\ctxt{L}}{\ctxtapply{\ctxt{LTT}'}{t}}
  \&[-25pt] 
  \&[-25pt] \ctxtapply{\ctxt{LTT}_2}{\termcont{\gamma}{\ctxtapply{\ctxt{LCC}_2}{\ctxtapply{\ctxt{LCC}}{\termname{\alpha'}{\termconc{\ctxtapply{\ctxt{LCT}}{\ctxtapply{\ctxt{L}}{t}}}{s}}}}}}
\end{tikzcd} \]
    \end{enumerate}
    
    \item $\rC$. There are two further cases.
    \begin{enumerate}
      \item It overlaps with $t$. Then, $t =
      \ctxtapply{\ctxt{LTC}_1}{\termrepl[\alpha]{\beta}{s'}{c}}$,
      $\ctxt{LTT} =
      \ctxtapply{\ctxt{LTC}_2}{\termrepl[\alpha']{\alpha}{s}{\ctxt{LCT}}}$
      and the LHS of the step $\rC$ is
      $\termrepl[\alpha']{\alpha}{s}{\ctxtapply{\ctxt{LCT}}{\ctxtapply{\ctxt{LTC}_1}{\termrepl[\alpha]{\beta}{s'}{c}}}}$.
      We conclude by setting $t' =
      \ctxtapply{\ctxt{LTT}_1}{\termcont{\gamma}{\ctxtapply{\ctxt{LCC}_1}{\termrepl[\alpha']{\beta}{\termpush{s'}{s}}{c}}}}$,
      $\ctxt{L}' = \ctxt{L}$ and $\ctxt{LTT}' =
      \ctxtapply{\ctxt{LTC}_2}{\ctxt{LCT}}$: \[
\begin{tikzcd}[ampersand replacement=\&]
\ctxtapply{\ctxt{L}}{\ctxtapply{\ctxt{LTT}}{t}} \arrow{d}[anchor=north,left]{\rC}
  \&[-25pt] 
  \&[-25pt] \ctxtapply{\ctxt{LTC}_2}{\termrepl[\alpha']{\alpha}{s}{\ctxtapply{\ctxt{LCT}}{\ctxtapply{\ctxt{L}}{\ctxtapply{\ctxt{LTC}_1}{\termrepl[\alpha]{\beta}{s'}{c}}}}}} \arrow[densely dashed]{d}[anchor=north,left]{\rC} \\ 
\ctxtapply{\ctxt{L}}{\ctxtapply{\ctxt{LTT}'}{t'}}
  \&[-25pt] 
  \&[-25pt] \ctxtapply{\ctxt{LTC}_2}{\ctxtapply{\ctxt{LCT}}{\ctxtapply{\ctxt{L}}{\ctxtapply{\ctxt{LTC}_1}{\termrepl[\alpha']{\beta}{\termpush{s'}{s}}{c}}}}}
\end{tikzcd} \]
      
      \item It does not overlap with $t$. Then, $\ctxt{LTT} =
      \ctxtapply{\ctxt{LTT}_2}{\termcont{\gamma}{\ctxtapply{\ctxt{LCC}_2}{\termrepl[\alpha']{\alpha}{s}{\ctxtapply{\ctxt{LCC}}{\termrepl[\alpha]{\beta}{s'}{\ctxt{LCT}}}}}}}$.
      We conclude with $t' = t$, $\ctxt{L}' = \ctxt{L}$ and $\ctxt{LTT}' =
      \ctxtapply{\ctxt{LTT}_2}{\termcont{\gamma}{\ctxtapply{\ctxt{LCC}_2}{\ctxtapply{\ctxt{LCC}}{\termrepl[\alpha']{\beta}{\termpush{s'}{s}}{\ctxt{LCT}}}}}}$: \[
\begin{tikzcd}[ampersand replacement=\&]
\ctxtapply{\ctxt{L}}{\ctxtapply{\ctxt{LTT}}{t}} \arrow{d}[anchor=north,left]{\rC}
  \&[-25pt] 
  \&[-25pt] \ctxtapply{\ctxt{LTT}_2}{\termcont{\gamma}{\ctxtapply{\ctxt{LCC}_2}{\termrepl[\alpha']{\alpha}{s}{\ctxtapply{\ctxt{LCC}}{\termrepl[\alpha]{\beta}{s'}{\ctxtapply{\ctxt{LCT}}{\ctxtapply{\ctxt{L}}{t}}}}}}}} \arrow[densely dashed]{d}[anchor=north,left]{\rC} \\ 
\ctxtapply{\ctxt{L}}{\ctxtapply{\ctxt{LTT}'}{t}}
  \&[-25pt] 
  \&[-25pt] \ctxtapply{\ctxt{LTT}_2}{\termcont{\gamma}{\ctxtapply{\ctxt{LCC}_2}{\ctxtapply{\ctxt{LCC}}{\termrepl[\alpha']{\beta}{\termpush{s'}{s}}{\ctxtapply{\ctxt{LCT}}{\ctxtapply{\ctxt{L}}{t}}}}}}}
\end{tikzcd} \]
    \end{enumerate}
  \end{enumerate}

  \item There are not other cases.
  \qedhere
\end{itemize}
\end{proof}


%% file: proofs/equivalence/lcc-bisimulation-v2.tex
\begin{proof}
  We focus on the first item, the second being similar.  
The proof proceeds by analysing all the overlapping cases
between the LHS of the $\rcan$-step and 
$\ctxtapply{\ctxt{R}}{\ctxtapply{\ctxt{LCC}}{c}}$:
\begin{itemize}
  \item The step is completely within $c$, \ie $c \reduce[\rcan]
  c'$. Then, it suffices to set $\ctxt{R}' = \ctxt{R}$ and $\ctxt{LCC}' = \ctxt{LCC}$ and the reduction sequences $d''\reducemany[\rM] d'$ and $e''\reducemany[\rM] e'$ are empty.
  
  \item The step is completely within $\ctxt{R}$, \ie $\ctxt{R}
  \reduce[\rcan] \ctxt{R}'$. Then, it suffices to set $c' = c$ and $\ctxt{LCC}' =
  \ctxt{LCC}$  and the reduction sequences $d''\reducemany[\rM] d'$ and $e''\reducemany[\rM] e'$ are empty. This case relies on the fact that all $\ctxt{R}$ contexts are also $\ctxt{LCC}$ contexts and that the latter are present in the patterns of the LHSs of rewrite rules defining $\rcan$.
  
  \item The step is completely within $\ctxt{LCC}$, \ie
  $\ctxt{LCC} \reduce[\rcan] \ctxt{LCC}'$. Then it suffices to set $c' = c$ and
  $\ctxt{R}' = \ctxt{R}$  and the reduction sequences $d''\reducemany[\rM] d'$ and $e''\reducemany[\rM] e'$ are empty. 
  
  \item The step overlaps with $\ctxt{LCC}$. There are three further cases depending on the reduction step applied  and the reduction sequences $d''\reducemany[\rM] d'$ and $e''\reducemany[\rM] e'$ are empty:
  \begin{enumerate}
    \item $\rB$. Then, $\ctxt{LCC} =
    \ctxtapply{\ctxt{LCC}_1}{\termname{\gamma}{\ctxtapply{\ctxt{LTC}_1}{\termapp{\ctxtapply{\ctxt{L}}{\termabs{x}{\ctxt{LTC}_2}}}{u}}}}$.
    We conclude with $c' = c$, $\ctxt{R}' = \ctxt{R}$ and $\ctxt{LCC}' =
    \ctxtapply{\ctxt{LCC}_1}{\termname{\gamma}{\ctxtapply{\ctxt{LTC}_1}{\ctxtapply{\ctxt{L}}{\termsubs{x}{u}{\ctxt{LTC}_2}}}}}$: \[
\begin{tikzcd}[ampersand replacement=\&]
\ctxtapply{\ctxt{R}}{\ctxtapply{\ctxt{LCC}}{c}} \arrow{d}[anchor=north,left]{\rB}
  \&[-25pt] 
  \&[-25pt] \ctxtapply{\ctxt{LCC}_1}{\termname{\gamma}{\ctxtapply{\ctxt{LTC}_1}{\termapp{\ctxtapply{\ctxt{L}}{\termabs{x}{\ctxtapply{\ctxt{LTC}_2}{\ctxtapply{\ctxt{R}}{c}}}}}{u}}}}\arrow[densely dashed]{d}[anchor=north,left]{\rB} \\ 
\ctxtapply{\ctxt{R}}{\ctxtapply{\ctxt{LCC}'}{c}}
  \&[-25pt] 
  \&[-25pt] \ctxtapply{\ctxt{LCC}_1}{\termname{\gamma}{\ctxtapply{\ctxt{LTC}_1}{\ctxtapply{\ctxt{L}}{\termsubs{x}{u}{\ctxtapply{\ctxt{LTC}_2}{\ctxtapply{\ctxt{R}}{c}}}}}}}
\end{tikzcd} \]
    
    \item $\rM$. Then, $\ctxt{LCC} =
    \ctxtapply{\ctxt{LCC}_1}{\termname{\gamma}{\ctxtapply{\ctxt{LTC}_1}{\termapp{\ctxtapply{\ctxt{L}}{\termcont{\alpha}{\ctxt{LTC}_2}}}{u}}}}$.
    We conclude with $c' = c$, $\ctxt{R}' = \ctxt{R}$ and $\ctxt{LCC}' =
    \ctxtapply{\ctxt{LCC}_1}{\termname{\gamma}{\ctxtapply{\ctxt{LTC}_1}{\ctxtapply{\ctxt{L}}{\termcont{\alpha'}{\termrepl[\alpha']{\alpha}{u}{\ctxt{LTC}_2}}}}}}$  and the reduction sequences $d''\reducemany[\rM] d'$ and $e''\reducemany[\rM] e'$ are empty: \[
\begin{tikzcd}[ampersand replacement=\&]
\ctxtapply{\ctxt{R}}{\ctxtapply{\ctxt{LCC}}{c}} \arrow{d}[anchor=north,left]{\rM}
  \&[-25pt] 
  \&[-25pt] \ctxtapply{\ctxt{LCC}_1}{\termname{\gamma}{\ctxtapply{\ctxt{LTC}_1}{\termapp{\ctxtapply{\ctxt{L}}{\termcont{\alpha}{\ctxtapply{\ctxt{LTC}_2}{\ctxtapply{\ctxt{R}}{c}}}}}{u}}}}\arrow[densely dashed]{d}[anchor=north,left]{\rM} \\ 
\ctxtapply{\ctxt{R}}{\ctxtapply{\ctxt{LCC}'}{c}}
  \&[-25pt] 
  \&[-25pt] \ctxtapply{\ctxt{LCC}_1}{\termname{\gamma}{\ctxtapply{\ctxt{LTC}_1}{\ctxtapply{\ctxt{L}}{\termcont{\alpha'}{\termrepl[\alpha']{\alpha}{u}{\ctxtapply{\ctxt{LTC}_2}{\ctxtapply{\ctxt{R}}{c}}}}}}}}
\end{tikzcd} \]
    
    \item $\rN$. Then  there are two further cases.
    \begin{enumerate}
      \item It overlaps with $c$. Then we have $c =
      \ctxtapply{\ctxt{LCC}_1}{\termname{\alpha}{t}}$,
      $\ctxt{LCC} =
      \ctxtapply{\ctxt{LCC}_3}{\termrepl[\alpha']{\alpha}{s}{\ctxt{LCC}_2}}$
      and the rule $\rN$ applies to 
      $\termrepl[\alpha']{\alpha}{s}{\ctxtapply{\ctxt{LCC}_2}{\ctxtapply{\ctxt{LCC}_1}{\termname{\alpha}{t}}}}$.
      We conclude by setting $c' =
      \ctxtapply{\ctxt{LCC}_1}{\termname{\alpha'}{\termconc{t}{s}}}$,
      $\ctxt{R}' = \ctxt{R}$ and $\ctxt{LCC}' =
      \ctxtapply{\ctxt{LCC}_3}{\ctxt{LCC}_2}$ and the reduction sequences $d''\reducemany[\rM] d'$ and $e''\reducemany[\rM] e'$ are empty: \[
\begin{tikzcd}[ampersand replacement=\&]
\ctxtapply{\ctxt{R}}{\ctxtapply{\ctxt{LCC}}{c}} \arrow{d}[anchor=north,left]{\rN}
  \&[-25pt] 
  \&[-25pt] \ctxtapply{\ctxt{LCC}_3}{\termrepl[\alpha']{\alpha}{s}{\ctxtapply{\ctxt{LCC}_2}{\ctxtapply{\ctxt{R}}{\ctxtapply{\ctxt{LCC}_1}{\termname{\alpha}{t}}}}}} \arrow[densely dashed]{d}[anchor=north,left]{\rN} \\ 
\ctxtapply{\ctxt{R}}{\ctxtapply{\ctxt{LCC}'}{c'}}
  \&[-25pt] 
  \&[-25pt] \ctxtapply{\ctxt{LCC}_3}{\ctxtapply{\ctxt{LCC}_2}{\ctxtapply{\ctxt{R}}{\ctxtapply{\ctxt{LCC}_1}{\termname{\alpha'}{\termconc{t}{s}}}}}}
\end{tikzcd} \]
      
      \item It does not overlap with $c$. Then, $\ctxt{LCC} =
      \ctxtapply{\ctxt{LCC}_3}{\termrepl[\alpha']{\alpha}{s}{\ctxtapply{\ctxt{LCC}_2}{\termname{\alpha}{\ctxt{LTC}}}}}$.
      We conclude with $c' = c$, $\ctxt{R}' = \ctxt{R}$ and $\ctxt{LCC}' =
      \ctxtapply{\ctxt{LCC}_3}{\ctxtapply{\ctxt{LCC}_2}{\termname{\alpha'}{\termconc{\ctxt{LTC}}{s}}}}$ and the reduction sequence $e''\reducemany[\rM] e'$ is empty: \[
\begin{tikzcd}[ampersand replacement=\&]
\ctxtapply{\ctxt{R}}{\ctxtapply{\ctxt{LCC}}{c}} \arrow{d}[anchor=north,left]{\rN}
  \&[-25pt] 
  \&[-25pt] \ctxtapply{\ctxt{LCC}_3}{\termrepl[\alpha']{\alpha}{s}{\ctxtapply{\ctxt{LCC}_2}{\termname{\alpha}{\ctxtapply{\ctxt{LTC}}{\ctxtapply{\ctxt{R}}{c}}}}}} \arrow[densely dashed]{d}[anchor=north,left]{\rN} \\ 
\ctxtapply{\ctxt{R}}{\ctxtapply{\ctxt{LCC}'}{c}}
  \&[-25pt] 
  \&[-25pt] \ctxtapply{\ctxt{LCC}_3}{\ctxtapply{\ctxt{LCC}_2}{\termname{\alpha'}{\termconc{\ctxtapply{\ctxt{LTC}}{\ctxtapply{\ctxt{R}}{c}}}{s}}}}
\end{tikzcd} \]
    \end{enumerate}
    
    \item $\rC$. Then there are two possibilities.
    \begin{enumerate}
      \item It overlaps with $c$. Then we have $c =
      \ctxtapply{\ctxt{LCC}_1}{\termrepl[\alpha]{\beta}{s'}{c_1}}$,
      $\ctxt{LCC} =
      \ctxtapply{\ctxt{LCC}_3}{\termrepl[\alpha']{\alpha}{s}{\ctxt{LCC}_2}}$
      and the rule $\rC$ applies to
      $\termrepl[\alpha']{\alpha}{s}{\ctxtapply{\ctxt{LCC}_2}{\ctxtapply{\ctxt{LCC}_1}{\termrepl[\alpha]{\beta}{s'}{c_1}}}}$.
      We conclude by setting $c' =
      \ctxtapply{\ctxt{LCC}_1}{\termrepl[\alpha']{\beta}{\termpush{s'}{s}}{c_1}}$,
      $\ctxt{R}' = \ctxt{R}$ and $\ctxt{LCC}' =
      \ctxtapply{\ctxt{LCC}_3}{\ctxt{LCC}_2}$  and the reduction sequences $d''\reducemany[\rM] d'$ and $e''\reducemany[\rM] e'$ are empty: \[
\begin{tikzcd}[ampersand replacement=\&]
\ctxtapply{\ctxt{R}}{\ctxtapply{\ctxt{LCC}}{c}} \arrow{d}[anchor=north,left]{\rC}
  \&[-25pt] 
  \&[-25pt] \ctxtapply{\ctxt{LCC}_3}{\termrepl[\alpha']{\alpha}{s}{\ctxtapply{\ctxt{LCC}_2}{\ctxtapply{\ctxt{R}}{\ctxtapply{\ctxt{LCC}_1}{\termrepl[\alpha]{\beta}{s'}{c_1}}}}}} \arrow[densely dashed]{d}[anchor=north,left]{\rC} \\ 
\ctxtapply{\ctxt{R}}{\ctxtapply{\ctxt{LCC}'}{c'}}
  \&[-25pt] 
  \&[-25pt] \ctxtapply{\ctxt{LCC}_3}{\ctxtapply{\ctxt{LCC}_2}{\ctxtapply{\ctxt{R}}{\ctxtapply{\ctxt{LCC}_1}{\termrepl[\alpha']{\beta}{\termpush{s'}{s}}{c_1}}}}}
\end{tikzcd} \]
      
      \item It does not overlap with $c$. Then, $\ctxt{LCC} =
      \ctxtapply{\ctxt{LCC}_3}{\termrepl[\alpha']{\alpha}{s}{\ctxtapply{\ctxt{LCC}_2}{\termrepl[\alpha]{\beta}{s'}{\ctxt{LCC}_1}}}}$.
      We conclude with $c' = c$, $\ctxt{R}' = \ctxt{R}$ and $\ctxt{LCC}' =
      \ctxtapply{\ctxt{LCC}_3}{\ctxtapply{\ctxt{LCC}_2}{\termrepl[\alpha']{\beta}{\termpush{s'}{s}}{\ctxt{LCC}_1}}}$  and the reduction sequences $d''\reducemany[\rM] d'$ and $e''\reducemany[\rM] e'$ are empty: \[
\begin{tikzcd}[ampersand replacement=\&]
\ctxtapply{\ctxt{R}}{\ctxtapply{\ctxt{LCC}}{c}} \arrow{d}[anchor=north,left]{\rC}
  \&[-25pt] 
  \&[-25pt] \ctxtapply{\ctxt{LCC}_3}{\termrepl[\alpha']{\alpha}{s}{\ctxtapply{\ctxt{LCC}_2}{\termrepl[\alpha]{\beta}{s'}{\ctxtapply{\ctxt{LCC}_1}{\ctxtapply{\ctxt{R}}{c}}}}}} \arrow[densely dashed]{d}[anchor=north,left]{\rC} \\ 
\ctxtapply{\ctxt{R}}{\ctxtapply{\ctxt{LCC}'}{c}}
  \&[-25pt] 
  \&[-25pt] \ctxtapply{\ctxt{LCC}_3}{\ctxtapply{\ctxt{LCC}_2}{\termrepl[\alpha']{\beta}{\termpush{s'}{s}}{\ctxtapply{\ctxt{LCC}_1}{\ctxtapply{\ctxt{R}}{c}}}}}
\end{tikzcd} \]
    \end{enumerate}
  \end{enumerate}
  
  \item The step overlaps with $\ctxt{R}$. Since the reduction step cannot be $\rB$ nor $\rM$, there are two further cases to consider.
  \begin{enumerate}

    \item $\rN$. Note that it cannot overlap  $\ctxt{LCC}$ too. Indeed, if it
    did then we have the context $\ctxt{R} =
    \ctxtapply{\ctxt{R}_1}{\termrepl[\alpha']{\alpha}{s}{\ctxtapply{\ctxt{R}_2}{\ctxtapply{\ctxt{LCC}_1}{\termname{\alpha}{\ctxtapply{\ctxt{LTC}_1}{\termcont{\delta}{\ctxt{LTC}_2}}}}}}}$.
    However, this is not allowed by the condition $\bn{\ctxt{R}} \notin
    \ctxt{LCC}$. There are two cases to consider.
    \begin{enumerate}
      \item The step overlaps with $c$. Then $d =
      \ctxtapply{\ctxt{R}_1}{\termrepl[\alpha']{\alpha}{s}{\ctxtapply{\ctxt{R}_2}{\ctxtapply{\ctxt{LCC}}{\ctxtapply{\ctxt{LCC}_1}{\termname{\alpha}{t}}}}}}$
      and $c = \ctxtapply{\ctxt{LCC}_1}{\termname{\alpha}{t}}$. We conclude
      with $c' = \ctxtapply{\ctxt{LCC}_1}{\termname{\alpha'}{\termconc{t}{s}}}$,
      $\ctxt{R}' = \ctxtapply{\ctxt{R}_1}{\ctxt{R}_2}$ and $\ctxt{LCC'} =
      \ctxt{LCC}$ and the reduction sequences $d'' \reducemany[\rM] d'$ and
      $e'' \reducemany[\rM] e'$ are empty: \[
\begin{tikzcd}[ampersand replacement=\&]
  \ctxtapply{\ctxt{R}_1}{\termrepl[\alpha']{\alpha}{s}{\ctxtapply{\ctxt{R}_2}{\ctxtapply{\ctxt{LCC}}{\ctxtapply{\ctxt{LCC}_1}{\termname{\alpha}{t}}}}}} \arrow{d}[anchor=north,left]{\rN}
  \&[-25pt] 
  \&[-25pt] \ctxtapply{\ctxt{LCC}}{\ctxtapply{\ctxt{R}_1}{\termrepl[\alpha']{\alpha}{s}{\ctxtapply{\ctxt{R}_2}{\ctxtapply{\ctxt{LCC}_1}{\termname{\alpha}{t}}}}}}\arrow[densely dashed]{d}[anchor=north,left]{\rN} \\ 
\ctxtapply{\ctxt{R}_1}{\ctxtapply{\ctxt{R}_2}{\ctxtapply{\ctxt{LCC}}{\ctxtapply{\ctxt{LCC}_1}{\termname{\alpha'}{\termconc{t}{s}}}}}}
  \&[-25pt] 
  \&[-25pt] \ctxtapply{\ctxt{LCC}}{\ctxtapply{\ctxt{R}_1}{\ctxtapply{\ctxt{R}_2}{\ctxtapply{\ctxt{LCC}_1}{\termname{\alpha'}{\termconc{t}{s}}}}}}
\end{tikzcd} \]

      \item The step does not overlap with $c$. Then $d =
      \ctxtapply{\ctxt{R}_1}{\termrepl[\alpha']{\alpha}{s}{\ctxtapply{\ctxt{R}_2}{\ntermren{\beta}{\alpha}{\ctxtapply{\ctxt{R_3}}{\ctxtapply{\ctxt{LCC}}{c}}}}}}$.
      We assume that $s=\termpush{u_1}{\termpush{\ldots}{u_n}}$. We conclude
      by considering $c' = c$, $\ctxt{R}' =
      \ctxtapply{\ctxt{R}_1}{\ctxtapply{\ctxt{R}_2}{\termname{\alpha'}{(\termcont{\beta_n}{\termrepl[\beta_n]{\beta_{n-1}}{u_n}{\termrepl[\beta_1]{\beta}{u_1}{\ctxt{R_3}}\ldots}})}}}$
      and $\ctxt{LCC'} = \ctxt{LCC}$. \[
\begin{tikzcd}[ampersand replacement=\&]
 \ctxtapply{\ctxt{R}_1}{\termrepl[\alpha']{\alpha}{s}{\ctxtapply{\ctxt{R}_2}{ \ntermren{\beta}{\alpha}{\ctxtapply{\ctxt{R_3}}{\ctxtapply{\ctxt{LCC}}{c}}}}}}\arrow{d}[anchor=north,left]{\rN}
  \&[-25pt] 
  \&[-25pt] \ctxtapply{\ctxt{LCC}}{\ctxtapply{\ctxt{R}_1}{\termrepl[\alpha']{\alpha}{s}{\ctxtapply{\ctxt{R}_2}{ \ntermren{\beta}{\alpha}{\ctxtapply{\ctxt{R_3}}{c}}}}}}\arrow[densely dashed]{d}[anchor=north,left]{\rN} \\ 
  \ctxtapply{\ctxt{R}_1}{\ctxtapply{\ctxt{R}_2}{\termname{\alpha'}{\termconc{(\termcont{\beta}{\ctxtapply{\ctxt{R_3}}{\ctxtapply{\ctxt{LCC}}{c}}})}{s}}}} \arrow[densely dashed,two heads]{d}[anchor=north,left]{\rM}
  \&[-25pt] 
  \&[-25pt] \ctxtapply{\ctxt{LCC}}{\ctxtapply{\ctxt{R}_1}{\ctxtapply{\ctxt{R}_2}{\termname{\alpha'}{\termconc{(\termcont{\beta}{\ctxtapply{\ctxt{R_3}}{c}})}{s}}}}}\arrow[densely dashed,two heads]{d}[anchor=north,left]{\rM}
  \\
  d'
  \&[-25pt] 
  \&[-25pt] e'
\end{tikzcd} \]
      where $d'$ and $e'$ are
      $\ctxtapply{\ctxt{R}_1}{\ctxtapply{\ctxt{R}_2}{\termname{\alpha'}{(\termcont{\beta_n}{\termrepl[\beta_n]{\beta_{n-1}}{u_n}{\termrepl[\beta_1]{\beta}{u_1}{\ctxtapply{\ctxt{R_3}}{\ctxtapply{\ctxt{LCC}}{c}}}\ldots}})}}}$
      and
      $e' = \ctxtapply{\ctxt{LCC}}{\ctxtapply{\ctxt{R}_1}{\ctxtapply{\ctxt{R}_2}{\termname{\alpha'}{(\termcont{\beta_n}{\termrepl[\beta_n]{\beta_{n-1}}{u_n}{\termrepl[\beta_1]{\beta}{u_1}{\ctxtapply{\ctxt{R_3}}{c}}\ldots}})}}}}$
      respectively.
    \end{enumerate}

    \item $\rC$. Note that it cannot overlap  $\ctxt{LCC}$ too. Indeed, if it
    did then we have the command $d =
    \ctxtapply{\ctxt{R}_1}{\termrepl[\alpha']{\alpha}{s}{\ctxtapply{\ctxt{R}_2}{\ctxtapply{\ctxt{LCC}_1}{\termrepl[\alpha]{\beta}{s'}{\ctxtapply{\ctxt{LCC}_2}{c}}}}}}$.
    However, this is not allowed by the condition $\bn{\ctxt{R}} \notin
    \ctxt{LCC}$. There are two cases to consider.
    \begin{enumerate}
      \item The step overlaps with $c$. Then $d =
      \ctxtapply{\ctxt{R}_1}{\termrepl[\alpha']{\alpha}{s}{\ctxtapply{\ctxt{R}_2}{\ctxtapply{\ctxt{LCC}}{\ctxtapply{\ctxt{LCC}_1}{\termrepl[\alpha]{\beta}{s'}{\ctxtapply{\ctxt{LCC}_2}{c'}}}}}}}$
      and $c =
      \ctxtapply{\ctxt{LCC}_1}{\termrepl[\alpha]{\beta}{s'}{\ctxtapply{\ctxt{LCC}_2}{c'}}}$
      and the reduction sequences $d'' \reducemany[\rM] d'$ and $e''
      \reducemany[\rM] e'$ are empty: \[\kern-1em
\begin{tikzcd}[ampersand replacement=\&]
  \ctxtapply{\ctxt{R}_1}{\termrepl[\alpha']{\alpha}{s}{\ctxtapply{\ctxt{R}_2}{\ctxtapply{\ctxt{LCC}}{\ctxtapply{\ctxt{LCC}_1}{\termrepl[\alpha]{\beta}{s'}{\ctxtapply{\ctxt{LCC}_2}{c'}}}}}}}\arrow{d}[anchor=north,left]{\rC}
  \&[-25pt] 
  \&[-25pt]   \ctxtapply{\ctxt{LCC}}{\ctxtapply{\ctxt{R}_1}{\termrepl[\alpha']{\alpha}{s}{\ctxtapply{\ctxt{R}_2}{\ctxtapply{\ctxt{LCC}_1}{\termrepl[\alpha]{\beta}{s'}{\ctxtapply{\ctxt{LCC}_2}{c'}}}}}}}\arrow[densely dashed]{d}[anchor=north,left]{\rC} \\ 
 \ctxtapply{\ctxt{R}_1}{\ctxtapply{\ctxt{R}_2}{\ctxtapply{\ctxt{LCC}}{\ctxtapply{\ctxt{LCC}_1}{\termrepl[\alpha]{\beta}{\termpush{s'}{s}}{\ctxtapply{\ctxt{LCC}_2}{c'}}}}}}
  \&[-25pt] 
  \&[-25pt]   \ctxtapply{\ctxt{LCC}}{\ctxtapply{\ctxt{R}_1}{\ctxtapply{\ctxt{R}_2}{\ctxtapply{\ctxt{LCC}_1}{\termrepl[\alpha]{\beta}{\termpush{s'}{s}}{\ctxtapply{\ctxt{LCC}_2}{c'}}}}}}
\end{tikzcd} \]

      \item The step does not overlap with $c$. Then $d \!=\!
      \ctxtapply{\ctxt{R}_1}{\termrepl[\alpha']{\alpha}{s}{\ctxtapply{\ctxt{R}_2}{\ctxtapply{\ctxt{R}_3}{\termrepl[\alpha]{\beta}{s'}{\ctxtapply{\ctxt{R}_4}{\ctxtapply{\ctxt{LCC}}{c}}}}}}}$.
      Then we set $c' = c$, $\ctxt{R}' =
      \ctxtapply{\ctxt{R}_1}{\ctxtapply{\ctxt{R}_2}{\ctxtapply{\ctxt{R}_3}{\termrepl[\alpha]{\beta}{\termpush{s'}{s}}{\ctxt{R}_4}}}}$
      and $\ctxt{LCC}' = \ctxt{LCC}$ and the reduction sequences $d''
      \reducemany[\rM] d'$ and $e'' \reducemany[\rM] e'$ are empty: \[
\begin{tikzcd}[ampersand replacement=\&]
  \ctxtapply{\ctxt{R}_1}{\termrepl[\alpha']{\alpha}{s}{\ctxtapply{\ctxt{R}_2}{\ctxtapply{\ctxt{R}_3}{\termrepl[\alpha]{\beta}{s'}{\ctxtapply{\ctxt{R}_4}{\ctxtapply{\ctxt{LCC}}{c}}}}}}}\arrow{d}[anchor=north,left]{\rC}
  \&[-25pt] 
  \&[-25pt]   \ctxtapply{\ctxt{LCC}}{\ctxtapply{\ctxt{R}_1}{\termrepl[\alpha']{\alpha}{s}{\ctxtapply{\ctxt{R}_2}{\ctxtapply{\ctxt{R}_3}{\termrepl[\alpha]{\beta}{s'}{\ctxtapply{\ctxt{R}_4}{c}}}}}}}\arrow[densely dashed]{d}[anchor=north,left]{\rC} \\ 
 \ctxtapply{\ctxt{R}_1}{\ctxtapply{\ctxt{R}_2}{\ctxtapply{\ctxt{R}_3}{\termrepl[\alpha]{\beta}{\termpush{s'}{s}}{\ctxtapply{\ctxt{R}_4}{\ctxtapply{\ctxt{LCC}}{c}}}}}}
  \&[-25pt] 
  \&[-25pt]   \ctxtapply{\ctxt{LCC}}{\ctxtapply{\ctxt{R}_1}{\ctxtapply{\ctxt{R}_2}{\ctxtapply{\ctxt{R}_3}{\termrepl[\alpha]{\beta}{\termpush{s'}{s}}{\ctxtapply{\ctxt{R}_4}{c}}}}}}
\end{tikzcd} \]
    \end{enumerate}
  \end{enumerate}

  \item There are no further cases.
  \qedhere
\end{itemize}
\end{proof}


%% file: proofs/equivalence/permute-v2.tex
\begin{proof}\mbox{}
\begin{enumerate}
  \item Let $\ctxtapply{\ctxt{L}}{\ctxtapply{\ctxt{LTT}}{t}} \in
  \Term{\calcLambdaM}$ with $\bv{\ctxt{L}} \notin \ctxt{LTT}$ and 
  $\fc{\ctxt{L}}{\ctxt{LTT}}$.
  By induction on the length of the longest
  $\rcan$ reduction sequence from $\ctxtapply{\ctxt{L}}{\ctxtapply{\ctxt{LTT}}{t}}$ to
  its normal form, resorting to
  Lemma~\ref{l:control:equivalence:ltt-bisimulation}.
  Suppose $d_0 = \ctxtapply{\ctxt{LTT}}{\ctxtapply{\ctxt{L}}{t}}
  \reducemany[\rcan] \fcan{\ctxtapply{\ctxt{LTT}}{\ctxtapply{\ctxt{L}}{t}}} =
  d_n$ is a longest reduction sequence consisting of $n$ steps. This is depicted on the
  left, in the figure below. Application of
  Lemma~\ref{l:control:equivalence:ltt-bisimulation} will produce the subdiagram at the top of the figure. The $\rcan$ reduction sequence $d_1
  \reducemany[\rM] d'_1 \reducemany[\rcan] d_n$ exists by confluence of $\rcan$
  and, moreover, it has at most $n-1$ steps since the reduction sequence from $d_0$
  was assumed to be a longest reduction. This allows us to apply the \ih to
  the reduction sequence $d'_1 \reducemany[\rcan] d_n$ (as indicated in the
  figure) to conclude:
\begin{center}
\begin{tikzcd}[ampersand replacement=\&]
  d_0 = \ctxtapply{\ctxt{LTT}}{\ctxtapply{\ctxt{L}}{t}} \arrow{d}[anchor=north,left]{\rcan}
  \arrow[phantom]{ddrrr}[description]{Lemma~\ref{l:control:equivalence:ltt-bisimulation}}
  \&[-25pt]
  \&[-25pt]
  \&[-25pt]
  \ctxtapply{\ctxt{L}}{\ctxtapply{\ctxt{LTT}}{t}} \arrow[densely dashed]{d}[anchor=north,left]{\rcan}
\\ 
d_1 \arrow[densely dashed,two heads]{dr}[anchor=north,left]{\rM} \arrow[two heads]{dd}[anchor=north,left]{\rcan}
\&[-25pt]
  \&[-25pt]
\&[-25pt]
\arrow[densely dashed,two heads]{d}[anchor=north,left]{\rM}
\\
\&[-25pt]
d'_1 = \ctxtapply{\ctxt{LTT}'}{\ctxtapply{\ctxt{L}'}{t'}} \arrow[densely dashed,two heads]{d}[anchor=north,left]{\rcan} \arrow[phantom]{drr}[description]{\ih}  
  \&[-25pt]
\&[-25pt]
\ctxtapply{\ctxt{L}'}{\ctxtapply{\ctxt{LTT}'}{t'}} \arrow[densely dashed, two heads]{d}[anchor=north,left]{\rcan}
\\
d_n = \fcan{\ctxtapply{\ctxt{LTT}}{\ctxtapply{\ctxt{L}}{t}}} \arrow[phantom]{r}[description]{=} 
\&[-25pt]
\fcan{\ctxtapply{\ctxt{LTT}'}{\ctxtapply{\ctxt{L}'}{t'}}}
\&[-25pt]
\eqsigma 
\&[-25pt]
\fcan{\ctxtapply{\ctxt{L}'}{\ctxtapply{\ctxt{LTT}'}{t'}}} \arrow[phantom]{r}[description]{=}
\&[-20pt] \fcan{\ctxtapply{\ctxt{L}}{\ctxtapply{\ctxt{LTT}}{t}}}
\end{tikzcd}
\end{center}

  \item Similar to the previous item but using Lemma~\ref{l:control:equivalence:lcc-bisimulation}.
  \qedhere
\end{enumerate}
\end{proof}


%% file: appendix-bisimulation.tex
\section{Strong Bisimulation Result}
\label{a:bisimulation}

\begin{lem}
Let $o,o' \in \Object{\calcLambdaM}$ such that $o \eqsigma[\ast] o'$ with
$\eqsigma[\ast]$ any rule from Figure~\ref{f:control:equivalence:eqsigma}. If $o
= \ctxtapply{\ctxt{LXC}}{c}$ for some $\ctxt{LXC}$ and some $c$, with $\ctxt{X}
\in \set{\ctxt{T}, \ctxt{C}}$, $\alpha \notin \ctxt{LXC}$,
$\fc{\alpha'}{\ctxt{LXC}}$ and $\fc{s}{\ctxt{LXC}}$, then there exist
$\ctxt{LXC}'$ and $c'$ such that $o' = \ctxtapply{\ctxt{LXC}'}{c'}$ with
$\alpha \notin \ctxt{LXC}'$, $\fc{\alpha'}{\ctxt{LXC}'}$, $\fc{s}{\ctxt{LXC}'}$
and $\fcan{\ctxtapply{\ctxt{LXC}}{\termrepl[\alpha']{\alpha}{s}{c}}} \eqsigma
\fcan{\ctxtapply{\ctxt{LXC}'}{\termrepl[\alpha']{\alpha}{s}{c'}}}$.
\label{l:control:bisimulation:ctxt-closure-rules}
\end{lem}

\input{proofs/bisimulation/ctxt-closure-rules}

\begin{lem}
Let $o, o' \in \Object{\calcLambdaM}$ such that $o \eqsigma o'$. Let $s$ be a
stack.
\begin{enumerate}
  \item\label{l:control:bisimulation:ctxt-closure-aux:conc} If $o, o' \in
  \Term{\calcLambdaM}$, then $\fcan{\termconc{o}{s}} \eqsigma
  \fcan{\termconc{o'}{s}}$.
  
  \item\label{l:control:bisimulation:ctxt-closure-aux:repl} If $o, o' \in
  \Command{\calcLambdaM}$, then $\fcan{\termrepl[\alpha']{\alpha}{s}{o}}
  \eqsigma \fcan{\termrepl[\alpha']{\alpha}{s}{o'}}$.
\end{enumerate}
\label{l:control:bisimulation:ctxt-closure-aux}
\end{lem}

\input{proofs/bisimulation/ctxt-closure-aux}

\gettoappendix{l:control:bisimulation:ctxt-closure}
\input{proofs/bisimulation/ctxt-closure}

\begin{lem}
Let $o, o' \in \Object{\calcLambdaM}$ and $u, u' \in
\Term{\calcLambdaM}$ such that $o \eqsigma o'$ and $u \eqsigma u'$.
\begin{enumerate}
  \item\label{l:control:bisimulation:meta-subs-eqsigma:o} Then,
  $\fcan{\subsapply{\subs{x}{u}}{o}} \eqsigma
  \fcan{\subsapply{\subs{x}{u}}{o'}}$.
  
  \item\label{l:control:bisimulation:meta-subs-eqsigma:u} Then,
  $\fcan{\subsapply{\subs{x}{u}}{o}} \eqsigma
  \fcan{\subsapply{\subs{x}{u'}}{o}}$.
\end{enumerate}
\label{l:control:bisimulation:meta-subs-eqsigma}
\end{lem}

\input{proofs/bisimulation/meta-subs-eqsigma}

\begin{lem}
Let $o, o', s, s' \in \Object{\calcLambdaM}$ with $s$ and $s'$ stacks such that
$o\eqsigma o'$ and  $s \eqsigma s'$.
\begin{enumerate}
  \item\label{l:control:bisimulation:meta-repl-eqsigma:o} Then,
  $\fcan{\replapply{\repl[\alpha']{\alpha}{s}}{o}} \eqsigma
  \fcan{\replapply{\repl[\alpha']{\alpha}{s}}{o'}}$.
  
  \item\label{l:control:bisimulation:meta-repl-eqsigma:s} Then,
  $\fcan{\replapply{\repl[\alpha']{\alpha}{s}}{o}} \eqsigma
  \fcan{\replapply{\repl[\alpha']{\alpha}{s'}}{o}}$.
\end{enumerate}
\label{l:control:bisimulation:meta-repl-eqsigma}
\end{lem}

\input{proofs/bisimulation/meta-repl-eqsigma}

\gettoappendix{l:control:bisimulation:meta-eqsigma}
\input{proofs/bisimulation/meta-eqsigma}

\gettoappendix{t:control:bisimulation}
\input{proofs/bisimulation/bisimulation}


%% file: proofs/bisimulation/ctxt-closure-rules.tex
\begin{proof}
By case analysis on $\ctxt{LXC}$. We only illustrate one of the base cases,
namely when $\ctxt{LXC} = \boxdot$, the others being similar. In the case that
$\ctxt{LXC} = \boxdot$, we must have $o = c$. There are possible rules for
commands: $\ruleEqsigExRepl$,  $\ruleEqsigExRen$, and $\ruleEqsigPopPop$. The
first two follow by using~\ref{l:control:equivalence:permute}
(\ref{l:control:equivalence:permute:r}), the last one is direct. We provide
details on the $\ruleEqsigPopPop$ case.
\begin{itemize}
  \item $\ruleEqsigPopPop$. Then, $o =
  \termname{\gamma'}{\termabs{x}{\termcont{\gamma}{\termname{\delta'}{\termabs{y}{\termcont{\delta}{c_0}}}}}}$ 
  and $o' =
  \termname{\delta'}{\termabs{y}{\termcont{\delta}{\termname{\gamma'}{\termabs{x}{\termcont{\gamma}{c_0}}}}}}$
  with $\delta \neq \gamma'$ and $\gamma \neq \delta'$. By
  $\alpha$-conversion we also assume $x \notin s$, $y \notin s$, $\gamma
  \neq \alpha'$, $\gamma \notin s$, $\delta \neq \alpha'$ and $\delta \notin
  s$. Then, there are four possible cases.
  \begin{enumerate}
    \item $\termrepl[\alpha']{\alpha}{s}{o} \reduce[\rN]
    \termname{\alpha'}{\termconc{(\termabs{x}{\termcont{\gamma}{\termname{\delta'}{\termabs{y}{\termcont{\delta}{c_0}}}}})}{s}}$
    (\ie $\alpha = \gamma'$, $\alpha \neq \delta'$ and $\gamma' \notin c_0$).
    Assume $s = \termpush{u}{s'}$ (the case $s = u$ is slightly
    simpler) and consider fresh names $\gamma'', \delta''$. Then,
    $\fcan{\termrepl[\alpha']{\alpha}{s}{o}} =
    \fcan{\termname{\alpha'}{\termsubs{x}{u}{(\termcont{\gamma''}{\termrepl[\gamma'']{\gamma}{s'}{(\termname{\delta'}{\termabs{y}{\termcont{\delta}{c_0}}})}})}}}$.
    Similarly, $\fcan{\termrepl[\alpha']{\alpha}{s}{o'}} =
    \fcan{\termname{\delta'}{\termabs{y}{\termcont{\delta}{\termname{\alpha'}{\termsubs{x}{u}{(\termcont{\gamma''}{\termrepl[\gamma'']{\gamma}{s'}{c_0}})}}}}}}$.
    In the case $s = u$ we simply avoid explicit replacement. Then,
    by Lemma~\ref{l:control:equivalence:permute}, we have on the one hand
    \[
      \fcan{\termrepl[\alpha']{\alpha}{s}{o}} \eqsigma
      \fcan{\ntermren{\gamma''}{\alpha'}{\termname{\delta'}{\termsubs{x}{u}{(\termabs{y}{\termcont{\delta}{\termrepl[\gamma'']{\gamma}{s'}{c_0}}})}}}}
    \]
    and on the other hand
    \[
      \fcan{\termrepl[\alpha']{\alpha}{s}{o'}} \eqsigma
      \fcan{\termname{\delta'}{\termsubs{x}{u}{(\termabs{y}{\termcont{\delta}{\ntermren{\gamma''}{\alpha'}{\termrepl[\gamma'']{\gamma}{s'}{c_0}}}})}}}
    \]
    resorting to item (\ref{l:control:equivalence:permute:l}) of the
    lemma that positions explicit substitutions and item
    (\ref{l:control:equivalence:permute:r}) on explicit replacement.
    Finally, applying $\ruleEqsigExRen$ we derive \[
\begin{array}{rcl}
\fcan{\termrepl[\alpha']{\alpha}{s}{o}}
  & \eqsigma        & \fcan{\ntermren{\gamma''}{\alpha'}{\termname{\delta'}{\termsubs{x}{u}{(\termabs{y}{\termcont{\delta}{\termrepl[\gamma'']{\gamma}{s'}{c_0}}})}}}} \\
  & =               & \ntermren{\gamma''}{\alpha'}{\termname{\delta'}{\termsubs{x}{\fcan{u}}{(\termabs{y}{\termcont{\delta}{\fcan{\termrepl[\gamma'']{\gamma}{s'}{c_0}}}})}}} \\
  & \ruleEqsigExRen & \termname{\delta'}{\termsubs{x}{\fcan{u}}{(\termabs{y}{\termcont{\delta}{\ntermren{\gamma''}{\alpha'}{\fcan{\termrepl[\gamma'']{\gamma}{s'}{c_0}}}}})}} \\
  & =               & \fcan{\termname{\delta'}{\termsubs{x}{u}{(\termabs{y}{\termcont{\delta}{\ntermren{\gamma''}{\alpha'}{\termrepl[\gamma'']{\gamma}{s'}{c_0}}}})}}} \\
  & \eqsigma        & \fcan{\termrepl[\alpha']{\alpha}{s}{o'}}
\end{array} \]
    We conclude with $\ctxt{LXC}' = \boxdot$ and $c' = o'$.
    
    \item $\termrepl[\alpha']{\alpha}{s}{o} \reduce[\rN]
    \termname{\gamma'}{\termabs{x}{\termcont{\gamma}{\termname{\alpha'}{\termconc{(\termabs{y}{\termcont{\delta}{c_0}})}{s}}}}}$
    (\ie $\alpha \neq \gamma'$, $\alpha = \delta'$ and $\delta' \notin c_0$).
    Similar to the previous case.
    
    \item $\termrepl[\alpha']{\alpha}{s}{o} \reduce[\rN,\rC]
    \termname{\gamma'}{\termabs{x}{\termcont{\gamma}{\termname{\delta'}{\termabs{y}{\termcont{\delta}{c_0'}}}}}}$
    (\ie $\fn[\alpha]{o} = 1$, $\alpha \neq \gamma'$ and $\alpha \neq
    \delta'$). Then, \[
\begin{array}{rcl}
\fcan{\termrepl[\alpha']{\alpha}{s}{o}}
  & =                 & \termname{\gamma'}{\termabs{x}{\termcont{\gamma}{\termname{\delta'}{\termabs{y}{\termcont{\delta}{\fcan{c_0'}}}}}}} \\
  & \ruleEqsigPopPop  & \termname{\delta'}{\termabs{y}{\termcont{\delta}{\termname{\gamma'}{\termabs{x}{\termcont{\gamma}{\fcan{c_0'}}}}}}} \\
  & =                 & \fcan{\termrepl[\alpha']{\alpha}{s}{o'}}
\end{array} \]
    We conclude with $\ctxt{LXC}' = \boxdot$ and $c' = o'$.
    
    \item Otherwise, $\fcan{\termrepl[\alpha']{\alpha}{s}{o}} =
    \termrepl[\alpha']{\alpha}{\fcan{s}}{o} \eqsigma
    \termrepl[\alpha']{\alpha}{\fcan{s}}{o'} =
    \fcan{\termrepl[\alpha']{\alpha}{s}{o'}}$, since $o \eqsigma o'$
    implies $\fcan{o} = o$ and $\fcan{o'} = o'$. We conclude with $\ctxt{LXC}'
    = \boxdot$ and $c' = o'$.
    \qedhere
  \end{enumerate}
\end{itemize}
\end{proof}


%% file: proofs/bisimulation/ctxt-closure-aux.tex
\begin{proof}
We prove both items by simultaneous induction on $o \eqsigma o'$. The cases
where $o \eqsigma o'$ holds by reflexivity, transitivity or symmetry are
straightforward. For congruence, we reason by induction on the context
$\ctxt{O}$ such that $o = \ctxtapply{\ctxt{O}}{p}$ and $o' =
\ctxtapply{\ctxt{O}}{p'}$ with $p \eqsigma[\ast] p'$, where $\eqsigma[\ast]$ is
an axiom in Figure~\ref{f:control:equivalence:eqsigma}.
\begin{itemize}
  \item $\ctxt{O} = \Box$. Then there are only two possible cases which are
  $\ruleEqsigExSubs$ and $\ruleEqsigTheta$. Then we conclude
  (\ref{l:control:bisimulation:ctxt-closure-aux:conc}) by
  Lemma~\ref{l:control:equivalence:permute} in the former case, and
  straightforwardly in the latter.
  
  \item $\ctxt{O} = \boxdot$. We conclude
  (\ref{l:control:bisimulation:ctxt-closure-aux:repl}) by
  Lemma~\ref{l:control:bisimulation:ctxt-closure-rules} and
  Lemma~\ref{l:control:equivalence:permute}.
  
  \item Cases $\ctxt{O} = \termapp{\ctxt{T}}{v}$, $\ctxt{O} =
  \termapp{t}{\ctxt{T}}$, $\ctxt{O} = \termsubs{x}{\ctxt{T}}{t}$ and $\ctxt{O}
  = \termrepl[\delta]{\gamma}{\ctxt{S}}{c}$ are immediate by resorting to the
  fact that $o$ and $o'$ are already in plain form; while case $\ctxt{O} =
  \termabs{x}{\ctxt{T}}$ requires an extra induction on $s$ to conclude.

  \item $\ctxt{O} = \termcont{\gamma}{\ctxt{C}}$. Here we conclude
  (\ref{l:control:bisimulation:ctxt-closure-aux:conc}) by resorting to \ih
  (\ref{l:control:bisimulation:ctxt-closure-aux:repl}).

  \item $\ctxt{O} = \termsubs{x}{v}{\ctxt{T}}$. In this case item
  (\ref{l:control:bisimulation:ctxt-closure-aux:conc}) follows immediately
  by \ih (\ref{l:control:bisimulation:ctxt-closure-aux:conc}).
  
  \item $\ctxt{O} =
  \termrepl[\delta]{\gamma}{s'}{\ctxt{C}}$. In this case we conclude
  (\ref{l:control:bisimulation:ctxt-closure-aux:repl}) by resorting to \ih
  (\ref{l:control:bisimulation:ctxt-closure-aux:repl}) and
  Lemma~\ref{l:control:equivalence:permute}.

  \item $\ctxt{O} = \termname{\delta}{\ctxt{T}}$. Then $o =
  \termname{\delta}{\ctxtapply{\ctxt{T}}{p}}$ and $o' =
  \termname{\delta}{\ctxtapply{\ctxt{T}}{p'}}$. Let $t =
  \ctxtapply{\ctxt{T}}{p}$ and $t' = \ctxtapply{\ctxt{T}}{p'}$. We then have $t
  \eqsigma t'$ and we consider three possible cases:
  \begin{enumerate}
    \item $\alpha = \delta$ and $\alpha \notin t$. Then
    (\ref{l:control:bisimulation:ctxt-closure-aux:repl}) follows immediately by
    \ih (\ref{l:control:bisimulation:ctxt-closure-aux:conc}).

    \item $\termrepl[\alpha']{\alpha}{s}{o} \reduce[\rN,\rC]
    \termname{\delta}{u}$ (\ie $\alpha \neq \delta$ and  $\alpha \in t$). There
    are two cases:
    \begin{enumerate}
      \item $u = \ctxtapply{\ctxt{T}'}{p}$ for some $\ctxt{T}'$.
      Similarly, we conclude
      (\ref{l:control:bisimulation:ctxt-closure-aux:repl}) by resorting to \ih
      (\ref{l:control:bisimulation:ctxt-closure-aux:conc}).
      
      \item $u = \ctxtapply{\ctxt{T}}{q}$ for some $q$. Then there is a linear
      context $\ctxt{LTC}$ such that $\ctxt{LTC} =
      \ctxtapply{\ctxt{LTX}}{\ctxt{LXC}}$, $\ctxt{T} = \ctxt{LTX}$ and $p =
      \ctxtapply{\ctxt{LXC}}{c}$ with $\alpha \neq \delta$, $\alpha \notin
      \ctxt{LTC}$ and $\alpha \in c$. By $\alpha$-conversion we assume
      $\fc{\alpha'}{\ctxt{LTC}}$ and $\fc{s}{\ctxt{LTC}}$, so that
      $\termrepl[\alpha']{\alpha}{s}{p} \reduce[\rN,\rC] q$. Then, by
      Lemma~\ref{l:control:bisimulation:ctxt-closure-rules} with $p
      \eqsigma[\ast] p'$ we have $p' = \ctxtapply{\ctxt{LXC}'}{c'}$ such that
      $\alpha \notin \ctxt{LXC}'$, $\fc{\alpha'}{\ctxt{LXC}'}$,
      $\fc{s}{\ctxt{LXC}'}$ and
      $\fcan{\ctxtapply{\ctxt{LXC}}{\termrepl[\alpha']{\alpha}{s}{c}}}
      \eqsigma
      \fcan{\ctxtapply{\ctxt{LXC}'}{\termrepl[\alpha']{\alpha}{s}{c'}}}$.
      Finally, $o = \termname{\delta}{\ctxtapply{\ctxt{T}}{p}} =
      \termname{\delta}{\ctxtapply{\ctxt{LTX}}{\ctxtapply{\ctxt{LXC}}{c}}}$ and
      $o' = \termname{\delta}{\ctxtapply{\ctxt{T}}{p'}} =
      \termname{\delta}{\ctxtapply{\ctxt{LTX}}{\ctxtapply{\ctxt{LXC}'}{c'}}}$,
      and by Lemma~\ref{l:control:equivalence:permute}
      (\ref{l:control:equivalence:permute:r}), we conclude
      (\ref{l:control:bisimulation:ctxt-closure-aux:repl}) as follows \[
\begin{array}{rcl}
\fcan{\termrepl[\alpha']{\alpha}{s}{o}}
  & \eqsigma  & \fcan{\termname{\delta}{\ctxtapply{\ctxt{LTX}}{\ctxtapply{\ctxt{LXC}}{\termrepl[\alpha']{\alpha}{s}{c}}}}} \\
  & =         & \termname{\delta}{\ctxtapply{\ctxt{LTX}}{\fcan{\ctxtapply{\ctxt{LXC}}{\termrepl[\alpha']{\alpha}{s}{c}}}}} \\
  & \eqsigma  & \termname{\delta}{\ctxtapply{\ctxt{LTX}}{\fcan{\ctxtapply{\ctxt{LXC}'}{\termrepl[\alpha']{\alpha}{s}{c'}}}}} \\
  & =         & \fcan{\termname{\delta}{\ctxtapply{\ctxt{LTX}}{\ctxtapply{\ctxt{LXC}'}{\termrepl[\alpha']{\alpha}{s}{c'}}}}} \\
  & \eqsigma  & \fcan{\termrepl[\alpha']{\alpha}{s}{o'}}
\end{array} \]
    \end{enumerate}
    
    \item Otherwise, $\fcan{\termrepl[\alpha']{\alpha}{s}{o}} =
    \termrepl[\alpha']{\alpha}{\fcan{s}}{o} \eqsigma
    \termrepl[\alpha']{\alpha}{\fcan{s}}{o'} =
    \fcan{\termrepl[\alpha']{\alpha}{s}{o'}}$, since $o \eqsigma o'$
    implies $\fcan{o} = o$ and  $\fcan{o'} = o'$.
  \end{enumerate}

  \item Cases $\ctxt{O} = \termpush{\ctxt{T}}{s'}$ and $\ctxt{O} =
  \termpush{t}{\ctxt{S}}$ do not apply as $o$ and $o'$ are not stacks by
  hypothesis.
  \qedhere
\end{itemize}
\end{proof}


%% file: proofs/bisimulation/ctxt-closure.tex
\begin{proof}
By induction on $\fsize{\ctxt{O}}$, where $\fsize{\_}$ measures the size of
contexts, \ie its number of constructors (excluding $\Box$ and $\boxdot$).
\begin{itemize}
  \item $\ctxt{O} = \Box$ and $\ctxt{O} = \boxdot$. Immediate since $o
  \eqsigma o'$ implies $\fcan{o} = o$ and $\fcan{o'} = o'$ by definition.

  \item $\ctxt{O} = \termapp{\ctxt{T}}{u}$. We conclude by the \ih and
  Lemma~\ref{l:control:bisimulation:ctxt-closure-aux}
  (\ref{l:control:bisimulation:ctxt-closure-aux:conc}).
  
  \item $\ctxt{O} = \termrepl[\alpha']{\alpha}{s}{\ctxt{C}}$. We conclude by
  the \ih and Lemma~\ref{l:control:bisimulation:ctxt-closure-aux}
  (\ref{l:control:bisimulation:ctxt-closure-aux:repl}).
  
  \item In all the other cases we conclude by the \ih
  \qedhere
\end{itemize}
\end{proof}


%% file: proofs/bisimulation/meta-subs-eqsigma.tex
\begin{proof}\hfill
\begin{enumerate}
  \item By induction on $o \eqsigma o'$ which, by definition, implies
  verifying the cases for reflexivity, transitivity, symmetry and
  congruence (\ie closure by contexts). All are straightforward but
  the latter. 

  For congruence we proceed by induction on the closure context
  $\ctxt{O}$ such that $o = \ctxtapply{\ctxt{O}}{p}$ and
  $o' = \ctxtapply{\ctxt{O}}{p'}$ with $p \eqsigma[\ast] p'$, where
  $\eqsigma[\ast]$ is any rule from 
  Figure~\ref{f:control:equivalence:eqsigma}.
  \begin{itemize}
    \item $\ctxt{O} = \Box$. Then, $o = p \eqsigma[\ast] p' =
    o'$. Moreover, $o, o' \in \Term{\calcLambdaM}$. Only two rules
    are applicable to terms:
    \begin{itemize}
      \item $\ruleEqsigExSubs$. Follows from Lemma~\ref{l:control:equivalence:permute}
      (\ref{l:control:equivalence:permute:l}).

      \item $\ruleEqsigTheta$. Straightfoward.
    \end{itemize}
      
    \item $\ctxt{O} = \boxdot$. Then, $o = p \eqsigma[\ast] p' =
    o'$. Moreover, we have $o, o' \in
    \Command{\calcLambdaM}$. Three rules are applicable to commands:
    \begin{itemize}
      \item $\ruleEqsigExRepl$ and $\ruleEqsigExRen$. Follows from 
      Lemma~\ref{l:control:equivalence:permute}
      (\ref{l:control:equivalence:permute:r}).

      \item $\ruleEqsigPopPop$. Straightforward.
    \end{itemize}
      
    \item $\ctxt{O} = \termapp{\ctxt{T}}{v}$ and $\ctxt{O} =
    \termapp{t}{\ctxt{T}}$. Both follow from the \ih and Lemma~\ref{l:control:bisimulation:ctxt-closure}.

    \item All the remaining cases follow from the \ih

  \end{itemize}

  \item By induction on $o$. All cases are by the \ih except for  $o =
  \termapp{t}{v}$ which resorts to
  Lemma~\ref{l:control:bisimulation:ctxt-closure}.
  \qedhere
\end{enumerate}
\end{proof}


%% file: proofs/bisimulation/meta-repl-eqsigma.tex
\begin{proof}\hfill
\begin{enumerate}
  \item By induction on $o \eqsigma o'$. By definition this implies
  verifying four conditions: reflexivity, transitivity, symmetry and
  congruence (\ie closure under contexts). All but the latter are
  immediate; we thus focus on the latter.

  Congruence is proved by induction on the closure context
  $\ctxt{O}$ such that $o = \ctxtapply{\ctxt{O}}{p}$ and
  $o' = \ctxtapply{\ctxt{O}}{p'}$ with $p \eqsigma[\ast] p'$, where 
  $\eqsigma[\ast]$ is any rule from 
  Figure~\ref{f:control:equivalence:eqsigma}.
  \begin{itemize}
    \item $\ctxt{O} = \Box$. Follows from Lemma~\ref{l:control:equivalence:permute}
    (\ref{l:control:equivalence:permute:l}).

    \item $\ctxt{O} = \boxdot$. Then, $o = p \eqsigma[\ast] p' = o'$. Moreover,
    $o, o' \in \Command{\calcLambdaM}$. Five rules are applicable to commands:
    \begin{itemize}
      \item $\ruleEqsigExRepl$. Follow from Lemma~\ref{l:control:equivalence:permute}
      (\ref{l:control:equivalence:permute:r}).

      \item $\ruleEqsigExRen$. Then, $o =
      \termname{\delta}{\termcont{\gamma}{\ctxtapply{\ctxt{LCC}}{c}}}$ and $o' =
      \ctxtapply{\ctxt{LCC}}{\termname{\delta}{\termcont{\gamma}{c}}}$, and $o$ and $o'$ are in plain form. Without loss of generality, we assume. There are two
      possible subcases:
      \begin{enumerate}
        \item $\delta = \alpha$. Then,
        $\replapply{\repl[\alpha']{\alpha}{s}}{o} =
        \termname{\alpha'}{\termconc{(\termcont{\gamma}{\ctxtapply{\ctxt{LCC}'}{ c'}})}{s}}$, where $\ctxt{LCC'} = \replapply{\repl[\alpha']{\alpha}{s}}{\ctxt{LCC}} $,  $c'= \replapply{\repl[\alpha']{\alpha}{s}}{c}$ and
        $\replapply{\repl[\alpha']{\alpha}{s}}{o'} =
        \replapply{\repl[\alpha']{\alpha}{s}}{\ctxtapply{\ctxt{LCC}}{\termname{\delta}{\termcont{\gamma}{c}}}}$. \[
        \begin{array}{rcll}
\fcan{\replapply{\repl[\alpha']{\alpha}{s}}{o}}
  & =         & \fcan{\termname{\alpha'}{\termconc{(\termcont{\gamma}{\replapply{\repl[\alpha']{\alpha}{s}}{\ctxtapply{\ctxt{LCC}}{c}}})}{s}}} \\
  & =         & \fcan{\termname{\alpha'}{\termcont{\gamma_n}{\termrepl[\gamma_n]{\gamma_{n-1}}{u_n}{\termrepl[\gamma_1]{\gamma}{u_1}{\replapply{\repl[\alpha']{\alpha}{s}}{\ctxtapply{\ctxt{LCC}}{c}}}\ldots}} }} \\
  & =         & \fcan{\termname{\alpha'}{\termcont{\gamma_n}{\termrepl[\gamma_n]{\gamma_{n-1}}{u_n}{\termrepl[\gamma_1]{\gamma}{u_1}{\ctxtapply{\ctxt{LCC}'}{ c'}}\ldots}} }} \\
  & \eqsigma  & \fcan{\ctxtapply{\ctxt{LCC}'}{\termname{\alpha'}{\termcont{\gamma_n}{\termrepl[\gamma_n]{\gamma_{n-1}}{u_n}{\termrepl[\gamma_1]{\gamma}{u_1}{c'}\ldots}}} }} & \text{(Lemma~\ref{l:control:equivalence:permute})}\\
  & =         & \fcan{\ctxtapply{\ctxt{LCC}'}{\termname{\alpha'}{\termconc{(\termcont{\gamma}{c'})}{s}}}} \\
  & =         & \fcan{\replapply{\repl[\alpha']{\alpha}{s}}{(\ctxtapply{\ctxt{LCC}}{\termname{\delta}{\termcont{\gamma}{c}}})}} \\ 
  & =         & \fcan{\replapply{\repl[\alpha']{\alpha}{s}}{o'}}
        \end{array} \]

        \item $\delta \neq \alpha$. Then,
        $\replapply{\repl[\alpha']{\alpha}{s}}{o} =
        \termname{\delta}{\termcont{\gamma}{\ctxtapply{\ctxt{LCC}'}{ c'}}}$, where $\ctxt{LCC'} = \replapply{\repl[\alpha']{\alpha}{s}}{\ctxt{LCC}} $,  $c'= \replapply{\repl[\alpha']{\alpha}{s}}{c}$ and
        $\replapply{\repl[\alpha']{\alpha}{s}}{o'} =
        \ctxtapply{\ctxt{LCC}'}{\termname{\delta}{\termcont{\gamma}{c'}}}$. \[
        \begin{array}{rcll}
\fcan{\replapply{\repl[\alpha']{\alpha}{s}}{o}}
  & =     & \fcan{\termname{\delta}{\termcont{\gamma}{\ctxtapply{\ctxt{LCC}'}{ c'}} }} \\
  & \eqsigma & \fcan{\ctxtapply{\ctxt{LCC}'}{\termname{\delta}{\termcont{\gamma}{c'}}}} & \text{(Lemma~\ref{l:control:equivalence:permute})}\\
  & =      & \fcan{\replapply{\repl[\alpha']{\alpha}{s}}{o'}}
        \end{array} \]
      \end{enumerate}

      \item $\ruleEqsigPopPop$. Then, $o =
      \termname{\gamma'}{\termabs{x}{\termcont{\gamma}{\termname{\delta'}{\termabs{y}{\termcont{\delta}{c}}}}}}$
      and $o' =
      \termname{\delta'}{\termabs{y}{\termcont{\delta}{\termname{\gamma'}{\termabs{x}{\termcont{\gamma}{c}}}}}}$
      with $\delta \neq \gamma'$ and $\gamma \neq \delta'$. Without
      loss of generality, we also assume $x \notin s$, $y \notin s$, $\gamma \neq
      \alpha'$, $\delta \neq \alpha'$, $\gamma \notin s$ and $\delta \notin s$.
      There are four possible subcases:
      \begin{enumerate}
        \item $\gamma' = \alpha$ and $\delta' = \alpha$. Then,
        $\replapply{\repl[\alpha']{\alpha}{s}}{o} =
        \termname{\alpha'}{\termconc{(\termabs{x}{\termcont{\gamma}{\termname{\alpha'}{\termconc{(\termabs{y}{\termcont{\delta}{c'}})}{s}}}})}{s}}$
        and $\replapply{\repl[\alpha']{\alpha}{s}}{o'} =
        \termname{\alpha'}{\termconc{(\termabs{y}{\termcont{\delta}{\termname{\alpha'}{\termconc{(\termabs{x}{\termcont{\gamma}{c'}})}{s}}}})}{s}}$
        with $c' = \replapply{\repl[\alpha']{\alpha}{s}}{c}$. Suppose $s =
        \termpush{u}{s'}$ (the case $s = u$ is slightly simpler) and
        consider fresh names $\gamma'', \delta''$. Then,
        \begin{center}
        $\fcan{\replapply{\repl[\alpha']{\alpha}{s}}{o}} =
        \fcan{\termname{\alpha'}{\termsubs{x}{u}{(\termcont{\gamma''}{\termrepl[\gamma'']{\gamma}{s'}{(\termname{\alpha'}{\termsubs{y}{u}{(\termcont{\delta''}{\termrepl[\delta'']{\delta}{s'}{c'}})}})}})}}}$
        \end{center}
        where the explicit substitutions result from contracting the
        $\rB$-redexes in $u$ and the explicit replacements result
        from contracting the $\rM$-redexes in $s'$ followed by the resulting
        $\rC$-redexes. If $s = u$, then there are no $\rM$-redexes to
        contract, and there are only explicit substitutions.
        Similarly, 
        \begin{center}
        $\fcan{\replapply{\repl[\alpha']{\alpha}{s}}{o'}} =
        \fcan{\termname{\alpha'}{\termsubs{y}{u}{(\termcont{\delta''}{\termrepl[\delta'']{\delta}{s'}{(\termname{\alpha'}{\termsubs{x}{u}{(\termcont{\gamma''}{\termrepl[\gamma'']{\gamma}{s'}{c'}})}})}})}}}$
        \end{center}
        Let $c'' =
        \termrepl[\delta'']{\delta}{s'}{\termrepl[\gamma'']{\gamma}{s'}{c'}}$.
        Then by Lemma~\ref{l:control:equivalence:permute}, on the one
        hand we have: 
        $\fcan{\replapply{\repl[\alpha']{\alpha}{s}}{o}} \eqsigma
        \fcan{\termname{\alpha'}{\termcont{\gamma''}{\termname{\alpha'}{\termsubs{y}{u}{\termsubs{x}{u}{(\termcont{\delta''}{c''})}}}}}}$,
        and on the other hand we have:
        $\fcan{\replapply{\repl[\alpha']{\alpha}{s}}{o'}} \eqsigma
        \fcan{\termname{\alpha'}{\termsubs{y}{u}{\termsubs{x}{u}{ (\termcont{\delta''}{\termname{\alpha'}{\termcont{\gamma''}{c''}}})}}}}$;
        resorting to item (\ref{l:control:equivalence:permute:l}) of
        that lemma to correctly place the explicit substitutions and item
        (\ref{l:control:equivalence:permute:r}). We then conclude by applying $\ruleEqsigExRen$: \[
\begin{array}{rcl}
\fcan{\replapply{\repl[\alpha']{\alpha}{s}}{o}}
  & \eqsigma      & \fcan{\termname{\alpha'}{\termcont{\gamma''}{\termname{\alpha'}{\termsubs{y}{u}{\termsubs{x}{u}{(\termcont{\delta''}{c''})}}}}}} \\
  & =             & \termname{\alpha'}{\termcont{\gamma''}{\termname{\alpha'}{\termsubs{y}{\fcan{u}}{\termsubs{x}{\fcan{u}}{(\termcont{\delta''}{\fcan{c''}})}}}}} \\
  & \ruleEqsigExRen & \termname{\alpha'}{\termsubs{y}{\fcan{u}}{\termsubs{x}{\fcan{u}}{(\termcont{\delta''}{\termname{\alpha'}{\termcont{\gamma''}{\fcan{c''}})}}}}} \\
  & =             & \fcan{\termname{\alpha'}{\termsubs{y}{u}{\termsubs{x}{u}{ (\termcont{\delta''}{\termname{\alpha'}{\termcont{\gamma''}{c''}}})}}}} \\
  & \eqsigma      & \fcan{\replapply{\repl[\alpha']{\alpha}{s}}{o'}}
\end{array} \]

        \item $\gamma' = \alpha$ and $\delta' \neq \alpha$, and the case $\gamma' \neq \alpha$ and $\delta' = \alpha$. Similar to the
        previous case.

        \item $\gamma' \neq \alpha$ and $\delta' \neq \alpha$. Straightforward.
      \end{enumerate}
    \end{itemize}

    \item $\ctxt{O} = \termapp{\ctxt{T}}{u}$ and $\ctxt{O} =
    \termapp{t}{\ctxt{T}}$. Follow from  Lemma~\ref{l:control:bisimulation:ctxt-closure}.

    \item $\ctxt{O} = \termabs{x}{\ctxt{T}}$, $\ctxt{O} =
    \termcont{\gamma}{\ctxt{C}}$,  $\ctxt{O} =
    \termsubs{x}{u}{\ctxt{T}}$, and  $\ctxt{O} = \termsubs{x}{\ctxt{T}}{t}$. Similar to the previous
    case.
    
    \item $\ctxt{O} = \termname{\gamma}{\ctxt{T}}$. If $\gamma\neq\alpha$, then
    we conclude from the \ih; otherwise we reason as follows. First note that
    $\fcan{\replapply{\repl[\alpha']{\alpha}{s}}{\ctxtapply{\ctxt{T}}{p}}}
    \eqsigma[\ih]
    \fcan{\replapply{\repl[\alpha']{\alpha}{s}}{\ctxtapply{\ctxt{T}}{p'}}}$.
    Then $\fcan{\replapply{\repl[\alpha']{\alpha}{s}}{o}} =
    \fcan{\termname{\alpha'}{\termconc{\fcan{(\replapply{\repl[\alpha']{\alpha}{s}}{\ctxtapply{\ctxt{T}}{p}})}}{s}}}
    \eqsigma
    \fcan{\termname{\alpha'}{\termconc{\fcan{(\replapply{\repl[\alpha']{\alpha}{s}}{\ctxtapply{\ctxt{T}}{p'}})}}{s}}}
    = \fcan{\replapply{\repl[\alpha']{\alpha}{s}}{o'}}$ by
    Lemma~\ref{l:control:bisimulation:ctxt-closure}. 

    \item $\ctxt{O} = \termrepl[\delta]{\gamma}{s'}{\ctxt{C}}$ and $\ctxt{O} = \termrepl[\delta]{\gamma}{\ctxt{S}}{c}$. By the \ih

    \item $\ctxt{O} = \termpush{\ctxt{T}}{s'}$ and $\ctxt{O} = \termpush{t}{\ctxt{S}}$. We use the \ih
  \end{itemize}

  \item By induction on $o$. All cases follow from the \ih and/or
  Lemma~\ref{l:control:bisimulation:ctxt-closure}.
  \qedhere
\end{enumerate}
\end{proof}


%% file: proofs/bisimulation/meta-eqsigma.tex
\begin{proof}
By Lemma~\ref{l:control:bisimulation:meta-subs-eqsigma}
and~\ref{l:control:bisimulation:meta-repl-eqsigma}, resp.
\end{proof}


%% file: proofs/bisimulation/bisimulation.tex
\begin{proof}
By induction on $o \eqsigma p$. By definition this requires verifying four
cases: reflexivity, transitivity, symmetry and congruence (\ie closure under
contexts). The cases where $o \eqsigma p$ holds by reflexivity or transitivity
are straightforward. We focus on the other two, which are dealt with
simultaneously by induction on the closure context $\ctxt{Q}$ such that $o =
\ctxtapply{\ctxt{Q}}{q}$ and $p = \ctxtapply{\ctxt{Q}}{q'}$ with $q
\eqsigma[\ast] q'$, where $\eqsigma[\ast]$ is any rule in
Figure~\ref{f:control:equivalence:eqsigma}. Note that $o \reducemean o'$ implies
$o = \ctxtapply{\ctxt{O}}{l}$ and $o' = \fcan{\ctxtapply{\ctxt{O}}{r}}$ with $l
\rrule{\ast} r$, $\ast \in \set{\rS, \rRm}$. We thus consider all possible
forms for $\ctxt{Q}$ and $\ctxt{O}$:
\begin{itemize}
  \item $\ctxt{Q} = \Box$. Then $o = q \eqsigma[\ast] q' = p$. Moreover, in
  this case $o, p \in \Term{\calcLambdaM}$. We only detail the cases
  where there is an overlap between the equivalence and the reduction rules,
  the others being immediate. Only two rules are applicable to terms:
  \begin{itemize}
    \item $\ruleEqsigExSubs$. Then, $o =
    \termsubs{x}{u}{\ctxtapply{\ctxt{LTT}}{t}}$ and $p =
    \ctxtapply{\ctxt{LTT}}{\termsubs{x}{u}{t}}$ with $x \notin \ctxt{LTT}$
    and $\fc{u}{\ctxt{LTT}}$. There are three further possible cases:
    \begin{enumerate}
      \item $\rS$-redex at the root. \[
\begin{tikzcd}
o = \termsubs{x}{u}{\ctxtapply{\ctxt{LTT}}{t}} \arrow[rightsquigarrow]{d}[anchor=north,left]{\rS}
  &[-30pt] \ruleEqsigExSubs
  &[-30pt] \ctxtapply{\ctxt{LTT}}{\termsubs{x}{u}{t}} = p \arrow[rightsquigarrow]{d}[anchor=north,left]{\rS} \\[-5pt]
o' = \fcan{\subsapply{\subs{x}{u}}{\ctxtapply{\ctxt{LTT}}{t}}}
  &[-30pt] =
  &[-30pt] \fcan{\ctxtapply{\ctxt{LTT}}{\subsapply{\subs{x}{u}}{t}}} = p'
\end{tikzcd} \]
      
      \item $\rS$-redex overlaps $\ctxt{LTT}$. Then we have the context
      $\ctxt{LTT} = \ctxtapply{\ctxt{LTT}_1}{\termsubs{y}{v}{\ctxt{LTT}_2}}$
      with commands $o' =
      \fcan{\termsubs{x}{u}{\ctxtapply{\ctxt{LTT}_1}{\subsapply{\subs{y}{v}}{\ctxtapply{\ctxt{LTT}_2}{t}}}}}
      =
      \fcan{\termsubs{x}{u}{\ctxtapply{\ctxt{LTT}_1}{\ctxtapply{\ctxt{LTT}'_2}{t'}}}}$
      and $p' = 
      \fcan{\ctxtapply{\ctxt{LTT}_1}{\ctxtapply{\ctxt{LTT}'_2}{\termsubs{x}{u}{t'}}}}$
      since $\fc{u}{\ctxt{LTT}}$ implies $y \notin u$. We conclude
      by Lemma~\ref{l:control:equivalence:permute}
      (\ref{l:control:equivalence:permute:l}): \[
\begin{tikzcd}
o = \termsubs{x}{u}{\ctxtapply{\ctxt{LTT}_1}{\termsubs{y}{v}{\ctxtapply{\ctxt{LTT}_2}{t}}}} \arrow[rightsquigarrow]{d}[anchor=north,left]{\rS}
  &[-30pt] \ruleEqsigExSubs
  &[-30pt] \ctxtapply{\ctxt{LTT}_1}{\termsubs{y}{v}{\ctxtapply{\ctxt{LTT}_2}{\termsubs{x}{u}{t}}}} = p \arrow[rightsquigarrow]{d}[anchor=north,left]{\rS} \\[-5pt]
o' = \fcan{\termsubs{x}{u}{\ctxtapply{\ctxt{LTT}_1}{\subsapply{\subs{y}{v}}{\ctxtapply{\ctxt{LTT}_2}{t}}}}}
  &[-30pt] \eqsigma
  &[-30pt] \fcan{\ctxtapply{\ctxt{LTT}_1}{\subsapply{\subs{y}{v}}{\ctxtapply{\ctxt{LTT}_2}{\termsubs{x}{u}{t}}}}} = p'
\end{tikzcd} \]
      
      \item $\rRm$-redex overlaps $\ctxt{LTT}$. Then we have the context
      $\ctxt{LTT} =
      \ctxtapply{\ctxt{LTC}}{\termrepl[\alpha']{\alpha}{s}{\ctxt{LCT}}}$ with
      commands $o' =
      \fcan{\termsubs{x}{u}{\ctxtapply{\ctxt{LTC}}{\replapply{\repl[\alpha']{\alpha}{s}}{\ctxtapply{\ctxt{LCT}}{t}}}}}
      =
      \fcan{\termsubs{x}{u}{\ctxtapply{\ctxt{LTC}}{\ctxtapply{\ctxt{LTC}'}{t'}}}}$
      and $p' = 
      \fcan{\ctxtapply{\ctxt{LTC}}{\ctxtapply{\ctxt{LTC}'}{\termsubs{x}{u}{t'}}}}$
      since $\fc{u}{\ctxt{LTT}}$ implies $\alpha \notin u$. We
      conclude by Lemma~\ref{l:control:equivalence:permute}
      (\ref{l:control:equivalence:permute:l}): \[
\begin{tikzcd}
o = \termsubs{x}{u}{\ctxtapply{\ctxt{LTC}}{\termrepl[\alpha']{\alpha}{s}{\ctxtapply{\ctxt{LCT}}{t}}}} \arrow[rightsquigarrow]{d}[anchor=north,left]{\rRm}
  &[-30pt] \ruleEqsigExSubs
  &[-30pt] \ctxtapply{\ctxt{LTC}}{\termrepl[\alpha']{\alpha}{s}{\ctxtapply{\ctxt{LCT}}{\termsubs{x}{u}{t}}}} = p \arrow[rightsquigarrow]{d}[anchor=north,left]{\rRm} \\[-5pt]
o' = \fcan{\termsubs{x}{u}{\ctxtapply{\ctxt{LTC}}{\replapply{\repl[\alpha']{\alpha}{s}}{\ctxtapply{\ctxt{LCT}}{t}}}}}
  &[-30pt] \eqsigma
  &[-30pt] \fcan{\ctxtapply{\ctxt{LTC}}{\replapply{\repl[\alpha']{\alpha}{s}}{\ctxtapply{\ctxt{LCT}}{\termsubs{x}{u}{t}}}}} = p'
\end{tikzcd} \]
    \end{enumerate}
    
    \item $\ruleEqsigTheta$. Then, $o =
    \termcont{\alpha}{\termname{\alpha}{t}}$ and $p = t$ with $\alpha \notin
    t$. This case is immediate since all reduction steps must be
    in $t$.
  \end{itemize}
  
  \item $\ctxt{Q} = \boxdot$. Then, $o = q \eqsigma[\ast] q' = p$. Moreover,
  in this case $o, p \in \Command{\calcLambdaM}$. We only detail the
  cases where there is an overlap between the equivalence and the reduction
  rules, the others being immediate. There are three rules
  applicable to commands:
  \begin{itemize}
    \item $\ruleEqsigExRepl$. Then, $o =
    \termrepl[\alpha']{\alpha}{s}{\ctxtapply{\ctxt{LCC}}{c}}$ and $p =
    \ctxtapply{\ctxt{LCC}}{\termrepl[\alpha']{\alpha}{s}{c}}$ with $\alpha
    \notin \ctxt{LCC}$, $\fc{\alpha'}{\ctxt{LCC}}$ and $\fc{s}{\ctxt{LCC}}$.
    There are three further possible cases:
    \begin{enumerate}
      \item $\rRm$-redex at the root. \[
\begin{tikzcd}
o = \termrepl[\alpha']{\alpha}{s}{\ctxtapply{\ctxt{LCC}}{c}} \arrow[rightsquigarrow]{d}[anchor=north,left]{\rRm}
  &[-30pt] \ruleEqsigExRepl
  &[-30pt]\ctxtapply{\ctxt{LCC}}{\termrepl[\alpha']{\alpha}{s}{c}} = p \arrow[rightsquigarrow]{d}[anchor=north,left]{\rRm} \\[-5pt]
o' = \fcan{\replapply{\repl[\alpha']{\alpha}{s}}{\ctxtapply{\ctxt{LCC}}{c}}}
  &[-30pt] =
  &[-30pt] \fcan{\ctxtapply{\ctxt{LCC}}{\replapply{\repl[\alpha']{\alpha}{s}}{c}}} = p'
\end{tikzcd} \]
      
      \item $\rS$-redex overlaps $\ctxt{LCC}$. Then we have the context
      $\ctxt{LCC} = \ctxtapply{\ctxt{LCT}}{\termsubs{x}{u}{\ctxt{LTC}}}$ with
      commands $o' =
      \fcan{\termrepl[\alpha']{\alpha}{s}{\ctxtapply{\ctxt{LCT}}{\subsapply{\subs{x}{u}}{\ctxtapply{\ctxt{LTC}}{c}}}}}
      =
      \fcan{\termrepl[\alpha']{\alpha}{s}{\ctxtapply{\ctxt{LCT}}{\ctxtapply{\ctxt{LTC}'}{c'}}}}$
      and $p' =
      \fcan{\ctxtapply{\ctxt{LCT}}{\ctxtapply{\ctxt{LTC}'}{\termrepl[\alpha']{\alpha}{s}{c'}}}}$
      since $\fc{s}{\ctxt{LCC}}$ implies $x \notin s$. We conclude by
      Lemma~\ref{l:control:equivalence:permute}
      (\ref{l:control:equivalence:permute:r}): \[
\begin{tikzcd}
o = \termrepl[\alpha']{\alpha}{s}{\ctxtapply{\ctxt{LCT}}{\termsubs{x}{u}{\ctxtapply{\ctxt{LTC}}{c}}}} \arrow[rightsquigarrow]{d}[anchor=north,left]{\rS}
  &[-30pt] \ruleEqsigExRepl
  &[-30pt] \ctxtapply{\ctxt{LCT}}{\termsubs{x}{u}{\ctxtapply{\ctxt{LTC}}{\termrepl[\alpha']{\alpha}{s}{c}}}} = p \arrow[rightsquigarrow]{d}[anchor=north,left]{\rS} \\[-5pt]
o' = \fcan{\termrepl[\alpha']{\alpha}{s}{\ctxtapply{\ctxt{LCT}}{\subsapply{\subs{x}{u}}{\ctxtapply{\ctxt{LTC}}{c}}}}}
  &[-30pt] \eqsigma
  &[-30pt] \fcan{\ctxtapply{\ctxt{LCT}}{\subsapply{\subs{x}{u}}{\ctxtapply{\ctxt{LTC}}{\termrepl[\alpha']{\alpha}{s}{c}}}}} = p'
\end{tikzcd} \]
      
      \item $\rRm$-redex overlaps $\ctxt{LCC}$. Then we have the context
      $\ctxt{LCC} =
      \ctxtapply{\ctxt{LCC}_1}{\termrepl[\delta]{\gamma}{s'}{\ctxt{LCC}_2}}$
      with commands $o' =
      \fcan{\termrepl[\alpha']{\alpha}{s}{\ctxtapply{\ctxt{LCC}_1}{\replapply{\repl[\delta]{\gamma}{s'}}{\ctxtapply{\ctxt{LCC}_2}{c}}}}}
      =
      \fcan{\termrepl[\alpha']{\alpha}{s}{\ctxtapply{\ctxt{LCC}_1}{\ctxtapply{\ctxt{LCC}'_2}{c'}}}}$
      and $p' = 
      \fcan{\ctxtapply{\ctxt{LCC}_1}{\ctxtapply{\ctxt{LCC}'_2}{\termrepl[\alpha']{\alpha}{s}{c'}}}}$
      since $\fc{\alpha}{\ctxt{LCC}}$ implies $\gamma \neq \alpha$, and
      $\fc{s}{\ctxt{LCC}}$ implies $\alpha \notin s$. We conclude by
      Lemma~\ref{l:control:equivalence:permute}
      (\ref{l:control:equivalence:permute:r}): \[
\begin{tikzcd}
o = \termrepl[\alpha']{\alpha}{s}{\ctxtapply{\ctxt{LCC}_1}{\termrepl[\delta]{\gamma}{s'}{\ctxtapply{\ctxt{LCC}_2}{c}}}} \arrow[rightsquigarrow]{d}[anchor=north,left]{\rRm}
  &[-30pt] \ruleEqsigExRepl
  &[-30pt] \ctxtapply{\ctxt{LCC}_1}{\termrepl[\delta]{\gamma}{s'}{\ctxtapply{\ctxt{LCC}_2}{\termrepl[\alpha']{\alpha}{s}{c}}}} = p \arrow[rightsquigarrow]{d}[anchor=north,left]{\rRm} \\[-5pt]
o' = \fcan{\termrepl[\alpha']{\alpha}{s}{\ctxtapply{\ctxt{LCC}_1}{\replapply{\repl[\delta]{\gamma}{s'}}{\ctxtapply{\ctxt{LCC}_2}{c}}}}}
  &[-30pt] \eqsigma
  &[-30pt] \fcan{\ctxtapply{\ctxt{LCC}_1}{\replapply{\repl[\delta]{\gamma}{s'}}{\ctxtapply{\ctxt{LCC}_2}{\termrepl[\alpha']{\alpha}{s}{c}}}}} = p'
\end{tikzcd} \]
\end{enumerate}

    \item $\ruleEqsigExRen$. Then, $o =
    \ntermren{\alpha}{\beta}{\ctxtapply{\ctxt{LCC}}{c}}$ and $p =
    \ctxtapply{\ctxt{LCC}}{\ntermren{\alpha}{\beta}{c}}$ with $\alpha \notin 
    \ctxt{LCC}, \fc{\beta}{\ctxt{LCC}}$. There are two further possible cases:
    \begin{enumerate}
      \item $\rS$-redex overlaps $\ctxt{LCC}$. Then we have the context
      $\ctxt{LCC} = \ctxtapply{\ctxt{LCT}}{\termsubs{x}{u}{\ctxt{LTC}}}$ with
      commands $o' =
      \fcan{\ntermren{\alpha}{\beta}{\ctxtapply{\ctxt{LCT}}{\subsapply{\subs{x}{u}}{\ctxtapply{\ctxt{LTC}}{c}}}}}
      =
      \fcan{\ntermren{\alpha}{\beta}{\ctxtapply{\ctxt{LCT}}{\ctxtapply{\ctxt{LTC}'}{c'}}}}$
      and $p' =
      \fcan{\ctxtapply{\ctxt{LCT}}{\ctxtapply{\ctxt{LTC}'}{\ntermren{\alpha}{\beta}{c'}}}}$.
      We conclude by Lemma~\ref{l:control:equivalence:permute}
      (\ref{l:control:equivalence:permute:r}): \[
\begin{tikzcd}
o =  \ntermren{\alpha}{\beta}{\ctxtapply{\ctxt{LCT}}{\termsubs{x}{u}{\ctxtapply{\ctxt{LTC}}{c}}}} \arrow[rightsquigarrow]{d}[anchor=north,left]{\rS}
  &[-30pt] \ruleEqsigExRen
  &[-30pt] \ctxtapply{\ctxt{LCT}}{\termsubs{x}{u}{\ctxtapply{\ctxt{LTC}}{\ntermren{\alpha}{\beta}{ c}}}} = p \arrow[rightsquigarrow]{d}[anchor=north,left]{\rS} \\[-5pt]
o' = \fcan{\ntermren{\alpha}{\beta}{\ctxtapply{\ctxt{LCT}}{\subsapply{\subs{x}{u}}{\ctxtapply{\ctxt{LTC}}{c}}}}}
  &[-30pt] \eqsigma
  &[-30pt] \fcan{\ctxtapply{\ctxt{LCT}}{\subsapply{\subs{x}{u}}{\ctxtapply{\ctxt{LTC}}{\ntermren{\alpha}{\beta}{ c}}}}} = p'
\end{tikzcd} \]
        
      \item $\rRm$-redex overlaps $\ctxt{LCC}$. Then we have the context
      $\ctxt{LCC} =
      \ctxtapply{\ctxt{LCC}_1}{\termrepl[\delta]{\gamma}{s}{\ctxt{LCC}_2}}$
      with commands $o' =
      \fcan{\ntermren{\alpha}{\beta}{\ctxtapply{\ctxt{LCC}_1}{\replapply{\repl[\delta]{\gamma}{s}}{\ctxtapply{\ctxt{LCC}_2}{c}}}}}
      =
      \fcan{\ntermren{\alpha}{\beta}{\ctxtapply{\ctxt{LCC}_1}{\ctxtapply{\ctxt{LCC}'_2}{c'}}}}$
      and $p' = 
      \fcan{\ctxtapply{\ctxt{LCC}_1}{\ctxtapply{\ctxt{LCC}'_2}{\ntermren{\alpha}{\beta}{c'}}}}$
      since $\fc{\beta}{\ctxt{LCC}}$ implies $\gamma \neq
      \beta$. We conclude by Lemma~\ref{l:control:equivalence:permute}
      (\ref{l:control:equivalence:permute:r}): \[
\begin{tikzcd}
o =  \ntermren{\alpha}{\beta}{\ctxtapply{\ctxt{LCC}_1}{\termrepl[\delta]{\gamma}{s}{\ctxtapply{\ctxt{LCC}_2}{c}}}} \arrow[rightsquigarrow]{d}[anchor=north,left]{\rRm}
  &[-30pt] \ruleEqsigExRen
  &[-30pt] \ctxtapply{\ctxt{LCC}_1}{\termrepl[\delta]{\gamma}{s}{\ctxtapply{\ctxt{LCC}_2}{\ntermren{\alpha}{\beta}{c}}}} = p \arrow[rightsquigarrow]{d}[anchor=north,left]{\rRm} \\[-5pt]
o' = \fcan{\ntermren{\alpha}{\beta}{\ctxtapply{\ctxt{LCC}_1}{\replapply{\repl[\delta]{\gamma}{s}}{\ctxtapply{\ctxt{LCC}_2}{c}}}}}
  &[-30pt] \eqsigma
  &[-30pt] \fcan{\ctxtapply{\ctxt{LCC}_1}{\replapply{\repl[\delta]{\gamma}{s}}{\ctxtapply{\ctxt{LCC}_2}{\ntermren{\alpha}{\beta}{c}}}}} = p'
\end{tikzcd} \]
    \end{enumerate}
    
    \item $\ruleEqsigPopPop$. Then, $o =
    \termname{\alpha'}{\termabs{z}{\termcont{\alpha}{\termname{\beta'}{\termabs{y}{\termcont{\beta}{c}}}}}}$
    and $p =
    \termname{\beta'}{\termabs{y}{\termcont{\beta}{\termname{\alpha'}{\termabs{z}{\termcont{\alpha}{c}}}}}}$
    with $\beta \neq \alpha'$ and $\alpha \neq \beta'$. This case is
    immediate since all possible reductions are in $c$.
  \end{itemize}
    
  \item $\ctxt{Q} = \termapp{\ctxt{T}}{u}$ or $\ctxt{Q} = \termapp{t}{\ctxt{T}}$.
  We use the \ih and Lemma~\ref{l:control:bisimulation:ctxt-closure}.
    
    \item $\ctxt{Q} = \termabs{x}{\ctxt{T}}$ and $\ctxt{Q} =
      \termcont{\alpha}{\ctxt{C}}$. By the \ih
    
    \item $\ctxt{Q} = \termsubs{x}{u}{\ctxt{T}}$. Then, $o =
    \termsubs{x}{u}{\ctxtapply{\ctxt{T}}{q}}$ and $p =
    \termsubs{x}{u}{\ctxtapply{\ctxt{T}}{q'}}$.  If the
    \reducemean[\rS] step executes the outermost explicit
    substitution, we use Lemma~\ref{l:control:bisimulation:meta-subs-eqsigma}
      (\ref{l:control:bisimulation:meta-subs-eqsigma:o}), otherwise we
      use Lemma~\ref{l:control:bisimulation:ctxt-closure} and the \ih
    
    \item $\ctxt{Q} = \termsubs{x}{\ctxt{T}}{t}$.  Then, $o =
    \termsubs{x}{\ctxtapply{\ctxt{T}}{q}}{t}$ and $p =
    \termsubs{x}{\ctxtapply{\ctxt{T}}{q'}}{t}$. If the
    \reducemean[\rS] step executes the outermost explicit
    substitution, we use  Lemma~\ref{l:control:bisimulation:meta-subs-eqsigma}
      (\ref{l:control:bisimulation:meta-subs-eqsigma:u}), otherwise we
      use Lemma~\ref{l:control:bisimulation:ctxt-closure} and the \ih
    
  \item $\ctxt{Q} = \termname{\alpha}{\ctxt{T}}$. By the \ih
    
    \item $\ctxt{Q} = \termrepl[\alpha']{\alpha}{s}{\ctxt{C}}$. Then, $o =
    \termrepl[\alpha']{\alpha}{s}{\ctxtapply{\ctxt{C}}{q}}$ and $p =
    \termrepl[\alpha']{\alpha}{s}{\ctxtapply{\ctxt{C}}{q'}}$. If the
    \reducemean[\rRm] step executes the outermost explicit
    replacement, we use Lemma~\ref{l:control:bisimulation:meta-repl-eqsigma}
      (\ref{l:control:bisimulation:meta-repl-eqsigma:o}), otherwise we
      use Lemma~\ref{l:control:bisimulation:ctxt-closure}  and the \ih
    
    \item $\ctxt{Q} = \termrepl[\alpha']{\alpha}{\ctxt{S}}{s}$. Then, $o =
    \termrepl[\alpha']{\alpha}{\ctxtapply{\ctxt{C}}{q}}{c}$ and $p =
    \termrepl[\alpha']{\alpha}{\ctxtapply{\ctxt{C}}{q'}}{c}$. If the
    \reducemean[\rRm] step executes the outermost explicit
    replacement, we use Lemma~\ref{l:control:bisimulation:meta-repl-eqsigma}
    (\ref{l:control:bisimulation:meta-repl-eqsigma:s}), otherwise we
    use Lemma~\ref{l:control:bisimulation:ctxt-closure}  and the \ih
    \qedhere
  \end{itemize}
\end{proof}


%% file: appendix-correspondence.tex
\section{Correspondence Results}
\label{app:correspondence}

\gettoappendix{l:eqnew-stack}
\input{proofs/new/eqnew-stack}

\gettoappendix{l:expan-aux-replacement}
\input{proofs/new/expan-aux-replacement.tex}

\gettoappendix{l:newrel-eqsigma}
\input{proofs/new/newrel-eqsigma.tex}

\gettoappendix{l:eqsigma-newrel}
\input{proofs/new/eqsigma-newrel.tex}


%% file: proofs/new/eqnew-stack.tex
\begin{proof} \mbox{}
\begin{enumerate}
  \item By induction on $s$.
  \begin{itemize}
  \item $s = v$. Then,
    \[\termconc{(\termapp{(\termabs{x}{t})}{u})}{s} 
=              \termapp{(\termapp{(\termabs{x}{t})}{u})}{v} 
\ruleEqnewApp  \termapp{(\termabs{x}{\termapp{t}{v}})}{u} 
=              \termapp{(\termabs{x}{\termconc{t}{s}})}{u}\]

  \item $s = \termpush{v}{s'}$. Then,
    \[
\begin{array}{llll}
\termconc{(\termapp{(\termabs{x}{t})}{u})}{s} & =             & \termconc{(\termapp{(\termapp{(\termabs{x}{t})}{u})}{v})}{s'} \\
&\ruleEqnewApp & \termconc{(\termapp{(\termabs{x}{\termapp{t}{v}})}{u})}{s'} \\
&\eqnew        & \termapp{(\termabs{x}{\termconc{(\termapp{t}{v})}{s'}})}{u} & (\ih\!) \\
&=             & \termapp{(\termabs{x}{\termconc{t}{s}})}{u}
\end{array} \]
  \end{itemize}

  \item By induction on $s$.
  \begin{itemize}
    \item $s = v$. By induction once again, this time on $s'$.
    \begin{itemize}
      \item $s' = w$. Then, \[
\begin{array}{llll}
\termname{\alpha'}{\termconc{(\termcont{\alpha}{\termname{\beta'}{\termconc{(\termcont{\beta}{c})}{s'}}})}{s}} &
=                   & \termname{\alpha'}{\termapp{(\termcont{\alpha}{\termname{\beta'}{\termapp{(\termcont{\beta}{c})}{w}}})}{v}} \\
&\ruleEqnewPushPush  & \termname{\beta'}{\termapp{(\termcont{\beta}{\termname{\alpha'}{\termapp{(\termcont{\alpha}{c})}{v}}})}{w}} \\
&=                   & \termname{\beta'}{\termconc{(\termcont{\beta}{\termname{\alpha'}{\termconc{(\termcont{\alpha}{c})}{s}}})}{s'}}
\end{array} \]

      \item $s' = \termpush{w}{s''}$. Then, \[
\begin{array}{llll}
 \termname{\alpha'}{\termconc{(\termcont{\alpha}{\termname{\beta'}{\termconc{(\termcont{\beta}{c})}{s'}}})}{s}} & 
=                   & \termname{\alpha'}{\termapp{(\termcont{\alpha}{\termname{\beta'}{\termconc{(\termapp{(\termcont{\beta}{c})}{w})}{s''}}})}{v}} \\
&\ruleEqnewTheta     & \termname{\alpha'}{\termapp{(\termcont{\alpha}{\termname{\beta'}{\termconc{(\termcont{\gamma}{\termname{\gamma}{\termapp{(\termcont{\beta}{c})}{w}}})}{s''}}})}{v}} & \text{($\gamma$ fresh)} \\
&\eqnew              & \termname{\beta'}{\termconc{(\termcont{\gamma}{\termname{\alpha'}{\termapp{(\termcont{\alpha}{\termname{\gamma}{\termapp{(\termcont{\beta}{c})}{w}}})}{v}}})}{s''}} & (\ih\!) \\
&\ruleEqnewPushPush  & \termname{\beta'}{\termconc{(\termcont{\gamma}{\termname{\gamma}{\termapp{(\termcont{\beta}{\termname{\alpha'}{\termapp{(\termcont{\alpha}{c})}{v}}})}{w}}})}{s''}} \\
&\ruleEqnewTheta     & \termname{\beta'}{\termconc{(\termapp{(\termcont{\beta}{\termname{\alpha'}{\termapp{(\termcont{\alpha}{c})}{v}}})}{w})}{s''}} \\
&=                   & \termname{\beta'}{\termconc{(\termcont{\beta}{\termname{\alpha'}{\termconc{(\termcont{\alpha}{c})}{s}}})}{s'}}
\end{array} \]
    \end{itemize}

    \item $s = \termpush{v}{s''}$. Then, \[
\begin{array}{llll}
\termname{\alpha'}{\termconc{(\termcont{\alpha}{\termname{\beta'}{\termconc{(\termcont{\beta}{c})}{s'}}})}{s}} & 
=                   & \termname{\alpha'}{\termconc{(\termapp{(\termcont{\alpha}{\termname{\beta'}{\termconc{(\termcont{\beta}{c})}{s'}}})}{v})}{s''}} \\
&\ruleEqnewTheta     & \termname{\alpha'}{\termconc{(\termcont{\gamma}{\termname{\gamma}{\termapp{(\termcont{\alpha}{\termname{\beta'}{\termconc{(\termcont{\beta}{c})}{s'}}})}{v}}})}{s''}} & \text{($\gamma$ fresh)} \\
&\eqnew              & \termname{\alpha'}{\termconc{(\termcont{\gamma}{\termname{\beta'}{\termconc{(\termcont{\beta}{\termname{\gamma}{\termapp{(\termcont{\alpha}{c})}{v}}})}{s'}}})}{s''}} & (\ih\!) \\
&\eqnew              & \termname{\beta'}{\termconc{(\termcont{\beta}{\termname{\alpha'}{\termconc{(\termcont{\gamma}{\termname{\gamma}{\termapp{(\termcont{\alpha}{c})}{v}}})}{s''}}})}{s'}} & (\ih\!) \\
&\ruleEqnewTheta     & \termname{\beta'}{\termconc{(\termcont{\beta}{\termname{\alpha'}{\termconc{(\termapp{(\termcont{\alpha}{c})}{v})}{s''}}})}{s'}} \\
&=                   & \termname{\beta'}{\termconc{(\termcont{\beta}{\termname{\alpha'}{\termconc{(\termcont{\alpha}{c})}{s}}})}{s'}}
\end{array} \]
  \end{itemize}

  \item By induction on $s$.
  \begin{itemize}
    \item $s = v$. Then, \[
\begin{array}{llll}
                   \termname{\alpha'}{\termconc{(\termcont{\alpha}{\termname{\beta'}{\termabs{x}{\termcont{\beta}{c}}}})}{s}} & 
=                 & \termname{\alpha'}{\termapp{(\termcont{\alpha}{\termname{\beta'}{\termabs{x}{\termcont{\beta}{c}}}})}{v}} \\
&\ruleEqnewPushPop & \termname{\beta'}{\termabs{x}{\termcont{\beta}{\termname{\alpha'}{\termapp{(\termcont{\alpha}{c})}{v}}}}} \\
&=                 & \termname{\beta'}{\termabs{x}{\termcont{\beta}{\termname{\alpha'}{\termconc{(\termcont{\alpha}{c})}{s}}}}}
\end{array} \]

    \item $s = \termpush{v}{s'}$. Then, \[
\begin{array}{llll}
                   \termname{\alpha'}{\termconc{(\termcont{\alpha}{\termname{\beta'}{\termabs{x}{\termcont{\beta}{c}}}})}{s}} &
=                 & \termname{\alpha'}{\termconc{(\termapp{(\termcont{\alpha}{\termname{\beta'}{\termabs{x}{\termcont{\beta}{c}}}})}{v})}{s'}} \\
&\ruleEqnewTheta   & \termname{\alpha'}{\termconc{(\termcont{\gamma}{\termname{\gamma}{\termapp{(\termcont{\alpha}{\termname{\beta'}{\termabs{x}{\termcont{\beta}{c}}}})}{v}}})}{s'}} & \text{($\gamma$ fresh)} \\
&\ruleEqnewPushPop & \termname{\alpha'}{\termconc{(\termcont{\gamma}{\termname{\beta'}{\termabs{x}{\termcont{\beta}{\termname{\gamma}{\termapp{(\termcont{\alpha}{c})}{v}}}}}})}{s'}} \\
&\eqnew            & \termname{\beta'}{\termabs{x}{\termcont{\beta}{\termname{\alpha'}{\termconc{(\termcont{\gamma}{\termname{\gamma}{\termapp{(\termcont{\alpha}{c})}{v}}})}{s'}}}}} & (\ih\!) \\
&\ruleEqnewTheta   & \termname{\beta'}{\termabs{x}{\termcont{\beta}{\termname{\alpha'}{\termconc{(\termapp{(\termcont{\alpha}{c})}{v})}{s'}}}}} \\
&=                 & \termname{\beta'}{\termabs{x}{\termcont{\beta}{\termname{\alpha'}{\termconc{(\termcont{\alpha}{c})}{s}}}}}
\end{array} \]
  \end{itemize}

  \item By induction on $s$.
  \begin{itemize}
    \item $s = v$. Then, \[
\begin{array}{llll}
  \termname{\alpha'}{\termconc{(\termcont{\alpha}{\ntermren{\beta}{\beta'}{c}})}{s}} &
=                 & \termname{\alpha'}{\termapp{(\termcont{\alpha}{\ntermren{\beta}{\beta'}{c}})}{v}} \\
&\ruleEqnewRenPush & \ntermren{\beta}{\beta'}{\termname{\alpha'}{\termapp{(\termcont{\alpha}{c})}{v}}} \\
&=                 & \ntermren{\beta}{\beta'}{\termname{\alpha'}{\termconc{(\termcont{\alpha}{c})}{s}}}
\end{array} \]

    \item $s = \termpush{v}{s'}$. Then, \[
\begin{array}{llll}
\termname{\alpha'}{\termconc{(\termcont{\alpha}{\ntermren{\beta}{\beta'}{c}})}{s}} & 
=                 & \termname{\alpha'}{\termconc{(\termapp{(\termcont{\alpha}{\ntermren{\beta}{\beta'}{c}})}{v})}{s'}} \\
&\ruleEqnewTheta   & \termname{\alpha'}{\termconc{(\termcont{\gamma}{\termname{\gamma}{\termapp{(\termcont{\alpha}{\ntermren{\beta}{\beta'}{c}})}{v}}})}{s'}} & \text{($\gamma$ fresh)} \\
&\ruleEqnewRenPush & \termname{\alpha'}{\termconc{(\termcont{\gamma}{\ntermren{\beta}{\beta'}{\termname{\gamma}{\termapp{(\termcont{\alpha}{c})}{v}}}})}{s'}} \\
&\eqnew            & \ntermren{\beta}{\beta'}{\termname{\alpha'}{\termconc{(\termcont{\gamma}{\termname{\gamma}{\termapp{(\termcont{\alpha}{c})}{v}}})}{s'}}} & (\ih\!) \\
&\ruleEqnewTheta   & \ntermren{\beta}{\beta'}{\termname{\alpha'}{\termconc{(\termapp{(\termcont{\alpha}{c})}{v})}{s'}}} \\
&=                 & \ntermren{\beta}{\beta'}{\termname{\alpha'}{\termconc{(\termcont{\alpha}{c})}{s}}}
\end{array} \]\vspace*{(-\baselineskip*2)-6pt}
  \end{itemize}
\end{enumerate}
\end{proof}


%% file: proofs/new/expan-aux-replacement.tex
\begin{proof}\hfill
\begin{enumerate}
  \item We address item (\ref{l:expan-aux-replacement:app}) first, by induction
  on the size of $\ctxt{LCC}$.
  \begin{itemize}
    \item $\ctxt{LCC} = \boxdot$. \[
\begin{array}{llll}
                 \fexp{\termrepl[\alpha']{\alpha}{s}{(\termname{\alpha}{t})}} &
=               & \termname{\alpha'}{\termcont{\alpha}{\termname{\alpha}{(\fexp{t}::\fexp{s})}}} \\
&\ruleEqnewTheta & \termname{\alpha'}{\fexp{t}::\fexp{s}} \\
&=               & \fexp{\termname{\alpha'}{\termconc{t}{s}}}
\end{array} \]

    \item $\ctxt{LCC} = \termname{\beta}{\ctxt{LTC}}$. We proceed by
    analyzing the shape of $\ctxt{LTC}$.
    \begin{itemize}
      \item $\ctxt{LTC} = \termcont{\gamma}{\ctxt{LCC'}}$. \[
\begin{array}{lll}
& \fexp{\termrepl[\alpha']{\alpha}{s}{(\termname{\beta}{\ctxtapply{\termcont{\gamma}{\ctxt{LCC'}}}{\termname{\alpha}{t}}})}} \\
=       & \termname{\alpha'}{\termconc{(\termcont{\alpha}{\termname{\beta}{\termcont{\gamma}{\fexp{\ctxtapply{\ctxt{LCC'}}{\termname{\alpha}{t}}}}}})}{\fexp{s}}} \\
\eqnew  & \termname{\beta}{\termcont{\gamma}{\termname{\alpha'}{\termconc{(\termcont{\alpha}{\fexp{\ctxtapply{\ctxt{LCC'}}{\termname{\alpha}{t}}}})}{\fexp{s}}}}} & \text{(Lemma~\ref{l:eqnew-stack}:\ref{l:eqnew-stack:ren})} \\
=       & \termname{\beta}{\termcont{\gamma}{\fexp{\termrepl[\alpha']{\alpha}{s}{\ctxtapply{\ctxt{LCC'}}{\termname{\alpha}{t}}}}}} \\
\eqnew  & \termname{\beta}{\termcont{\gamma}{\fexp{\ctxtapply{\ctxt{LCC'}}{\termname{\alpha'}{\termconc{t}{s}}}}}}  & (\ih\!) \\
=       & \fexp{\termname{\beta}{\termcont{\gamma}{\ctxtapply{\ctxt{LCC'}}{\termname{\alpha'}{\termconc{t}{s}}}}}} \\
=       & \fexp{\ctxtapply{\ctxt{LCC}}{\termname{\alpha'}{\termconc{t}{s}}}}
\end{array} \]
       
      \item $\ctxt{LTC} = \termapp{\ctxt{LTC'}}{w}$. \[
\begin{array}{lll}
& \fexp{\termrepl[\alpha']{\alpha}{s}{(\termname{\beta}{\termapp{\ctxtapply{\ctxt{LTC'}}{\termname{\alpha}{t}}}{w}})}} \\
=               & \termname{\alpha'}{\termconc{(\termcont{\alpha}{\termname{\beta}{\termapp{\fexp{\ctxtapply{\ctxt{LTC'}}{\termname{\alpha}{t}}}}{\fexp{w}}}})}{\fexp{s}}} \\
\ruleEqnewTheta & \termname{\alpha'}{\termconc{(\termcont{\alpha}{\termname{\beta}{\termapp{(\termcont{\gamma}{\termname{\gamma}{\fexp{\ctxtapply{\ctxt{LTC'}}{\termname{\alpha}{t}}}}})}{\fexp{w}}}})}{\fexp{s}}} & \text{($\gamma$ fresh)} \\
\eqnew          & \termname{\beta}{\termapp{(\termcont{\gamma}{\termname{\alpha'}{\termconc{(\termcont{\alpha}{\termname{\gamma}{\fexp{\ctxtapply{\ctxt{LTC'}}{\termname{\alpha}{t}}}}})}{\fexp{s}}}})}{\fexp{w}}} & \text{(Lemma~\ref{l:eqnew-stack}:\ref{l:eqnew-stack:push})} \\
=               & \termname{\beta}{\termapp{(\termcont{\gamma}{\fexp{\termrepl[\alpha']{\alpha}{s}{(\termname{\gamma}{\ctxtapply{\ctxt{LTC'}}{\termname{\alpha}{t}}})}}})}{\fexp{w}}} \\
\eqnew          & \termname{\beta}{\termapp{(\termcont{\gamma}{\fexp{\termname{\gamma}{ \ctxtapply{\ctxt{LTC'}}{\termname{\alpha'}{\termconc{t}{s}}}}}})}{\fexp{w}}} & (\ih\!) \\
=               & \termname{\beta}{\termapp{(\termcont{\gamma}{\termname{\gamma}{\fexp{ \ctxtapply{\ctxt{LTC'}}{\termname{\alpha'}{\termconc{t}{s}}}}}})}{\fexp{w}}} \\
\ruleEqnewTheta & \termname{\beta}{\termapp{\fexp{\ctxtapply{\ctxt{LTC'}}{\termname{\alpha'}{\termconc{t}{s}}}}}{\fexp{w}}} \\
=               & \fexp{\termname{\beta}{\termapp{\ctxtapply{\ctxt{LTC'}}{\termname{\alpha'}{\termconc{t}{s}}}}{w}}} \\
=               & \fexp{\ctxtapply{\ctxt{LCC}}{\termname{\alpha'}{\termconc{t}{s}}}}
\end{array} \]
      Note that the \ih applies since $\fsize{\termname{\gamma}{\ctxt{LTC'}}} < \fsize{\ctxt{LCC}}$.

      \item $\ctxt{LTC} = \termabs{x}{\ctxt{LTC'}}$. \[
\begin{array}{lll}
& \fexp{\termrepl[\alpha']{\alpha}{s}{(\termname{\beta}{\ctxtapply{\termabs{x}{\ctxt{LTC}'}}{\termname{\alpha}{t}}})}} \\ 
=               & \termname{\alpha'}{\termconc{(\termcont{\alpha}{\termname{\beta}{\termabs{x}{\fexp{\ctxtapply{\ctxt{LTC'}}{\termname{\alpha}{t}}}}}})}{\fexp{s}}} \\
\ruleEqnewTheta & \termname{\alpha'}{\termconc{(\termcont{\alpha}{\termname{\beta}{\termabs{x}{\termcont{\gamma}{\termname{\gamma}{\fexp{\ctxtapply{\ctxt{LTC'}}{\termname{\alpha}{t}}}}}}}})}{\fexp{s}}} & \text{($\gamma$ fresh)} \\
\eqnew          & \termname{\beta}{\termabs{x}{\termcont{\gamma}{\termname{\alpha'}{\termconc{(\termcont{\alpha}{\termname{\gamma}{\fexp{\ctxtapply{\ctxt{LTC'}}{\termname{\alpha}{t}}}}})}{\fexp{s}}}}}} & \text{(Lemma~\ref{l:eqnew-stack}:\ref{l:eqnew-stack:pop})} \\
=               & \termname{\beta}{\termabs{x}{\termcont{\gamma}{\fexp{\termrepl[\alpha']{\alpha}{s}{(\termname{\gamma}{\ctxtapply{\ctxt{LTC'}}{\termname{\alpha}{t}}})}}}}} \\
\eqnew          & \termname{\beta}{\termabs{x}{\termcont{\gamma}{\fexp{\termname{\gamma}{\ctxtapply{\ctxt{LTC'}}{\termname{\alpha'}{\termconc{t}{s}}}}}}}} & (\ih\!) \\
=               & \termname{\beta}{\termabs{x}{\termcont{\gamma}{\termname{\gamma}{\fexp{\ctxtapply{\ctxt{LTC'}}{\termname{\alpha'}{\termconc{t}{s}}}}}}}} \\
\ruleEqnewTheta & \termname{\beta}{\termabs{x}{\fexp{\ctxtapply{\ctxt{LTC'}}{\termname{\alpha'}{\termconc{t}{s}}}}}} \\
=               & \fexp{\termname{\beta}{\termabs{x}{\ctxtapply{\ctxt{LTC'}}{\termname{\alpha'}{\termconc{t}{s}}}}}} \\
=               & \fexp{\ctxtapply{\ctxt{LCC}}{\termname{\alpha'}{\termconc{t}{s}}}} 
\end{array} \]
      Note that the \ih applies since $\fsize{\termname{\gamma}{\ctxt{LTC'}}} < \fsize{\ctxt{LCC}}$.

      \item $\ctxt{LTC} = \termsubs{x}{u}{\ctxt{LTC'}}$. \[
\begin{array}{lll}
   &              \fexp{\termrepl[\alpha']{\alpha}{s}{(\termname{\beta}{\termsubs{x}{t}{\ctxtapply{\ctxt{LTC'}}{\termname{\alpha}{t}}}})}} \\
=               & \termname{\alpha'}{\termconc{(\termcont{\alpha}{\fexp{\termname{\beta}{\termsubs{x}{t}{\ctxtapply{\ctxt{LTC'}}{\termname{\alpha}{t}}}}}})}{\fexp{s}}} \\
=               & \termname{\alpha'}{\termconc{(\termcont{\alpha}{\termname{\beta}{\termapp{(\termabs{x}{\fexp{\ctxtapply{\ctxt{LTC'}}{\termname{\alpha}{t}}}})}{\fexp{u}}}})}{\fexp{s}}} \\
\ruleEqnewTheta & \termname{\alpha'}{\termconc{(\termcont{\alpha}{\termname{\beta}{\termapp{(\termcont{\gamma}{\termname{\gamma}{\termabs{x}{\fexp{\ctxtapply{\ctxt{LTC'}}{\termname{\alpha}{t}}}}}})}{\fexp{u}}}})}{\fexp{s}}} & \text{($\gamma$ fresh)} \\
\eqnew          & \termname{\beta}{\termapp{(\termcont{\gamma}{\termname{\alpha'}{\termconc{(\termcont{\alpha}{\termname{\gamma}{\termabs{x}{\fexp{\ctxtapply{\ctxt{LTC'}}{\termname{\alpha}{t}}}}}})}{\fexp{s}}}})}{\fexp{u}}} & \text{(Lemma~\ref{l:eqnew-stack}:\ref{l:eqnew-stack:push})} \\
=               & \termname{\beta}{\termapp{(\termcont{\gamma}{\fexp{\termrepl[\alpha']{\alpha}{s}{(\termname{\gamma}{\termabs{x}{\ctxtapply{\ctxt{LTC'}}{\termname{\alpha}{t}}}})}}})}{\fexp{u}}} \\
\eqnew          & \termname{\beta}{\termapp{(\termcont{\gamma}{\fexp{\termname{\gamma}{\termabs{x}{\ctxtapply{\ctxt{LTC'}}{\termname{\alpha'}{\termconc{t}{s}}}}}}})}{\fexp{u}}} & (\ih\!) \\
=               & \termname{\beta}{\termapp{(\termcont{\gamma}{\termname{\gamma}{\termabs{x}{\fexp{\ctxtapply{\ctxt{LTC'}}{\termname{\alpha'}{\termconc{t}{s}}}}}}})}{\fexp{u}}} \\
\ruleEqnewTheta & \termname{\beta}{\termapp{(\termabs{x}{\fexp{\ctxtapply{\ctxt{LTC'}}{\termname{\alpha'}{\termconc{t}{s}}}}})}{\fexp{u}}} \\
=               & \fexp{\termname{\beta}{\termsubs{x}{u}{\ctxtapply{\ctxt{LTC'}}{\termname{\alpha'}{\termconc{t}{s}}}}}} \\
=               & \fexp{\ctxtapply{\ctxt{LCC}}{\termname{\alpha'}{\termconc{t}{s}}}}
\end{array} \]
      Note that the \ih applies since $\fsize{\termname{\gamma}{\termabs{x}{\ctxt{LTC'}}}} < \fsize{\ctxt{LCC}}$.
    \end{itemize}

    \item $\ctxt{LCC} = \termrepl[\gamma']{\gamma}{s'}{\ctxt{LCC'}}$. \[
\begin{array}{lll}
& \fexp{\termrepl[\alpha']{\alpha}{s}{(\termrepl[\gamma']{\gamma}{s'}{\ctxtapply{\ctxt{LCC'}}{\termname{\alpha}{t}}})}} \\ 
=       & \termname{\alpha'}{\termconc{(\termcont{\alpha}{\fexp{\termrepl[\gamma']{\gamma}{s'}{\ctxtapply{\ctxt{LCC'}}{\termname{\alpha}{t}}}}})}{\fexp{s}}} \\
=       & \termname{\alpha'}{\termconc{(\termcont{\alpha}{\termname{\gamma'}{\termconc{(\termcont{\gamma}{\fexp{\ctxtapply{\ctxt{LCC'}}{\termname{\alpha}{t}}}})}{\fexp{s'}}}})}{\fexp{s}}} \\
\eqnew  & \termname{\gamma'}{\termconc{(\termcont{\gamma}{\termname{\alpha'}{\termconc{(\termcont{\alpha}{\fexp{\ctxtapply{\ctxt{LCC'}}{\termname{\alpha}{t}}}})}{\fexp{s}}}})}{\fexp{s'}}} & \text{(Lemma~\ref{l:eqnew-stack}:\ref{l:eqnew-stack:push})} \\
=       & \termname{\gamma'}{\termconc{(\termcont{\gamma}{\fexp{\termrepl[\alpha']{\alpha}{s}{\ctxtapply{\ctxt{LCC'}}{\termname{\alpha}{t}}}}})}{\fexp{s'}}} \\
\eqnew  & \termname{\gamma'}{\termconc{(\termcont{\gamma}{\fexp{\ctxtapply{\ctxt{LCC'}}{\termname{\alpha'}{\termconc{t}{s}}}}})}{\fexp{s'}}} & (\ih\!)\\
=       & \fexp{\termrepl[\gamma']{\gamma}{s'}{\ctxtapply{\ctxt{LCC'}}{\termname{\alpha'}{\termconc{t}{s}}}}} \\
=       & \fexp{\ctxtapply{\ctxt{LCC}}{\termname{\alpha'}{\termconc{t}{s}}}}
\end{array} \]
  \end{itemize}

  \item We now address item (\ref{l:expan-aux-replacement:conc}), also by
  induction on the size of $\ctxt{LCC}$.
  \begin{itemize}
    \item $\ctxt{LCC} = \boxdot$. \[
\begin{array}{llll}
 \fexp{\termrepl[\alpha']{\alpha}{s}{\termrepl[\alpha]{\beta}{s'}{c}}} &
=               & \termname{\alpha'}{\termconc{(\termcont{\alpha}{\termname{\alpha}{\termconc{(\termcont{\beta}{\fexp{c}})}{\fexp{s'}}}})}{\fexp{s}}} \\
&\ruleEqnewTheta & \termname{\alpha'}{\termconc{\termconc{(\termcont{\beta}{\fexp{c}})}{\fexp{s'}}}{\fexp{s}}} \\
&=               & \fexp{\termrepl[\alpha']{\beta}{\termpush{s'}{s}}{c}}
\end{array} \]

    \item $\ctxt{LCC} = \termname{\delta}{\ctxt{LTC}}$. We proceed by
    analyzing the shape of $\ctxt{LTC}$.
    \begin{itemize}
      \item $\ctxt{LTC} = \termcont{\gamma}{\ctxt{LCC'}}$. \[
\begin{array}{lll}
& \fexp{\termrepl[\alpha']{\alpha}{s}{(\termname{\delta}{\termcont{\gamma}{\ctxtapply{\ctxt{LCC'}}{\termrepl[\alpha]{\beta}{s'}{c}}}})}} \\
=       & \termname{\alpha'}{\termconc{(\termcont{\alpha}{\fexp{\termname{\delta}{\termcont{\gamma}{\ctxtapply{\ctxt{LCC'}}{\termrepl[\alpha]{\beta}{s'}{c}}}}}})}{\fexp{s}}} \\
=       & \termname{\alpha'}{\termconc{(\termcont{\alpha}{\termname{\delta}{\termcont{\gamma}{\fexp{\ctxtapply{\ctxt{LCC'}}{\termrepl[\alpha]{\beta}{s'}{c}}}}}})}{\fexp{s}}} \\
\eqnew  & \termname{\delta}{\termcont{\gamma}{\termname{\alpha'}{\termconc{(\termcont{\alpha}{\fexp{\ctxtapply{\ctxt{LCC'}}{\termrepl[\alpha]{\beta}{s'}{c}}}})}{\fexp{s}}}}} & \text{(Lemma~\ref{l:eqnew-stack}:\ref{l:eqnew-stack:ren})} \\
=       & \termname{\delta}{\termcont{\gamma}{\fexp{\termrepl[\alpha']{\alpha}{s}{\ctxtapply{\ctxt{LCC'}}{\termrepl[\alpha]{\beta}{s'}{c}}}}}} \\
\eqnew  & \termname{\delta}{\termcont{\gamma}{\fexp{\ctxtapply{\ctxt{LCC'}}{\termrepl[\alpha']{\beta}{\termpush{s'}{s}}{c}}}}} & (\ih\!) \\
=       & \fexp{\termname{\delta}{\termcont{\gamma}{\ctxtapply{\ctxt{LCC'}}{\termrepl[\alpha']{\beta}{\termpush{s'}{s}}{c}}}}} \\
=       & \fexp{\ctxtapply{\ctxt{LCC}}{\termrepl[\alpha']{\beta}{\termpush{s'}{s}}{c}}}
\end{array} \]

      \item $\ctxt{LTC} = \termapp{\ctxt{LTC'}}{w}$. \[
\begin{array}{lll}
& \fexp{\termrepl[\alpha']{\alpha}{s}{(\termname{\delta}{\termapp{\ctxtapply{\ctxt{LTC'}}{\termrepl[\alpha]{\beta}{s'}{c}}}{w}})}} \\ 
=               & \termname{\alpha'}{\termconc{(\termcont{\alpha}{\termname{\delta}{\fexp{\termapp{\ctxtapply{\ctxt{LTC'}}{\termrepl[\alpha]{\beta}{s'}{c}}}{w}}}})}{\fexp{s}}} \\
\ruleEqnewTheta & \termname{\alpha'}{\termconc{(\termcont{\alpha}{\termname{\delta}{\termapp{(\termcont{\gamma}{\termname{\gamma}{\fexp{\ctxtapply{\ctxt{LTC'}}{\termrepl[\alpha]{\beta}{s'}{c}}}}})}{\fexp{w}}}})}{\fexp{s}}} & \text{($\gamma$ fresh)} \\
\eqnew          & \termname{\delta}{\termapp{(\termcont{\gamma}{\termname{\alpha'}{\termconc{(\termcont{\alpha}{\termname{\gamma}{\fexp{\ctxtapply{\ctxt{LTC'}}{\termrepl[\alpha]{\beta}{s'}{c}}}}})}{\fexp{s}}}})}{\fexp{w}}} & \text{(Lemma~\ref{l:eqnew-stack}:\ref{l:eqnew-stack:push})} \\
=               & \termname{\delta}{\termapp{(\termcont{\gamma}{\fexp{\termrepl[\alpha']{\alpha}{s}{(\termname{\gamma}{\ctxtapply{\ctxt{LTC'}}{\termrepl[\alpha]{\beta}{s'}{c}}})}}})}{\fexp{w}}} \\
\eqnew          & \termname{\delta}{\termapp{(\termcont{\gamma}{\fexp{\termname{\gamma}{\ctxtapply{\ctxt{LTC'}}{\termrepl[\alpha']{\beta}{\termpush{s'}{s}}{c}}}}})}{\fexp{w}}} & (\ih\!) \\
=               & \termname{\delta}{\termapp{(\termcont{\gamma}{\termname{\gamma}{\fexp{\ctxtapply{\ctxt{LTC'}}{\termrepl[\alpha']{\beta}{\termpush{s'}{s}}{c}}}}})}{\fexp{w}}} \\
\ruleEqnewTheta & \termname{\delta}{\termapp{(\fexp{\ctxtapply{\ctxt{LTC'}}{\termrepl[\alpha']{\beta}{\termpush{s'}{s}}{c}}})}{\fexp{w}}} \\
=               & \fexp{\termname{\delta}{\termapp{\ctxtapply{\ctxt{LTC'}}{\termrepl[\alpha']{\beta}{\termpush{s'}{s}}{c}}}{w}}} \\
=               & \fexp{\ctxtapply{\ctxt{LCC}}{\termrepl[\alpha']{\beta}{\termpush{s'}{s}}{c}}}
\end{array} \]
      Note that the \ih applies since $\fsize{\termname{\gamma}{\ctxt{LTC'}}} < \fsize{\ctxt{LCC}}$.

      \item $\ctxt{LTC} = \termabs{x}{\ctxt{LTC'}}$. \[
\begin{array}{cll}
& \fexp{\termrepl[\alpha']{\alpha}{s}{(\termname{\delta}{\termabs{x}{\ctxtapply{\ctxt{LTC'}}{\termrepl[\alpha]{\beta}{s'}{c}}}})}} \\ 
=               & \termname{\alpha'}{\termconc{(\termcont{\alpha}{\termname{\delta}{\fexp{\termabs{x}{\ctxtapply{\ctxt{LTC'}}{\termrepl[\alpha]{\beta}{s'}{c}}}}}})}{\fexp{s}}} \\
\ruleEqnewTheta & \termname{\alpha'}{\termconc{(\termcont{\alpha}{\termname{\delta}{\termabs{x}{\termcont{\gamma}{\termname{\gamma}{\fexp{\ctxtapply{\ctxt{LTC'}}{\termrepl[\alpha]{\beta}{s'}{c}}}}}}}})}{\fexp{s}}} & \text{($\gamma$ fresh)} \\
\eqnew          & \termname{\delta}{\termabs{x}{\termcont{\gamma}{\termname{\alpha'}{\termconc{(\termcont{\alpha}{\termname{\gamma}{\fexp{\ctxtapply{\ctxt{LTC'}}{\termrepl[\alpha]{\beta}{s'}{c}}}}})}{\fexp{s}}}}}} & \text{(Lemma~\ref{l:eqnew-stack}:\ref{l:eqnew-stack:pop})} \\
=               & \termname{\delta}{\termabs{x}{\termcont{\gamma}{\fexp{\termrepl[\alpha']{\alpha}{s}{(\termname{\gamma}{\ctxtapply{\ctxt{LTC'}}{\termrepl[\alpha]{\beta}{s'}{c}}})}}}}} \\
\eqnew          & \termname{\delta}{\termabs{x}{\termcont{\gamma}{\fexp{\termname{\gamma}{\ctxtapply{\ctxt{LTC'}}{\termrepl[\alpha']{\beta}{\termpush{s'}{s}}{c}}}}}}} & (\ih\!) \\
=               & \termname{\delta}{\termabs{x}{\termcont{\gamma}{\termname{\gamma}{\fexp{\ctxtapply{\ctxt{LTC'}}{\termrepl[\alpha']{\beta}{\termpush{s'}{s}}{c}}}}}}} \\
\ruleEqnewTheta & \termname{\delta}{\termabs{x}{\fexp{\ctxtapply{\ctxt{LTC'}}{\termrepl[\alpha']{\beta}{\termpush{s'}{s}}{c}}}}} \\
=               & \fexp{\termname{\delta}{\termabs{x}{\ctxtapply{\ctxt{LTC'}}{\termrepl[\alpha']{\beta}{\termpush{s'}{s}}{c}}}}} \\
=               & \fexp{\ctxtapply{\ctxt{LCC}}{\termrepl[\alpha']{\beta}{\termpush{s'}{s}}{c}}}
\end{array} \]
      Note that the \ih applies since $\fsize{\termname{\gamma}{\ctxt{LTC'}}} < \fsize{\ctxt{LCC}}$.

      \item $\ctxt{LTC} = \termsubs{x}{u}{\ctxt{LTC'}}$. \[
\begin{array}{lll}
                & \fexp{\termrepl[\alpha']{\alpha}{s}{(\termname{\delta}{\termsubs{x}{u}{\ctxtapply{\ctxt{LTC'}}{\termrepl[\alpha]{\beta}{s'}{c}}}})}} \\
=               & \termname{\alpha'}{\termconc{(\termcont{\alpha}{\termname{\delta}{\fexp{\termsubs{x}{u}{\ctxtapply{\ctxt{LTC'}}{\termrepl[\alpha]{\beta}{s'}{c}}}}}})}{\fexp{s}}} \\
=               & \termname{\alpha'}{\termconc{(\termcont{\alpha}{\termname{\delta}{\termapp{(\termabs{x}{\fexp{\ctxtapply{\ctxt{LTC'}}{\termrepl[\alpha]{\beta}{s'}{c}}}})}{\fexp{u}}}})}{\fexp{s}}} \\
\ruleEqnewTheta & \termname{\alpha'}{\termconc{(\termcont{\alpha}{\termname{\delta}{\termapp{(\termcont{\gamma}{\termname{\gamma}{\termabs{x}{\fexp{\ctxtapply{\ctxt{LTC'}}{\termrepl[\alpha]{\beta}{s'}{c}}}}}})}{\fexp{u}}}})}{\fexp{s}}} & \text{($\gamma$ fresh)} \\
\eqnew          & \termname{\delta}{\termapp{(\termcont{\gamma}{\termname{\alpha'}{\termconc{(\termcont{\alpha}{\termname{\gamma}{\termabs{x}{\fexp{\ctxtapply{\ctxt{LTC'}}{\termrepl[\alpha]{\beta}{s'}{c}}}}}})}{\fexp{s}}}})}{\fexp{u}}} & \text{(Lemma~\ref{l:eqnew-stack}:\ref{l:eqnew-stack:push})} \\
=               & \termname{\delta}{\termapp{(\termcont{\gamma}{\fexp{\termrepl[\alpha']{\alpha}{s}{(\termname{\gamma}{\termabs{x}{\ctxtapply{\ctxt{LTC'}}{\termrepl[\alpha]{\beta}{s'}{c}}}})}}})}{\fexp{u}}} \\
\eqnew          & \termname{\delta}{\termapp{(\termcont{\gamma}{\fexp{\termname{\gamma}{\termabs{x}{\ctxtapply{\ctxt{LTC'}}{\termrepl[\alpha']{\beta}{\termpush{s'}{s}}{c}}}}}})}{\fexp{u}}} & (\ih\!) \\
=               & \termname{\delta}{\termapp{(\termcont{\gamma}{\termname{\gamma}{\termabs{x}{\fexp{\ctxtapply{\ctxt{LTC'}}{\termrepl[\alpha']{\beta}{\termpush{s'}{s}}{c}}}}}})}{\fexp{u}}} \\
\ruleEqnewTheta & \termname{\delta}{\termapp{(\termabs{x}{\fexp{\ctxtapply{\ctxt{LTC'}}{\termrepl[\alpha']{\beta}{\termpush{s'}{s}}{c}}}})}{\fexp{u}}} \\
=               & \fexp{\termname{\delta}{\termsubs{x}{u}{\ctxtapply{\ctxt{LTC'}}{\termrepl[\alpha']{\beta}{\termpush{s'}{s}}{c}}}}} \\
=               & \fexp{\ctxtapply{\ctxt{LCC}}{\termrepl[\alpha']{\beta}{\termpush{s'}{s}}{c}}}
\end{array} \]
    \end{itemize}
    Note that the \ih applies since $\fsize{\termname{\gamma}{\termabs{x}{\ctxt{LTC'}}}} < \fsize{\ctxt{LCC}}$.

    \item $\ctxt{LCC} = \termrepl[\gamma']{\gamma}{s''}{\ctxt{LCC'}}$. \[
\begin{array}{cll}
        & \fexp{\termrepl[\alpha']{\alpha}{s}{\termrepl[\gamma']{\gamma}{s''}{\ctxtapply{\ctxt{LCC'}}{\termrepl[\alpha]{\beta}{s'}{c}}}}}\\
=       & \termname{\alpha'}{\termconc{(\termcont{\alpha}{\fexp{\termrepl[\gamma']{\gamma}{s''}{\ctxtapply{\ctxt{LCC'}}{\termrepl[\alpha]{\beta}{s'}{c}}}}})}{\fexp{s}}} \\
=       & \termname{\alpha'}{\termconc{(\termcont{\alpha}{\termname{\gamma'}{\termconc{(\termcont{\gamma}{\fexp{\ctxtapply{\ctxt{LCC'}}{\termrepl[\alpha]{\beta}{s'}{c}}}})}{\fexp{s''}}}})}{\fexp{s}}} \\
\eqnew  & \termname{\gamma'}{\termconc{(\termcont{\gamma}{\termname{\alpha'}{\termconc{(\termcont{\alpha}{\fexp{\ctxtapply{\ctxt{LCC'}}{\termrepl[\alpha]{\beta}{s'}{c}}}})}{\fexp{s}}}})}{\fexp{s''}}} & \text{(Lemma~\ref{l:eqnew-stack}:\ref{l:eqnew-stack:push})} \\
=       & \termname{\gamma'}{\termconc{(\termcont{\gamma}{\fexp{\termrepl[\alpha']{\alpha}{s}{\ctxtapply{\ctxt{LCC'}}{\termrepl[\alpha]{\beta}{s'}{c}}}}})}{\fexp{s''}}} \\
\eqnew  & \termname{\gamma'}{\termconc{(\termcont{\gamma}{\fexp{\ctxtapply{\ctxt{LCC'}}{\termrepl[\alpha']{\beta}{\termpush{s'}{s}}{c}}}})}{\fexp{s''}}} & (\ih\!) \\
=       & \fexp{\termrepl[\gamma']{\gamma}{s''}{\ctxtapply{\ctxt{LCC'}}{\termrepl[\alpha']{\beta}{\termpush{s'}{s}}{c}}}} \\
=       & \fexp{\ctxtapply{\ctxt{LCC}}{\termrepl[\alpha']{\beta}{\termpush{s'}{s}}{c}}}
\end{array} \]\vspace*{(-\baselineskip*2)}
  \end{itemize}
\end{enumerate}
\end{proof}


%% file: proofs/new/newrel-eqsigma.tex
\begin{proof}
By induction on $\eqnew$. We first analyse the base cases.
\begin{itemize}
  \item $o = \termapp{(\termabs{y}{\termabs{x}{t}})}{v} \eqnew
  \termabs{x}{\termapp{(\termabs{y}{t})}{v}} = p$, where $x \notin {v}$. Then
  $\fcan{o} = \fcan{\termsubs{y}{v}{(\termabs{x}{t})}} \eqsigma
  \fcan{\termabs{x}{\termsubs{y}{v}{t}}} = \fcan{p}$ holds by
  Lemma~\ref{l:control:equivalence:permute}:\ref{l:control:equivalence:permute:l}.

  \item $o = \termapp{(\termabs{x}{\termapp{t}{v}})}{u} \eqnew
  \termapp{\termapp{(\termabs{x}{t})}{u}}{v} = p$, where $x \notin {v}$. Then
  $\fcan{o} \!=\! \fcan{\termsubs{y}{v}{(\termapp{t}{u})}} \!\eqsigma\!
  \fcan{\termapp{\termsubs{y}{v}{t}}{u}} \!=\! \fcan{p}$ holds by
  Lemma~\ref{l:control:equivalence:permute}:\ref{l:control:equivalence:permute:l}.
     
  \item $o = \termapp{(\termabs{x}{\termcont{\alpha}{\termname{\beta}{u}}})}{v}
  \eqnew
  \termcont{\alpha}{\termname{\beta}{\termapp{(\termabs{x}{u})}{v}}} = p$,
  where $\alpha \notin {v}$. Then $\fcan{o} =
  \fcan{\termsubs{x}{v}{(\termcont{\alpha}{\termname{\beta}{u}})}} \\ \eqsigma
  \fcan{\termcont{\alpha}{\termname{\beta}{\termsubs{x}{v}{u}}}} = \fcan{p}$
  holds by
  Lemma~\ref{l:control:equivalence:permute}:\ref{l:control:equivalence:permute:l}.

  \item $o =
  \termname{\alpha'}{\termapp{(\termcont{\alpha}{\termname{\beta'}{\termapp{(\termcont{\beta}{c})}{w}}})}{v}}
  \eqnew
  \termname{\beta'}{\termapp{(\termcont{\beta}{\termname{\alpha'}{\termapp{(\termcont{\alpha}{c})}{v}}})}{w}}
  = p$, where $ \alpha \notin {w}, \beta \notin {v}, \beta \neq \alpha', \alpha
  \neq \beta'$. Then $\fcan{o} =
  \fcan{\termname{\alpha'}{\termcont{\alpha''}{\termrepl[\alpha'']{\alpha}{v}{(\termname{\beta'}{\termcont{\beta''}{\termrepl[\beta'']{\beta}{w}{c}}})}}}}
  \eqsigma$\\
  $\fcan{\termname{\beta'}{\termcont{\beta''}{\termrepl[\beta'']{\beta}{w}{(\termname{\alpha'}{\termcont{\alpha''}{\termrepl[\alpha'']{\alpha}{v}{c}}})}}}}
  = \fcan{p}$ holds by
  Lemma~\ref{l:control:equivalence:permute}:\ref{l:control:equivalence:permute:r}.

  \item $o =
  \termname{\alpha'}{\termapp{(\termcont{\alpha}{\termname{\beta'}{\termabs{x}{\termcont{\beta}{c}}}})}{v}}
  \eqnew
  \termname{\beta'}{\termabs{x}{\termcont{\beta}{\termname{\alpha'}{\termapp{(\termcont{\alpha}{c})}{v}}}}}
  = p$, where $x \notin {v}, \beta \notin {v}, \beta \neq \alpha', \alpha \neq
  \beta'$. Then $\fcan{o} =
  \fcan{\termname{\alpha'}{\termcont{\alpha''}{\termrepl[\alpha'']{\alpha}{v}{(\termname{\beta'}{\termabs{x}{\termcont{\beta}{c}}})}}}}
  \eqsigma
  \fcan{\termname{\beta'}{\termabs{x}{\termcont{\beta}{\termname{\alpha'}{\termcont{\alpha''}{\termrepl[\alpha'']{\alpha}{v}{c}}}}}}}
  = \fcan{p}$ holds by
  Lemma~\ref{l:control:equivalence:permute}:\ref{l:control:equivalence:permute:r}.

  \item $o =
  \termname{\alpha'}{\termabs{x}{\termcont{\alpha}{\termname{\beta'}{\termabs{y}{\termcont{\beta}{c}}}}}}
  \eqnew
  \termname{\beta'}{\termabs{y}{\termcont{\beta}{\termname{\alpha'}{\termabs{x}{\termcont{\alpha}{c}}}}}}
  = p$, where $\beta \neq \alpha', \alpha \neq \beta'$. Then $\fcan{o} =
  \termname{\alpha'}{\termabs{x}{\termcont{\alpha}{\termname{\beta'}{\termabs{y}{\termcont{\beta}{\fcan{c}}}}}}}
  \ruleEqsigPopPop 
  \termname{\beta'}{\termabs{y}{\termcont{\beta}{\termname{\alpha'}{\termabs{x}{\termcont{\alpha}{\fcan{c}}}}}}}
  = \fcan{p}$.

  \item $o =
  \termcont{\alpha}{\termname{\alpha}{v}} \eqnew v$, where $\alpha \notin {v}$.
  Then $\fcan{o} = \termcont{\alpha}{\termname{\alpha}{\fcan{v}}}
  \ruleEqsigTheta \fcan{v} = \fcan{p}$.

  \item $o =
  \termname{\alpha'}{\termcont{\alpha}{\termname{\beta'}{\termcont{\beta}{c}}}}
  \eqnew
  \termname{\beta'}{\termcont{\beta}{\termname{\alpha'}{\termcont{\alpha}{c}}}}
  = p$, where $\beta' \neq \alpha$ and $\alpha'\neq \beta$. Then $\fcan{o} =
  \termname{\alpha'}{\termcont{\alpha}{\termname{\beta'}{\termcont{\beta}{\fcan{c}}}}}
  \ruleEqsigExRen
  \termname{\beta'}{\termcont{\beta}{\termname{\alpha'}{\termcont{\alpha}{\fcan{c}}}}}
  = \fcan{p}$.

  \item $o =
  \termname{\alpha'}{\termcont{\alpha}{\termname{\beta'}{\termabs{x}{\termcont{\beta}{c}}}}}
  \eqnew
  \termname{\beta'}{\termabs{x}{\termcont{\beta}{\termname{\alpha'}{\termcont{\alpha}{c}}}}}
  = p$, where $\beta' \neq \alpha$ and $\alpha'\neq \beta$. Then $\fcan{o} =
  \termname{\alpha'}{\termcont{\alpha}{\termname{\beta'}{\termabs{x}{\termcont{\beta}{\fcan{c}}}}}}
  \ruleEqsigExRen
  \termname{\beta'}{\termabs{x}{\termcont{\beta}{\termname{\alpha'}{\termcont{\alpha}{\fcan{c}}}}}}
  = \fcan{p}$.

  \item $o =
  \termname{\alpha'}{\termcont{\alpha}{\termname{\beta'}{(\termcont{\beta}{c})w}}}
  \eqnew
  \termname{\beta'}{(\termcont{\beta}{\termname{\alpha'}{\termcont{\alpha}{c}}})w}
  = p$, where $\beta' \neq \alpha$ and $\alpha'\neq \beta$. Then $\fcan{o} =
  \fcan{\termname{\alpha'}{\termcont{\alpha}{\termname{\beta'}{\termcont{\beta''}{\termrepl[\beta'']{\beta}{w}{c}}}}}}
  \eqsigma
  \fcan{\termname{\beta'}{\termcont{\beta''}{\termrepl[\beta'']{\beta}{w}{(\termname{\alpha'}{\termcont{\alpha}{c}})}}}}
  = \fcan{p}$ holds by
  Lemma~\ref{l:control:equivalence:permute}:\ref{l:control:equivalence:permute:r}.
\end{itemize}
    
For the inductive cases, let consider $o = \ctxtapply{\ctxt{O}}{o'} \eqnew
\ctxtapply{\ctxt{O}}{p'} = p$, where $o' \eqnew p'$. The \ih\ gives $\fcan{o'}
\eqsigma \fcan{p'}$, and Lemma~\ref{l:control:bisimulation:ctxt-closure} gives
$\fcan{\ctxtapply{\ctxt{O}}{\fcan{o'}}} \eqsigma
\fcan{\ctxtapply{\ctxt{O}}{\fcan{p'}}} $, so that we conclude by the fact that
$\fcan{\ctxtapply{\ctxt{O}}{\fcan{o'}}} = \fcan{\ctxtapply{\ctxt{O}}{o'}}$ and
$\fcan{\ctxtapply{\ctxt{O}}{\fcan{p'}}} = \fcan{\ctxtapply{\ctxt{O}}{p'}}$.   
\end{proof}


%% file: proofs/new/eqsigma-newrel.tex
\begin{proof}
The cases where $o \eqsigma p$ holds by reflexivity, transitivity or symmetry
are straightforward. For congruence, we reason by induction on the context
$\ctxt{O}$ such that $o = \ctxtapply{\ctxt{O}}{l}$ and $p =
\ctxtapply{\ctxt{O}}{r}$ with $l \eqsigma[\ast] r$, where $\eqsigma[\ast]$ is
an axiom in Figure~\ref{f:control:equivalence:eqsigma}.
\begin{itemize}
  \item $\ctxt{O} = \Box$. There are two possible rules:
  \begin{enumerate}
    \item $\ruleEqsigExSubs$. Then, $o =
    \termsubs{x}{u}{\ctxtapply{\ctxt{LTT}}{t}}$ and $p =
    \ctxtapply{\ctxt{LTT}}{\termsubs{x}{u}{t}}$ with $x \notin \ctxt{LTT}$ and
    $\fc{u}{\ctxt{LTT}}$. We proceed by induction on the size of $\ctxt{LTT}$:
    \begin{itemize}
      \item $\ctxt{LTT} = \Box$. This case is immediate since $o = p$.

      \item $\ctxt{LTT} = \termapp{\ctxt{LTT}'}{v}$. Then, \[
\begin{array}{llll}
\fexp{\termsubs{x}{u}{(\termapp{\ctxtapply{\ctxt{LTT}'}{t}}{v})}}
  & =             & \termapp{(\termabs{x}{\termapp{\fexp{\ctxtapply{\ctxt{LTT}'}{t}}}{\fexp{v}}})}{\fexp{u}} \\
  & \ruleEqnewApp & \termapp{\termapp{(\termabs{x}{\fexp{\ctxtapply{\ctxt{LTT}'}{t}}})}{\fexp{u}}}{\fexp{v}} \\
  & =             & \termapp{\fexp{\termsubs{x}{u}{\ctxtapply{\ctxt{LTT}'}{t}}}}{\fexp{v}} \\
  & \eqnew        & \termapp{\fexp{\ctxtapply{\ctxt{LTT}'}{\termsubs{x}{u}{t}}}}{\fexp{v}} & (\ih\!) \\
  & =             & \fexp{\termapp{\ctxtapply{\ctxt{LTT}'}{\termsubs{x}{u}{t}}}{v}}
\end{array} \] 

      \item $\ctxt{LTT} = \termabs{y}{\ctxt{LTT}'}$. Then, \[
\begin{array}{llll}
\fexp{\termsubs{x}{u}{(\termabs{y}{\ctxtapply{\ctxt{LTT}'}{t}})}}
  & =             & \termapp{(\termabs{x}{\termabs{y}{\fexp{\ctxtapply{\ctxt{LTT}'}{t}}}})}{\fexp{u}} \\
  & \ruleEqnewAbs & \termabs{y}{\termapp{(\termabs{x}{\fexp{\ctxtapply{\ctxt{LTT}'}{t}}})}{\fexp{u}}} \\
  & =             & \termabs{y}{\fexp{\termsubs{x}{u}{\ctxtapply{\ctxt{LTT}'}{t}}}} \\
  & \eqnew        & \termabs{y}{\fexp{\ctxtapply{\ctxt{LTT}'}{\termsubs{x}{u}{t}}}} & (\ih\!) \\
  & =             & \fexp{\termabs{y}{\ctxtapply{\ctxt{LTT}'}{\termsubs{x}{u}{t}}}}
\end{array} \]

      \item $\ctxt{LTT} = \termcont{\alpha}{\ctxt{LCT}}$. We proceed by
      analyzing the shape of $\ctxt{LCT}$:
      \begin{itemize}
        \item $\ctxt{LCT} = \termname{\delta}{\ctxt{LTT}'}$. Then, \[
\begin{array}{llll}
\fexp{\termsubs{x}{u}{(\termcont{\alpha}{\termname{\delta}{\ctxtapply{\ctxt{LTT}'}{t}}})}}
  & =               & \termapp{(\termabs{x}{\termcont{\alpha}{\termname{\delta}{\fexp{\ctxtapply{\ctxt{LTT}'}{t}}}}})}{\fexp{u}} \\
  & \ruleEqnewCont  & \termcont{\alpha}{\termname{\delta}{\termapp{(\termabs{x}{\fexp{\ctxtapply{\ctxt{LTT}'}{t}}})}{\fexp{u}}}} \\
  & =               & \termcont{\alpha}{\termname{\delta}{\fexp{\termsubs{x}{u}{\ctxtapply{\ctxt{LTT}'}{t}}}}} \\
  & \eqnew          & \termcont{\alpha}{\termname{\delta}{\fexp{\ctxtapply{\ctxt{LTT}'}{\termsubs{x}{u}{t}}}}} & (\ih\!) \\
  & =               & \fexp{\termcont{\alpha}{\termname{\delta}{\ctxtapply{\ctxt{LTT}'}{\termsubs{x}{u}{t}}}}}
\end{array} \] 

        \item $\ctxt{LCT} = \termrepl[\delta]{\gamma}{s}{\ctxt{LCT}'}$. Then, \[
\begin{array}{llll}
\fexp{\termsubs{x}{u}{(\termcont{\alpha}{\termrepl[\delta]{\gamma}{s}{\ctxtapply{\ctxt{LCT}'}{t}}})}}
  & =               & \termapp{(\termabs{x}{\termcont{\alpha}{\termname{\delta}{\termconc{(\termcont{\gamma}{\fexp{\ctxtapply{\ctxt{LCT}'}{t}}})}{\fexp{s}}}}})}{\fexp{u}} \\
  & \ruleEqnewCont  & \termcont{\alpha}{\termname{\delta}{\termapp{(\termabs{x}{\termconc{(\termcont{\gamma}{\fexp{\ctxtapply{\ctxt{LCT}'}{t}}})}{\fexp{s}}})}{\fexp{u}}}} \\
  & \eqnew          & \termcont{\alpha}{\termname{\delta}{\termconc{(\termapp{(\termabs{x}{\termcont{\gamma}{\fexp{\ctxtapply{\ctxt{LCT}'}{t}}}})}{\fexp{u}})}{\fexp{s}}}}  & (\text{L.~\ref{l:eqnew-stack}:\ref{l:eqnew-stack:app}}) \\
  & =               & \termcont{\alpha}{\termname{\delta}{\termconc{\fexp{\termsubs{x}{u}{(\termcont{\gamma}{\ctxtapply{\ctxt{LCT}'}{t}})}}}{\fexp{s}}}} \\
  & \eqnew          & \termcont{\alpha}{\termname{\delta}{\termconc{\fexp{\termcont{\gamma}{\ctxtapply{\ctxt{LCT}'}{\termsubs{x}{u}{t}}}}}{\fexp{s}}}}                      & (\ih\!) \\
  & =               & \termcont{\alpha}{\termname{\delta}{\termconc{(\termcont{\gamma}{\fexp{\ctxtapply{\ctxt{LCT}'}{\termsubs{x}{u}{t}}}})}{\fexp{s}}}} \\
  & =               & \termcont{\alpha}{\fexp{\termrepl[\delta]{\gamma}{s}{\ctxtapply{\ctxt{LCT}'}{\termsubs{x}{u}{t}}}}} \\
  & =               & \fexp{\termcont{\alpha}{\termrepl[\delta]{\gamma}{s}{\ctxtapply{\ctxt{LCT}'}{\termsubs{x}{u}{t}}}}}
\end{array} \] 
      \end{itemize}

      \item $\ctxt{LTT} = \termsubs{y}{v}{\ctxt{LTT}'}$. Then, \[
\begin{array}{llll}
\fexp{\termsubs{x}{u}{\termsubs{y}{v}{\ctxtapply{\ctxt{LTT}'}{t}}}}
  & =             & \termapp{(\termabs{x}{\termapp{(\termabs{y}{\fexp{\ctxtapply{\ctxt{LTT}'}{t}}})}{\fexp{v}}})}{\fexp{u}} \\
  & \ruleEqnewApp & \termapp{\termapp{(\termabs{x}{\termabs{y}{\fexp{\ctxtapply{\ctxt{LTT}'}{t}}}})}{\fexp{u}}}{\fexp{v}} \\
  & \ruleEqnewAbs & \termapp{(\termabs{y}{\termapp{(\termabs{x}{\fexp{\ctxtapply{\ctxt{LTT}'}{t}}})}{\fexp{u}}})}{\fexp{v}} \\
  & =             & \termapp{(\termabs{y}{\fexp{\termsubs{x}{u}{\ctxtapply{\ctxt{LTT}'}{t}}}})}{\fexp{v}} \\
  & \eqnew        & \termapp{(\termabs{y}{\fexp{\ctxtapply{\ctxt{LTT}'}{\termsubs{x}{u}{t}}}})}{\fexp{v}} & (\ih\!) \\
  & =             & \fexp{\termsubs{y}{v}{\ctxtapply{\ctxt{LTT}'}{\termsubs{x}{u}{t}}}}
\end{array} \] 
    \end{itemize}

    \item $\ruleEqsigTheta$. Then, $o =
    \termcont{\alpha}{\termname{\alpha}{t}}$ and $p = t$ with $\alpha \notin
    t$. Moreover, $\fexp{\termcont{\alpha}{\termname{\alpha}{t}}} =
    \termcont{\alpha}{\termname{\alpha}{\fexp{t}}}$ by definition. We conclude
    by $\ruleEqnewTheta$.
  \end{enumerate}
  
  \item $\ctxt{O} = \boxdot$. There are three possible rules:
  \begin{enumerate}
    \item $\ruleEqsigExRepl$. Then, $o =
    \termrepl[\beta]{\alpha}{s}{\ctxtapply{\ctxt{LCC}}{c}}$ and $p =
    \ctxtapply{\ctxt{LCC}}{\termrepl[\beta]{\alpha}{s}{c}}$ with $\alpha \notin
    \ctxt{LCC}$, $\fc{\beta}{\ctxt{LCC}}$ and $\fc{s}{\ctxt{LCC}}$. We proceed
    by induction on the size of $\ctxt{LCC}$:
    \begin{itemize}
      \item $\ctxt{LCC} = \boxdot$. This case is immediate since $o = p$.

      \item $\ctxt{LCC} = \termname{\delta}{\ctxt{LTC}}$. We proceed by
      analyzing the shape of $\ctxt{LTC}$:
      \begin{itemize}
        \item $\termapp{\ctxt{LTC}'}{u}$. Then, \[
\begin{array}{lll}
                    & \fexp{\termrepl[\beta]{\alpha}{s}{(\termname{\delta}{\termapp{\ctxtapply{\ctxt{LTC}'}{c}}{u}})}} \\
=                   & \termname{\beta}{\termconc{(\termcont{\alpha}{\termname{\delta}{\termapp{\fexp{\ctxtapply{\ctxt{LTC}'}{c}}}{\fexp{u}}}})}{\fexp{s}}} \\
\ruleEqnewTheta     & \termname{\beta}{\termconc{(\termcont{\alpha}{\termname{\delta}{\termapp{(\termcont{\delta'}{\termname{\delta'}{\fexp{\ctxtapply{\ctxt{LTC}'}{c}}}})}{\fexp{u}}}})}{\fexp{s}}}  & \text{($\delta'$ fresh)} \\
\eqnew              & \termname{\delta}{\termapp{(\termcont{\delta'}{\termname{\beta}{\termconc{(\termcont{\alpha}{\termname{\delta'}{\fexp{\ctxtapply{\ctxt{LTC}'}{c}}}})}{\fexp{s}}}})}{\fexp{u}}}  & (\text{L.~\ref{l:eqnew-stack}:\ref{l:eqnew-stack:push}}) \\
=                   & \termname{\delta}{\termapp{(\termcont{\delta'}{\fexp{\termrepl[\beta]{\alpha}{s}{(\termname{\delta'}{\ctxtapply{\ctxt{LTC}'}{c}})}}})}{\fexp{u}}} \\
\eqnew              & \termname{\delta}{\termapp{(\termcont{\delta'}{\fexp{\termname{\delta'}{\ctxtapply{\ctxt{LTC}'}{\termrepl[\beta]{\alpha}{s}{c}}}}})}{\fexp{u}}}                                 & (\ih\!) \\
=                   & \termname{\delta}{\termapp{(\termcont{\delta'}{\termname{\delta'}{\fexp{\ctxtapply{\ctxt{LTC}'}{\termrepl[\beta]{\alpha}{s}{c}}}}})}{\fexp{u}}} \\
\ruleEqnewTheta     & \termname{\delta}{\termapp{\fexp{\ctxtapply{\ctxt{LTC}'}{\termrepl[\beta]{\alpha}{s}{c}}}}{\fexp{u}}} \\
=                   & \fexp{\termname{\delta}{\termapp{\ctxtapply{\ctxt{LTC}'}{\termrepl[\beta]{\alpha}{s}{c}}}{u}}}
\end{array} \]

        \item $\termabs{x}{\ctxt{LTC}'}$. Then, \[
\begin{array}{lll}
                  & \fexp{\termrepl[\beta]{\alpha}{s}{(\termname{\delta}{\termabs{x}{\ctxtapply{\ctxt{LTC}'}{c}}})}} \\
=                 & \termname{\beta}{\termconc{(\termcont{\alpha}{\termname{\delta}{\termabs{x}{\fexp{\ctxtapply{\ctxt{LTC}'}{c}}}}})}{\fexp{s}}} \\
\ruleEqnewTheta   & \termname{\beta}{\termconc{(\termcont{\alpha}{\termname{\delta}{\termabs{x}{\termcont{\delta'}{\termname{\delta'}{\fexp{\ctxtapply{\ctxt{LTC}'}{c}}}}}}})}{\fexp{s}}} & \text{($\delta'$ fresh)} \\
\eqnew            & \termname{\delta}{\termabs{x}{\termcont{\delta'}{\termname{\beta}{\termconc{(\termcont{\alpha}{\termname{\delta'}{\fexp{\ctxtapply{\ctxt{LTC}'}{c}}}})}{\fexp{s}}}}}} & (\text{L.~\ref{l:eqnew-stack}:\ref{l:eqnew-stack:pop}}) \\
=                 & \termname{\delta}{\termabs{x}{\termcont{\delta'}{\fexp{\termrepl[\beta]{\alpha}{s}{(\termname{\delta'}{\ctxtapply{\ctxt{LTC}'}{c}})}}}}} \\
\eqnew            & \termname{\delta}{\termabs{x}{\termcont{\delta'}{\fexp{\termname{\delta'}{\ctxtapply{\ctxt{LTC}'}{\termrepl[\beta]{\alpha}{s}{c}}}}}}}                                & (\ih\!) \\
=                 & \termname{\delta}{\termabs{x}{\termcont{\delta'}{\termname{\delta'}{\fexp{\ctxtapply{\ctxt{LTC}'}{\termrepl[\beta]{\alpha}{s}{c}}}}}}} \\
\ruleEqnewTheta   & \termname{\delta}{\termabs{x}{\fexp{\ctxtapply{\ctxt{LTC}'}{\termrepl[\beta]{\alpha}{s}{c}}}}} \\
=                 & \fexp{\termname{\delta}{\termabs{x}{\ctxtapply{\ctxt{LTC}'}{\termrepl[\beta]{\alpha}{s}{c}}}}}
\end{array} \]

        \item $\termcont{\gamma}{\ctxt{LCC}'}$. Then, \[
\begin{array}{lll}
                  & \fexp{\termrepl[\beta]{\alpha}{s}{(\ntermren{\gamma}{\delta}{\ctxtapply{\ctxt{LCC}'}{c}})}} \\
=                 & \termname{\beta}{\termconc{(\termcont{\alpha}{\ntermren{\gamma}{\delta}{\fexp{\ctxtapply{\ctxt{LCC}'}{c}}}})}{\fexp{s}}} \\
\eqnew            & \ntermren{\gamma}{\delta}{\termname{\beta}{\termconc{(\termcont{\alpha}{\fexp{\ctxtapply{\ctxt{LCC}'}{c}}})}{\fexp{s}}}}  & (\text{L.~\ref{l:eqnew-stack}:\ref{l:eqnew-stack:ren}}) \\
=                 & \ntermren{\gamma}{\delta}{\fexp{\termrepl[\beta]{\alpha}{s}{(\ctxtapply{\ctxt{LCC}'}{c})}}} \\
\eqnew            & \ntermren{\gamma}{\delta}{\fexp{\ctxtapply{\ctxt{LCC}'}{\termrepl[\beta]{\alpha}{s}{c}}}}                                 & (\ih\!) \\
=                 & \fexp{\ntermren{\gamma}{\delta}{\ctxtapply{\ctxt{LCC}'}{\termrepl[\beta]{\alpha}{s}{c}}}}
\end{array} \]

        \item $\termsubs{x}{u}{\ctxt{LTC}'}$. Then, \[
\begin{array}{lll}
                    & \fexp{\termrepl[\beta]{\alpha}{s}{(\termname{\delta}{\termsubs{x}{u}{\ctxtapply{\ctxt{LTC}'}{c}}})}} \\
=                   & \termname{\beta}{\termconc{(\termcont{\alpha}{\termname{\delta}{\termapp{(\termabs{x}{\fexp{\ctxtapply{\ctxt{LTC}'}{c}}})}{\fexp{u}}}})}{\fexp{s}}} \\
\ruleEqnewTheta     & \termname{\beta}{\termconc{(\termcont{\alpha}{\termname{\delta}{\termapp{(\termcont{\delta'}{\termname{\delta'}{\termabs{x}{\fexp{\ctxtapply{\ctxt{LTC}'}{c}}}}})}{\fexp{u}}}})}{\fexp{s}}} & \text{($\delta'$ fresh)} \\
\eqnew              & \termname{\delta}{\termapp{(\termcont{\delta'}{\termname{\beta}{\termconc{(\termcont{\alpha}{\termname{\delta'}{\termabs{x}{\fexp{\ctxtapply{\ctxt{LTC}'}{c}}}}})}{\fexp{s}}}})}{\fexp{u}}} & (\text{L.~\ref{l:eqnew-stack}:\ref{l:eqnew-stack:push}}) \\
=                   & \termname{\delta}{\termapp{(\termcont{\delta'}{\fexp{\termrepl[\beta]{\alpha}{s}{(\termname{\delta'}{\termabs{x}{\ctxtapply{\ctxt{LTC}'}{c}}})}}})}{\fexp{u}}} \\
\eqnew              & \termname{\delta}{\termapp{(\termcont{\delta'}{\fexp{\termname{\delta'}{\termabs{x}{\ctxtapply{\ctxt{LTC}'}{\termrepl[\beta]{\alpha}{s}{c}}}}}})}{\fexp{u}}}                                & (\ih\!) \\
=                   & \termname{\delta}{\termapp{(\termcont{\delta'}{\termname{\delta'}{\fexp{\termabs{x}{\ctxtapply{\ctxt{LTC}'}{\termrepl[\beta]{\alpha}{s}{c}}}}}})}{\fexp{u}}} \\
\ruleEqnewTheta     & \termname{\delta}{\termapp{\fexp{\termabs{x}{\ctxtapply{\ctxt{LTC}'}{\termrepl[\beta]{\alpha}{s}{c}}}}}{\fexp{u}}} \\
=                   & \fexp{\termname{\delta}{\termsubs{x}{u}{\ctxtapply{\ctxt{LTC}'}{\termrepl[\beta]{\alpha}{s}{c}}}}}
\end{array} \]
      \end{itemize}

      \item $\ctxt{LCC} = \termrepl[\delta]{\gamma}{s'}{\ctxt{LCC}'}$. Then, \[\kern-5em
\begin{array}{lll}
                    & \fexp{\termrepl[\beta]{\alpha}{s}{\termrepl[\delta]{\gamma}{s'}{\ctxtapply{\ctxt{LTC}'}{c}}}} \\
=                   & \termname{\beta}{\termconc{(\termcont{\alpha}{\termname{\delta}{\termconc{(\termcont{\gamma}{\fexp{\ctxtapply{\ctxt{LTC}'}{c}}})}{\fexp{s'}}}})}{\fexp{s}}} \\
\eqnew              & \termname{\delta}{\termconc{(\termcont{\gamma}{\termname{\beta}{\termconc{(\termcont{\alpha}{\fexp{\ctxtapply{\ctxt{LTC}'}{c}}})}{\fexp{s}}}})}{\fexp{s'}}} & (\text{L.~\ref{l:eqnew-stack}:\ref{l:eqnew-stack:push}}) \\
=                   & \termname{\delta}{\termconc{(\termcont{\gamma}{\fexp{\termrepl[\beta]{\alpha}{s}{\ctxtapply{\ctxt{LTC}'}{c}}}})}{\fexp{s'}}} \\
\eqnew              & \termname{\delta}{\termconc{(\termcont{\gamma}{\fexp{\ctxtapply{\ctxt{LTC}'}{\termrepl[\beta]{\alpha}{s}{c}}}})}{\fexp{s'}}}                                & (\ih\!) \\
=                   & \fexp{\termrepl[\delta]{\gamma}{s'}{\termrepl[\beta]{\alpha}{s}{\ctxtapply{\ctxt{LTC}'}{c}}}}
\end{array} \] 
    \end{itemize}

    \item $\ruleEqsigExRen$. Then, $o =
    \ntermren{\alpha}{\beta}{\ctxtapply{\ctxt{LCC}}{c}}$ and $p =
    \ctxtapply{\ctxt{LCC}}{\ntermren{\alpha}{\beta}{c}}$ with $\alpha \notin
    \ctxt{LCC}$ and $\fc{\beta}{\ctxt{LCC}}$. We proceed by induction on the
    size of $\ctxt{LCC}$:
    \begin{itemize}
      \item $\ctxt{LCC} = \boxdot$. This case is immediate since $o = p$.

      \item $\ctxt{LCC} = \termname{\delta}{\ctxt{LTC}}$. We proceed by
      analyzing the shape of $\ctxt{LTC}$:
      \begin{itemize}
        \item $\termapp{\ctxt{LTC}'}{u}$. Then, \[
\begin{array}{lll}
                  & \fexp{\ntermren{\alpha}{\beta}{\termname{\delta}{\termapp{\ctxtapply{\ctxt{LTC}'}{c}}{u}}}} \\
=                 & \ntermren{\alpha}{\beta}{\termname{\delta}{\termapp{\fexp{\ctxtapply{\ctxt{LTC}'}{c}}}{\fexp{u}}}} \\
\ruleEqnewTheta   & \ntermren{\alpha}{\beta}{\termname{\delta}{\termapp{(\termcont{\delta'}{\termname{\delta'}{\fexp{\ctxtapply{\ctxt{LTC}'}{c}}}})}{\fexp{u}}}} & \text{($\delta'$ fresh)} \\
\ruleEqnewRenPush & \termname{\delta}{\termapp{(\termcont{\delta'}{\ntermren{\alpha}{\beta}{\termname{\delta'}{\fexp{\ctxtapply{\ctxt{LTC}'}{c}}}}})}{\fexp{u}}} \\
=                 & \termname{\delta}{\termapp{(\termcont{\delta'}{\fexp{\ntermren{\alpha}{\beta}{\termname{\delta'}{\ctxtapply{\ctxt{LTC}'}{c}}}}})}{\fexp{u}}} \\
\eqnew            & \termname{\delta}{\termapp{(\termcont{\delta'}{\fexp{\termname{\delta'}{\ctxtapply{\ctxt{LTC}'}{\ntermren{\alpha}{\beta}{c}}}}})}{\fexp{u}}} & (\ih\!) \\
=                 & \termname{\delta}{\termapp{(\termcont{\delta'}{\termname{\delta'}{\fexp{\ctxtapply{\ctxt{LTC}'}{\ntermren{\alpha}{\beta}{c}}}}})}{\fexp{u}}} \\
\ruleEqnewTheta   & \termname{\delta}{\termapp{\fexp{\ctxtapply{\ctxt{LTC}'}{\ntermren{\alpha}{\beta}{c}}}}{\fexp{u}}} \\
=                 & \fexp{\termname{\delta}{\termapp{\ctxtapply{\ctxt{LTC}'}{\ntermren{\alpha}{\beta}{c}}}{u}}} \\
\end{array} \] 

        \item $\termabs{x}{\ctxt{LTC}'}$. Then, \[
\begin{array}{lll}
                  & \fexp{\ntermren{\alpha}{\beta}{\termname{\delta}{\termabs{x}{\ctxtapply{\ctxt{LTC}'}{c}}}}} \\
=                 & \ntermren{\alpha}{\beta}{\termname{\delta}{\termabs{x}{\fexp{\ctxtapply{\ctxt{LTC}'}{c}}}}} \\
\ruleEqnewTheta   & \ntermren{\alpha}{\beta}{\termname{\delta}{\termabs{x}{\termcont{\delta'}{\termname{\delta'}{\fexp{\ctxtapply{\ctxt{LTC}'}{c}}}}}}} & \text{($\delta'$ fresh)} \\
\ruleEqnewRenPop  & \termname{\delta}{\termabs{x}{\termcont{\delta'}{\ntermren{\alpha}{\beta}{\termname{\delta'}{\fexp{\ctxtapply{\ctxt{LTC}'}{c}}}}}}} \\
=                 & \termname{\delta}{\termabs{x}{\termcont{\delta'}{\fexp{\ntermren{\alpha}{\beta}{\termname{\delta'}{\ctxtapply{\ctxt{LTC}'}{c}}}}}}} \\
\eqnew            & \termname{\delta}{\termabs{x}{\termcont{\delta'}{\fexp{\termname{\delta'}{\ctxtapply{\ctxt{LTC}'}{\ntermren{\alpha}{\beta}{c}}}}}}} & (\ih\!) \\
=                 & \termname{\delta}{\termabs{x}{\termcont{\delta'}{\termname{\delta'}{\fexp{\ctxtapply{\ctxt{LTC}'}{\ntermren{\alpha}{\beta}{c}}}}}}} \\
\ruleEqnewTheta   & \termname{\delta}{\termabs{x}{\fexp{\ctxtapply{\ctxt{LTC}'}{\ntermren{\alpha}{\beta}{c}}}}} \\
=                 & \fexp{\termname{\delta}{\termabs{x}{\ctxtapply{\ctxt{LTC}'}{\ntermren{\alpha}{\beta}{c}}}}} \\
\end{array} \] 

        \item $\termcont{\gamma}{\ctxt{LCC}'}$. Then, \[
\begin{array}{lll}
                  & \fexp{\ntermren{\alpha}{\beta}{\ntermren{\gamma}{\delta}{\ctxtapply{\ctxt{LCC}'}{c}}}} \\
=                 & \ntermren{\alpha}{\beta}{\ntermren{\gamma}{\delta}{\fexp{\ctxtapply{\ctxt{LCC}'}{c}}}} \\
\ruleEqnewRenRen  & \ntermren{\gamma}{\delta}{\ntermren{\alpha}{\beta}{\fexp{\ctxtapply{\ctxt{LCC}'}{c}}}} \\
=                 & \ntermren{\gamma}{\delta}{\fexp{\ntermren{\alpha}{\beta}{\ctxtapply{\ctxt{LCC}'}{c}}}} \\
\eqnew            & \ntermren{\gamma}{\delta}{\fexp{\ctxtapply{\ctxt{LCC}'}{\ntermren{\alpha}{\beta}{c}}}} & (\ih\!) \\
=                 & \ntermren{\gamma}{\delta}{\fexp{\ctxtapply{\ctxt{LCC}'}{\ntermren{\alpha}{\beta}{c}}}} \\
=                 & \fexp{\ntermren{\gamma}{\delta}{\ctxtapply{\ctxt{LCC}'}{\ntermren{\alpha}{\beta}{c}}}} \\
\end{array} \] 

        \item $\termsubs{x}{u}{\ctxt{LTC}'}$. Then, \[
\begin{array}{lll}
                  & \fexp{\ntermren{\alpha}{\beta}{\termname{\delta}{\termsubs{x}{u}{\ctxtapply{\ctxt{LTC}'}{c}}}}} \\
=                 & \ntermren{\alpha}{\beta}{\termname{\delta}{\termapp{(\termabs{x}{\fexp{\ctxtapply{\ctxt{LTC}'}{c}}})}{\fexp{u}}}} \\
\ruleEqnewTheta   & \ntermren{\alpha}{\beta}{\termname{\delta}{\termapp{(\termcont{\delta'}{\termname{\delta'}{\termabs{x}{\fexp{\ctxtapply{\ctxt{LTC}'}{c}}}}})}{\fexp{u}}}} & \text{($\delta'$ fresh)} \\
\ruleEqnewRenPush & \termname{\delta}{\termapp{(\termcont{\delta'}{\ntermren{\alpha}{\beta}{\termname{\delta'}{\termabs{x}{\fexp{\ctxtapply{\ctxt{LTC}'}{c}}}}}})}{\fexp{u}}} \\
=                 & \termname{\delta}{\termapp{(\termcont{\delta'}{\fexp{\ntermren{\alpha}{\beta}{\termname{\delta'}{\termabs{x}{\ctxtapply{\ctxt{LTC}'}{c}}}}}})}{\fexp{u}}} \\
\eqnew            & \termname{\delta}{\termapp{(\termcont{\delta'}{\fexp{\termname{\delta'}{\termabs{x}{\ctxtapply{\ctxt{LTC}'}{\ntermren{\alpha}{\beta}{c}}}}}})}{\fexp{u}}} & (\ih\!) \\
=                 & \termname{\delta}{\termapp{(\termcont{\delta'}{\termname{\delta'}{\fexp{\termabs{x}{\ctxtapply{\ctxt{LTC}'}{\ntermren{\alpha}{\beta}{c}}}}}})}{\fexp{u}}} \\
\ruleEqnewTheta   & \termname{\delta}{\termapp{\fexp{\termabs{x}{\ctxtapply{\ctxt{LTC}'}{\ntermren{\alpha}{\beta}{c}}}}}{\fexp{u}}} \\
=                 & \fexp{\termname{\delta}{\termsubs{x}{u}{\ctxtapply{\ctxt{LTC}'}{\ntermren{\alpha}{\beta}{c}}}}} \\
\end{array} \] 
      \end{itemize}

      \item $\ctxt{LCC} = \termrepl[\delta]{\gamma}{s'}{\ctxt{LCC}'}$. Then, \[
\begin{array}{lll}
                  & \fexp{\ntermren{\alpha}{\beta}{\termrepl[\delta]{\gamma}{s'}{\ctxtapply{\ctxt{LCC}'}{c}}}} \\
=                 & \ntermren{\alpha}{\beta}{\termname{\delta}{\termconc{(\termcont{\gamma}{\fexp{\ctxtapply{\ctxt{LCC}'}{c}}})}{\fexp{s'}}}} \\
\eqnew            & \termname{\delta}{\termconc{(\termcont{\gamma}{\ntermren{\alpha}{\beta}{\fexp{\ctxtapply{\ctxt{LCC}'}{c}}}})}{\fexp{s'}}} & (\text{L.~\ref{l:eqnew-stack}:\ref{l:eqnew-stack:ren}}) \\
=                 & \termname{\delta}{\termconc{(\termcont{\gamma}{\fexp{\ntermren{\alpha}{\beta}{\ctxtapply{\ctxt{LCC}'}{c}}}})}{\fexp{s'}}} \\
\eqnew            & \termname{\delta}{\termconc{(\termcont{\gamma}{\fexp{\ctxtapply{\ctxt{LCC}'}{\ntermren{\alpha}{\beta}{c}}}})}{\fexp{s'}}} & (\ih\!) \\
=                 & \fexp{\termrepl[\delta]{\gamma}{s'}{\ctxtapply{\ctxt{LCC}'}{\ntermren{\alpha}{\beta}{c}}}}
\end{array} \] 
    \end{itemize}

    \item $\ruleEqsigPopPop$. Then, $o =
    \termname{\alpha'}{\termabs{x}{\termcont{\alpha}{\termname{\beta'}{\termabs{y}{\termcont{\beta}{c}}}}}}$
    and $p =
    \termname{\beta'}{\termabs{y}{\termcont{\beta}{\termname{\alpha'}{\termabs{x}{\termcont{\alpha}{c}}}}}}$
    with $\beta \neq \alpha'$ and $\alpha \neq \beta'$. Moreover, by definition we have $\fexp{o} =
    \termname{\alpha'}{\termabs{x}{\termcont{\alpha}{\termname{\beta'}{\termabs{y}{\termcont{\beta}{\fexp{c}}}}}}}$
    and $\fexp{p} =
    \termname{\beta'}{\termabs{y}{\termcont{\beta}{\termname{\alpha'}{\termabs{x}{\termcont{\alpha}{\fexp{c}}}}}}}$. We conclude by $\ruleEqnewPopPop$.
  \end{enumerate}

  \item All the remaining cases are straightforward by using the \ih
  \qedhere
\end{itemize}
\end{proof}
